\documentclass{aa}  

\usepackage{graphicx}
\usepackage{txfonts}
\usepackage{subcaption}         
\usepackage{lscape}             
\usepackage{placeins}           
\usepackage{mathcomp}
\usepackage{bm}

\usepackage{refcount}

\begin{document}

   \title{Millimeter dust continuum and polarization in protoplanetary disks with scattering: A slab model}



   \author{Naoya Kitade\inst{1,2} 
        \and Akimasa Kataoka\inst{1,2} 
        }

   \institute{National Astronomical Observatory of Japan, National Institutes of Natural Sciences, 2-21-1 Osawa, Mitaka, Tokyo 181-8588, Japan\\
             \email{naoya.kitade.astro@gmail.com}
            \and Astronomical Science Program, Graduate Institute for Advanced Studies, SOKENDAI, 2-21-1 Osawa, Mitaka, Tokyo 181-8588, Japan\\ }


 
  \abstract
   {Millimeter continuum emission and self-scattering polarization from protoplanetary disks are widely used to constrain dust properties. Interpreting these observations requires practical prescriptions for the disk emission. However, only approximate formulae are available for the continuum emission, and no widely applicable formula has yet been established for the polarized emission.}
   {We aim (i) to assess the validity of commonly used analytic approximations for the (sub)millimeter continuum emission from protoplanetary disks, and (ii) to derive realistic prescriptions for the disk emission for both the continuum and the polarization.} 
   {We numerically solve the radiative transfer equation in an isothermal, constant-density plane-parallel slab, including dust absorption, emission, and self-scattering with full Stokes parameters. }
   {We find that commonly used analytic approximations for the continuum emission are systematically about 10 to $15\%$ lower than our numerical solutions. Consequently, SED analyses of (sub)millimeter observations that adopt these formulae are likely to overestimate the optical depth (and thus the disk mass) and the dust temperature, and underestimate the albedo (and thus altering the inferred constraints on grain size). We also provide empirical fitting formulae that reproduce our numerical results for the continuum emission and polarization fraction. These formulae will enable observational data analyses to be carried out more accurately and efficiently than with the conventional approaches.}
   {For the analysis of (sub)millimeter observations, we recommend using our new empirical formulae or interpolation of our numerical results, rather than commonly used approximations.}


    \authorrunning{Kitade \& Kataoka}
    \titlerunning{Millimeter dust continuum and polarization in protoplanetary disks with scattering: A slab model}
   \maketitle
   \nolinenumbers

\section{Introduction} \label{sec:introduction}
Constraining dust properties in protoplanetary disks is essential for revealing the planet formation process. Planets form in these disks, and their earliest stage begins with dust growth, yet the underlying physics remains uncertain, with key obstacles such as radial drift and fragmentation \citep[e.g.,][]{Adachi1976, Weidenschilling1977, Chokshi1993, Brauer2008}. Importantly, the impact of these obstacles depends on dust properties: the radial drift speed depends on dust size and porosity, while the fragmentation velocity depends on dust composition.

Constraints on dust properties have been derived from millimeter observations. The method used to derive these constraints differs between the optically thin and thick regimes. In the optically thin limit, the absorption opacity index $\beta$, defined by $\kappa_\mathrm{abs} \propto \nu^\beta$, can be constrained from the observed spectral index $\alpha$ of the continuum intensity ($I_\nu \propto \nu^\alpha$); in the Rayleigh–Jeans and optically thin limits, $\beta = \alpha - 2$ \citep[e.g.,][]{1983QJRAS..24..267H, 1991ApJ...381..250B, 2005ApJ...631.1134A}.
The inferred index $\beta$ is commonly used to constrain dust properties. For example, \citet{Ricci2010} constrained the maximum grain size to be millimeter-sized based on the inferred $\beta$, under the assumption of a power-law grain size distribution between a minimum and a maximum grain size. However, porosity is difficult to constrain from $\beta$ because $\beta$ is only weakly sensitive to porosity \citep[e.g.,][]{2014A&A...568A..42K}.

In the optically thick regime, multi-wavelength SED fitting has been used to constrain dust properties \citep[e.g.,][]{Carrasco-Gonzalez_2019, Macias2021, Ueda_2020, Ueda_2021, Ueda_2022, 2025ApJ...990..183U, Sierra_2021, Guerra-Alvarado2024}. In this regime, the simple relation $\beta = \alpha - 2$ no longer holds because scattering contributes significantly to the disk emission. Instead, multi-wavelength observations can be fitted using an approximate formula for disk emission that accounts for dust self-scattering \citep[e.g.,][]{1979rpa..book.....R, MIYAKE199320, 2018ApJ...869L..45B,  2019ApJ...876....7S, Carrasco-Gonzalez_2019, 2019ApJ...877L..18Z}. With this approach, the dust albedo and the extinction opacity can be inferred, thereby providing constraints on dust properties such as size, porosity, and composition \citep[e.g.,][]{MIYAKE199320, 2014A&A...568A..42K}. For example, \citet{Carrasco-Gonzalez_2019} performed SED fitting using ALMA Bands 7, 6, and 4 and VLA Ka- and Q-band observations of HL Tau, constraining the maximum grain size to be a few millimeters.

Millimeter polarization observations provide an additional way to constrain dust properties. Protoplanetary disks show polarized emission through various mechanisms, one of which is dust self-scattering \citep[e.g.,][]{2015ApJ...809...78K, 2016ApJ...820...54K, 2017ApJ...844L...5K, 2016MNRAS.456.2794Y, 2017MNRAS.472..373Y}. For compact grains, the millimeter polarization fraction produced by self-scattering is strongest when the maximum grain size $a_{\mathrm{max}}$ is on the order of the wavelength $\lambda$ divided by $2\pi$ \citep[$a_{\mathrm{max}} \sim \lambda / 2\pi$;][]{2015ApJ...809...78K}. Thus, a peak in the polarization fraction at a particular wavelength can be used to constrain the maximum grain size. In contrast, if dust grains are porous, millimeter polarization fraction remains significant over a wide range of wavelengths, allowing dust porosity to be constrained with multi-wavelength polarization observations \citep[e.g.,][]{2019ApJ...885...52T, 2023ApJ...953...96Z, 2024NatAs...8.1148U}. For example, \citet{2024NatAs...8.1148U} used ALMA Band 6 and 7 polarization observations of IM Lup to constrain the dust porosity to approximately 0.8.

Despite recent advances in deriving dust-properties from millimeter continuum and polarization observations, two key challenges remain.
The first challenge is that the validity of the approximate formulae commonly used in SED fitting of continuum emission has not been systematically tested. 
These formulae rely on assumptions such as the Eddington approximation, isotropic scattering, and the two-stream approximation \citep[e.g.,][]{MIYAKE199320, 2018ApJ...869L..45B, 2019ApJ...876....7S, Carrasco-Gonzalez_2019, 2019ApJ...877L..18Z}. However, these assumptions may not hold in realistic disk environments. As a result, these approximate formulae may be invalid under realistic disk conditions, which could cause dust properties inferred from SED fitting to deviate from the true values.

The second challenge is that a widely applicable formula for the emergent polarization from disks has not been developed. This makes it difficult to analyze multi-wavelength polarization data in a manner analogous to continuum SED fitting. As an alternative to formula-based analyses, Monte Carlo radiative transfer simulations can, in principle, be used to interpret multi-wavelength polarization observations. However, because the polarized intensity is typically only a few percent of the total intensity, achieving sufficient signal-to-noise in Monte Carlo simulations often requires a large number of photon packets and hence substantial computational time, making it impractical to explore a broad range of dust parameters. Accordingly, formulae that describe the emergent polarization from disks are needed to interpret multi-wavelength polarization observations.

In this study, we address these two challenges, (i) the untested validity of commonly used approximations for disk emission and (ii) the lack of general formulae for the emergent polarization, in the following three steps. First, we numerically solve the radiative transfer equation in a one-dimensional plane-parallel slab, including dust absorption, emission, and self-scattering with the full Stokes parameters. Second, to evaluate the validity of existing approximate formulae used for SED fitting, we compare these approximate formulae with our numerical solutions for the millimeter continuum. Finally, based on our numerical solutions, we develop a fitting formula for the emergent polarized emission from disks to enable efficient interpretation of multi-wavelength polarization observations. If we find that the commonly used approximations for the continuum emission are not valid, we will also develop an empirical fitting formula for the continuum emission.

This paper is organized as follows. In Section \ref{sec:method}, we present our method for numerically solving the radiative transfer equation. In Section \ref{sec:NumericalResults}, we show the numerical solutions. In Section \ref{sec:Comparison}, we compare our numerical solutions with existing approximate formulae commonly used for SED fitting. In Section \ref{sec:FittingFormulae}, we provide fitting formulae that reproduce our numerical solutions for both the emergent continuum and polarization. In section \ref{subsec:Mie}, we estimate Mie-scattering disk emission from our Rayleigh-scattering results. In Section \ref{sec:Discussion}, we discuss the validity of our numerical method, the approximations employed, and the application to observations. In Section \ref{sec:conclusion}, we conclude this paper.

\section{Method} \label{sec:method}
In this section, we describe our numerical method for solving the radiative transfer equation in a one-dimensional plane-parallel slab, including dust absorption, emission, and self‐scattering with full Stokes parameters.

\subsection{Setup}

\begin{figure}[htbp]
\resizebox{\hsize}{!}{\includegraphics{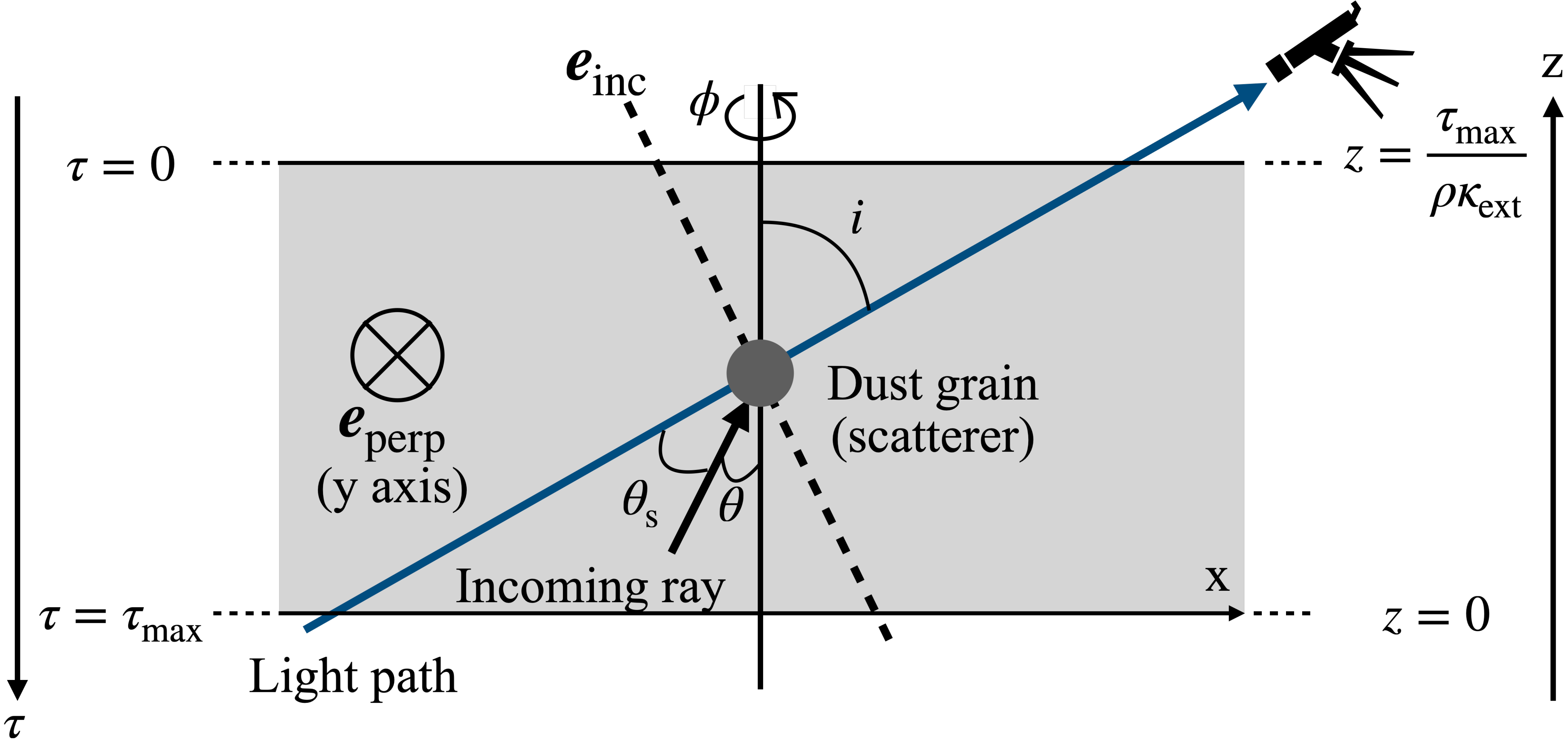}}
\caption{Schematic representation of the plane-parallel slab and the light path. The slab extends infinitely in the x- and y-direction, but has a finite extent in the z-direction. Note that $z$ and $\tau$ increase in opposite directions. The direction of the incoming ray at a dust grain is specified by $\theta$ and $\phi$; we set $\phi = 0$ in this figure. The dashed line represents $\boldsymbol{e}_\mathrm{inc}$, and $\boldsymbol{e}_\mathrm{perp}$ is perpendicular to the page.}
\label{fig:coordinate}
\end{figure}

\begin{figure}[htbp]
\resizebox{\hsize}{!}{\includegraphics{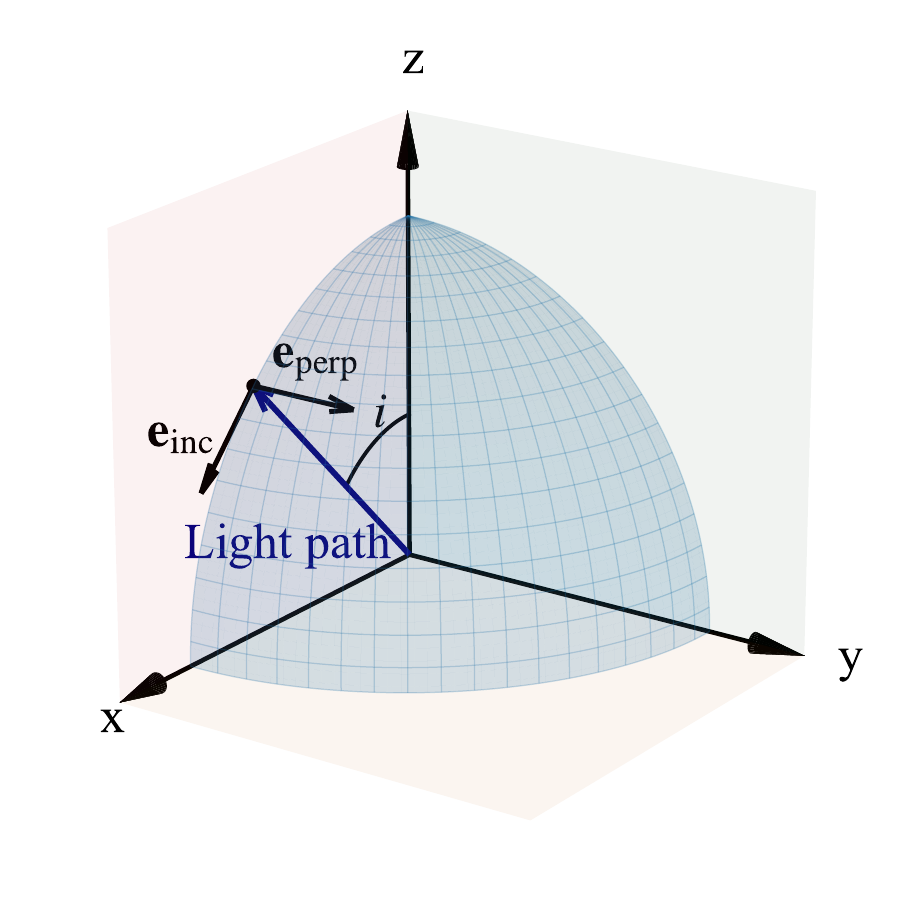}}
\caption{Schematic defining $\boldsymbol{e}_\mathrm{inc}$ and $\boldsymbol{e}_\mathrm{perp}$. $\boldsymbol{e}_\mathrm{perp}$ is defined as the direction perpendicular to both the $z$-axis and the line of sight, and $\boldsymbol{e}_\mathrm{inc}$ as the direction perpendicular to both the line of sight and $\boldsymbol{e}_\mathrm{perp}$.}
\label{fig:coordinate2}
\end{figure}

We model the system as an isothermal, constant–density, plane-parallel slab. We adopt a Cartesian coordinate system $(x,y,z)$ such that the $z$-axis is normal to the slab (see Fig. \ref{fig:coordinate}).
We define the emergent Stokes $Q$ such that $Q>0$ corresponds to polarization aligned with $\boldsymbol{e}_\mathrm{inc}$, and $Q<0$ corresponds to polarization aligned with $\boldsymbol{e}_\mathrm{perp}$.
$\boldsymbol{e}_\mathrm{perp}$ is defined as the direction perpendicular to both the $z$-axis and the line of sight, and $\boldsymbol{e}_\mathrm{inc}$ as the direction perpendicular to both the line of sight and $\boldsymbol{e}_\mathrm{perp}$ (see Fig. \ref{fig:coordinate2}). With this choice of coordinate system, the emergent Stokes $U$ vanishes identically.

We show the polarized radiative transfer equation for a plane-parallel slab and define the notation. The radiative transfer equation for the full Stokes parameters, including absorption, thermal emission, and self-scattering, can be written as
\begin{equation}
\begin{split}
    \label{eq:RT_fullStokes}
    \mu
    \frac{d}{d\tau} 
    \begin{pmatrix}
        I(\mu,\tau)\\
        Q(\mu,\tau)\\
        U(\mu,\tau)\\
        V(\mu,\tau)
    \end{pmatrix}
    &=
    \begin{pmatrix}
        I(\mu,\tau)\\
        Q(\mu,\tau)\\
        U(\mu,\tau)\\
        V(\mu,\tau)
    \end{pmatrix} 
    -
    (1 - \omega) 
    \begin{pmatrix}
        B(T)\\
        0\\
        0\\
        0
    \end{pmatrix} \\
    &- \frac{1}{\kappa_\mathrm{abs} + \kappa_\mathrm{sca}}
    \int 
    \bm{M} 
    \bm{Z}
    \bm{M'}
    \begin{pmatrix}
        I(\mu',\tau)\\
        Q(\mu',\tau)\\
        U(\mu',\tau)\\
        V(\mu',\tau)
    \end{pmatrix}
    d\Omega
    .
\end{split}
\end{equation}
Here, $\mu$ is defined as $\cos{i}$, where $i$ is the angle between the line of sight and the $z$-axis; hereafter we refer to $i$ as the inclination. We define $\mu>0$ for rays propagating toward increasing $z$. $\tau$ is the vertical extinction optical depth from the surface, defined by $d\tau = -\rho \kappa_{\mathrm{ext}}dz$, where $\rho$ is the dust mass density, and $\kappa_{\mathrm{ext}}$ is the mass extinction opacity.
Throughout this paper, optical depth refers to the extinction optical depth.
$I, Q, U, V$ are the Stokes parameters, which have the same units as the specific intensity. We omit the subscript $\nu$ on the Stokes parameters and the opacity. The dust albedo is $\omega \equiv \kappa_{\mathrm{sca}} / (\kappa_{\mathrm{abs}} + \kappa_{\mathrm{sca}})$, where $\kappa_{\mathrm{abs}}$ and $\kappa_{\mathrm{sca}}$ are the mass absorption and scattering opacities, respectively. $B(T)$ is the Planck function at dust temperature $T$. The matrices $\bm{M}$ and $\bm{M'}$ are the coordinate transformation matrices,
\begin{equation} \label{eq:Matrix_M}
\bm{M} \equiv 
    \begin{pmatrix}
        1 & 0 & 0 & 0 \\
        0 & \cos{2B_\mathrm{rot}} & \sin{2B_\mathrm{rot}} & 0 \\
        0 & - \sin{2B_\mathrm{rot}} & \cos{2B_\mathrm{rot}} & 0 \\
        0 & 0 & 0 & 1
    \end{pmatrix}
\end{equation}
and 
\begin{equation} \label{eq:Matrix_Mdash}
\bm{M'} \equiv 
    \begin{pmatrix}
        1 & 0 & 0 & 0 \\
        0 & \cos{2A_\mathrm{rot}} & \sin{2A_\mathrm{rot}} & 0 \\
        0 & - \sin{2A_\mathrm{rot}} & \cos{2A_\mathrm{rot}} & 0 \\
        0 & 0 & 0 & 1
    \end{pmatrix},
\end{equation}
where the rotation angles $A_\mathrm{rot}$ and $B_\mathrm{rot}$ are described in Appendix \ref{Appendix:derivation_RTeq}. The matrix $\bm{Z}$ is the scattering matrix. All scattering matrices are normalized such that $\int_{4\pi} Z_{11} d\Omega = \kappa_\mathrm{sca}$, where $Z_{11}$ is the (1,1) element of the scattering matrix $\bm{Z}$. $\mu'$ is defined as $\cos{\theta}$ where $\theta$ is the polar angle of the incoming ray at a scattering grain. Here, the incoming ray refers not to external background radiation, but to radiation emitted by dust grains within the slab surrounding the scattering grain. The differential solid angle is $d\Omega = \sin{\theta} d\theta d\phi$, where $\phi$ is defined as the azimuthal angle around the $z$-axis; $\phi = 0$ is along the positive direction of $x$-axis and increases clockwise when viewed from the positive direction of $z$-axis. 
Fig. \ref{fig:coordinate} shows the definitions of $i$, $\theta$, and the directions of $\tau$ and $z$.
The derivation of Eq. (\ref{eq:RT_fullStokes}) is presented in Appendix \ref{Appendix:derivation_RTeq}.

We mainly use the Rayleigh scattering matrix as the scattering matrix. The Rayleigh scattering matrix is given by 
\begin{equation} \label{eq:Rayleigh}
\bm{Z}^\mathrm{R} \equiv 
    \frac{3 \kappa_\mathrm{sca}}{8\pi}
    \begin{pmatrix}
    \frac{1}{2}(1+\cos^2{\theta_\mathrm{s}}) & \frac{1}{2}(\cos^2{\theta_\mathrm{s}} - 1) & 0 & 0 \\
    \frac{1}{2}(\cos^2{\theta_\mathrm{s}} - 1) & \frac{1}{2}(1+\cos^2{\theta_\mathrm{s}}) & 0 & 0 \\
    0 & 0 & \cos{\theta_\mathrm{s}} & 0 \\
    0 & 0 & 0 & \cos{\theta_\mathrm{s}}
    \end{pmatrix},
\end{equation}
where $\theta_\mathrm{s}$ is the scattering angle, defined as the angle between the incoming ray and the scattered ray. 
Although dust grains in protoplanetary disks may exhibit Mie scattering at millimeter wavelengths, we use this scattering matrix to illustrate the underlying physics in a simplified framework.
We discuss the numerical results obtained using the Mie scattering matrix in Section \ref{subsec:Mie}. 

We assume no external incident background radiation at the slab boundaries. Accordingly, we impose $(I, Q, U, V) = (0,0,0,0)$ at $\tau = \tau_\mathrm{max}$ for incident radiation with $\mu > 0$ and $\tau = 0$ for incident radiation with $\mu < 0$, where $\tau_{\mathrm{max}}$ denotes the total vertical optical depth of the slab.

We also assume that the polarization arises solely from dust self-scattering, neglecting any contribution from grain alignment or other mechanisms.

\subsection{Numerical method}
To compute Eq. (\ref{eq:RT_fullStokes}) numerically, we discretize the radiation field in $(\tau, \mu, \phi)$ on a grid.
The vertical optical depth from the surface $\tau$ is discretized uniformly over $0\leq\tau\leq\tau_\mathrm{max}$ with spacing $d\tau = \tau_\mathrm{max}/\mathrm{N_{\tau}}$. The direction cosine $\mu$ and $\mu'$ are discretized using $\mathrm{N}_\mu$ Gauss–Legendre nodes. The azimuthal angle $\phi$ $(0\leq\phi<2\pi)$ is mapped to $\phi'$ $(-1\leq\phi'<1)$ and discretized using $\mathrm{N}_{\phi'}$ Gauss-Legendre nodes. These discretizations allow us to evaluate the scattering integral in Eq. (\ref{eq:RT_fullStokes}) using Gauss-Legendre quadrature.

Since evaluating the scattering term requires the full radiation field inside the slab, we use the lambda iteration method to numerically solve Eq. (\ref{eq:RT_fullStokes}) \citep{1978stat.book.....M}. 
We first set appropriate initial guesses for the Stokes parameters in the scattering term. Using the initial guesses, we numerically solve the right-hand side of Eq. (\ref{eq:RT_fullStokes}) and obtain the full radiation field inside the slab. We then recompute the scattering term from the updated radiation field and solve Eq. (\ref{eq:RT_fullStokes}) again to update the radiation field. This procedure is repeated until the radiation field converges. See also \citet{10.1093/mnras/stac753}.

As the initial guess for the Stokes parameters in the lambda iteration method, we adopt the semi-analytic solution.
For the Stokes $I$ component, we adopt
\begin{equation} \label{eq:analyticalIntensity}
\begin{split}
I &= \frac{\omega B}{e^{-\sqrt{3}\epsilon \tau_{\mathrm{max}}} (\epsilon - 1) - (\epsilon + 1) } \\
&\times
\left(
\frac{e^{-\sqrt{3}\epsilon\tau} - e^{(-\sqrt{3}\epsilon - \frac{1}{\mu}) \tau_{\mathrm{max}} + \frac{\tau}{\mu}}}{\sqrt{3}\epsilon \mu + 1} 
- \frac{e^{\sqrt{3}\epsilon (\tau - \tau_{\mathrm{max}})} - e^{\frac{\tau - \tau_{\mathrm{max}}}{\mu}}}{\sqrt{3} \epsilon \mu - 1}
\right) \\
&+
B 
(1 - e^{\frac{\tau - \tau_{\mathrm{max}}}{\mu}}),
\end{split}
\end{equation}
where $\epsilon$ is defined as $\sqrt{1-\omega}$.
This equation is obtained by solving the isotropic-scattering radiative transfer equation,
\begin{equation} \label{eq:RT_iso}
\mu \frac{dI}{d\tau} = I - (1-\omega)B - \omega J,
\end{equation} 
using the mean intensity $J$ given by 
\begin{equation} \label{eq:meanIntensity}
J = B \left( 1 + \frac{e^{-\sqrt{3(1-\omega)} \tau} + e^{\sqrt{3(1-\omega)}(\tau - \tau_{\mathrm{max}})}}{e^{-\sqrt{3(1-\omega)}\tau_{\mathrm{max}}} (\sqrt{1-\omega} - 1) - (\sqrt{1-\omega} + 1)} \right)  
\end{equation}
from \citet{MIYAKE199320}.
We set the initial guesses for the Stokes parameters $Q, U,$ and $V$ to zero.

We describe the parameter ranges and how we choose the discretization.
We explore $0.01 \leq \tau_\mathrm{max} \leq 15$, $0 < \mu < 1$, and $0 \leq \omega \leq 0.9$. Although small dust grains in the Rayleigh‐scattering regime do not reach such high albedos, we adopt these values to investigate the effect of scattering on the Stokes parameters.
$\mathrm{N}_\tau, \mathrm{N}_\mu,$ and $\mathrm{N}_{\phi'}$ are chosen such that, at $i=45 \tcdegree$, the relative difference in the Stokes parameters between $(\mathrm{N}_\tau, \mathrm{N}_\mu, \mathrm{N}_{\phi'})$ and $(2\mathrm{N}_\tau, 2\mathrm{N}_\mu, 2\mathrm{N}_{\phi'})$ is below 0.2\%. Note that the computed values may become unreliable for inclination angles approaching 90 degrees. This is primarily because, near edge-on views, the light path length through the medium increases, requiring a larger number of $\mathrm{N}_\tau$. In addition, the contribution from incident radiation with $\mu' \sim 0$ (grazing incidence) becomes significant, requiring a larger angular quadrature $\mathrm{N}_\mu$.
The specific values for $\mathrm{N}_\tau$, $\mathrm{N}_\mu$ and $\mathrm{N}_{\phi'}$ are listed in Table \ref{tab:Ntau}. These values are kept fixed for all $\omega$.
For the iterative solution, the number of iterations $\mathrm{N_{iter}}$ is determined such that the relative error between $\bm{S}_\mathrm{N}$ and $\bm{S}_{\mathrm{N+1}}$ is less than 0.01\%. The specific $\mathrm{N_{iter}}$ value is also listed in Table \ref{tab:Nite}.

\begin{table}
\caption{}
\label{tab:Ntau}
\centering
\begin{tabular}{cccc}
$\tau_{\mathrm{max}}$ & $\mathrm{N}_\tau$ & $\mathrm{N}_{\mu}$ & $\mathrm{N}_{\phi'}$ \\
\hline \hline
0.01 & 700 & 3200 & 16 \\
0.03 & 700 & 800 & 16 \\
0.05 & 700 & 800 & 16 \\
0.1 & 700 & 400 & 16 \\
0.2 & 700 & 400 & 16 \\
0.3 & 700 & 400 & 16 \\
0.4 & 700 & 400 & 16 \\
0.5 & 700 & 100 & 16 \\
0.6 & 840 & 100 & 16 \\
0.7 & 980 & 100 & 16 \\
0.8 & 1120 & 100 & 16 \\
0.9 & 1260 & 100 & 16 \\
1.0 & 1400 & 100 & 16 \\
1.2 & 1680 & 100 & 16 \\
1.4 & 1960 & 100 & 16 \\
1.6 & 2240 & 100 & 16 \\
1.8 & 2520 & 100 & 16 \\
2.0 & 2800 & 100 & 16 \\
3.0 & 4200 & 100 & 16 \\
4.0 & 5600 & 100 & 16 \\
6.0 & 8400 & 100 & 16 \\
8.0 & 11200 & 100 & 16 \\
10.0 & 14000 & 100 & 16 \\
12.0 & 16800 & 100 & 16 \\
15.0 & 21000 & 100 & 16 \\
\hline
\end{tabular}
\tablefoot{$\tau_\mathrm{max}$ dependence of the grid point counts $\mathrm{N}_\tau, \mathrm{N}_\mu$ and $\mathrm{N_\phi}$ used in the numerical calculations. $\mathrm{N}_{\mu'}$ is set equal to $\mathrm{N}_\mu$.}
\end{table}

\begin{table}
\caption{}
\label{tab:Nite}
\centering
\begin{tabular}{ccc}
$\omega$ & $\mathrm{N_{iter}} (\tau_\mathrm{max} \leq 10)$ & $\mathrm{N_{iter}} (\tau_\mathrm{max} \geq 12)$ \\
\hline \hline 
0.0 & 20 & 40\\
0.1 & 20 & 40\\
0.2 & 20 & 40\\
0.3 & 20 & 40\\
0.4 & 20 & 40\\
0.5 & 20 & 40\\
0.6 & 20 & 40\\
0.7 & 20 & 40\\
0.8 & 30 & 50\\
0.9 & 30 & 50\\
\hline
\end{tabular}
\tablefoot{$\omega$ dependence of the number of iterations $\mathrm{N_{iter}}$ used in the numerical calculations.}
\end{table}

\section{Results} \label{sec:NumericalResults}
In this section, we present radiative-transfer results for plane-parallel slabs by numerically solving Eq. (\ref{eq:RT_fullStokes}). In Section \ref{subsec:Emergent_Stokes}, we first summarize how Stokes parameters of emergent intensity depend on the total vertical optical depth $\tau_\mathrm{max}$, the dust albedo $\omega$, and the inclination $i$. In section \ref{subsec:inseide_slab}, we discuss the physical origin of the scattering effect. Specifically, we explain the scattering-induced surface attenuation of Stokes $I$ (the surface-layer effect) seen in Section \ref{subsec:Emergent_Stokes}, and demonstrate that scattering of the incoming polarized component can make a non-negligible contribution to the emergent polarization. Owing to symmetry in a plane-parallel slab, Stokes $U$ and $V$ vanish and are not discussed. Therefore, the absolute value of Stokes $Q$ represents the polarized intensity. The $\tau$-dependence of the Stokes parameters inside the slab is presented in Appendix \ref{sec:Variation_inside_slab}. Analytic approximations for the Stokes parameters in the optically thin limit and their comparison with the numerical results are given in Appendix~\ref{Appendix:approximation}.

\subsection{Stokes parameters of emergent intensity}
\label{subsec:Emergent_Stokes}
\subsubsection{Dependence on optical depth and albedo}
\label{subsubsec:Dependence_on_albedo}

\begin{figure}[htbp]
\resizebox{\hsize}{!}{\includegraphics{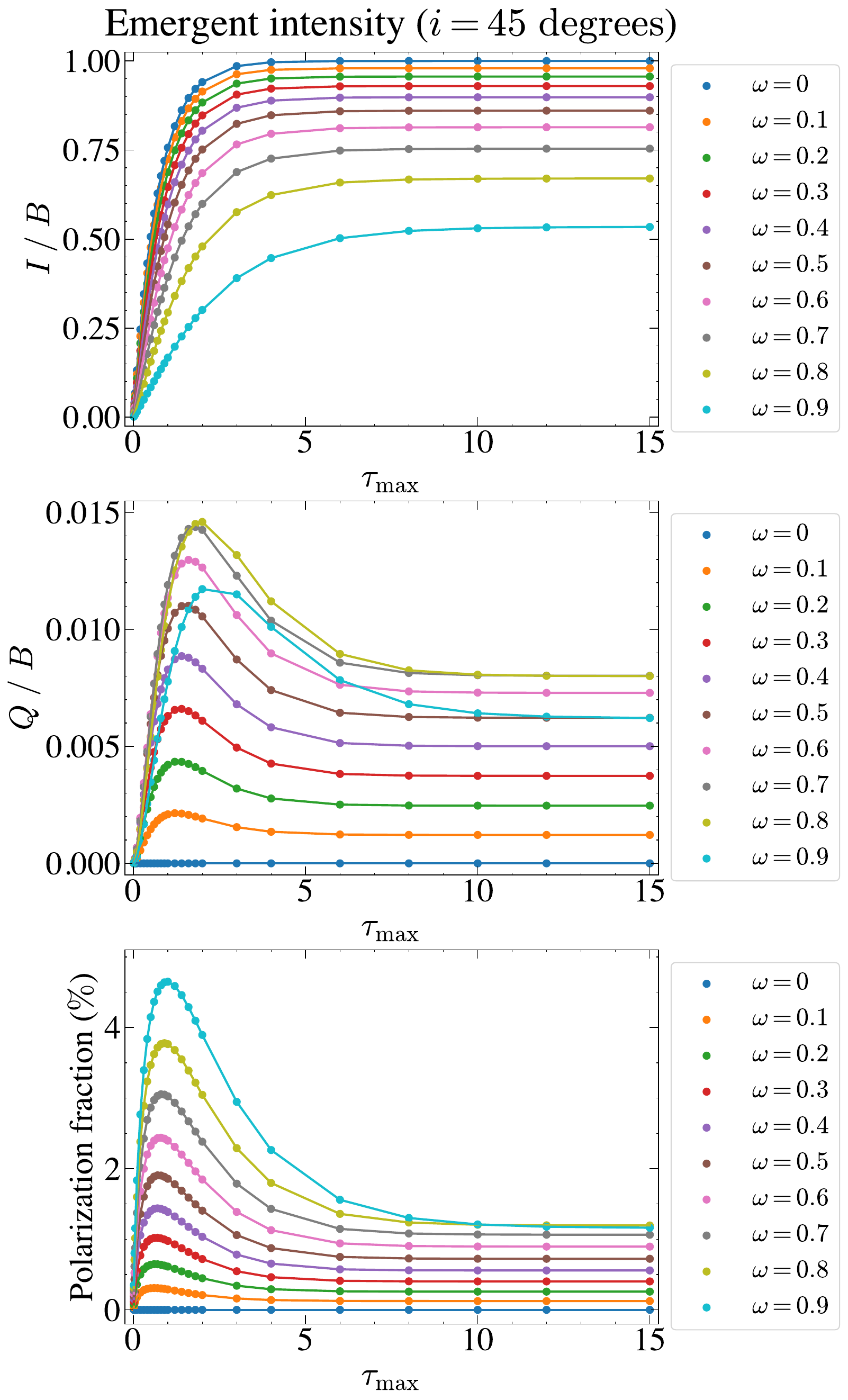}}
\caption{Emergent Stokes $I$, Stokes $Q$, and polarization fraction ($\equiv Q/I$) as functions of the total vertical optical depth $\tau_\mathrm{max}$ for a plane-parallel slab at $i=45\tcdegree$, obtained by numerically solving Eq. (\ref{eq:RT_fullStokes}). Stokes $I$ and $Q$ are normalized by the Planck function, $B$. Each curve represents a different albedo value, $\omega$.}
\label{fig:IQPF_inc45}
\end{figure}

In Section \ref{subsubsec:Dependence_on_albedo}, we investigate how Stokes $I$, $Q$, and the polarization fraction ($\equiv Q/I$) of the emergent intensity depend on the total optical depth $\tau_\mathrm{max}$ and dust albedo $\omega$. Fig. \ref{fig:IQPF_inc45} shows these results for a fiducial inclination of $i = 45 \tcdegree$.
In the top panel of Fig. \ref{fig:IQPF_inc45}, Stokes $I$ increases with $\tau_\mathrm{max}$ and saturates for $\tau_\mathrm{max} \gg 1$, and decreases as $\omega$ increases. The decrease of Stokes $I$ with increasing $\omega$ in the optically thick regime is consistent with the scattering-induced attenuation effects discussed by \citet{2019ApJ...877L..18Z}. The detailed mechanism underlying this effect is described later in Section \ref{subsubsec:surface-layer}.

In the middle panel of Fig. \ref{fig:IQPF_inc45}, Stokes $Q$ increases with $\tau_\mathrm{max}$ for $\tau_\mathrm{max} \lesssim 1$, peaks around $\tau_\mathrm{max} \simeq 1-2$, and then decreases, approaching an asymptotic value for $\tau_\mathrm{max} \gg 1$ for all values of $\omega$. This behavior can be understood as follows. The strength of self-scattering polarization (here represented by Stokes $Q$) is set by (i) the strength of the radiation field and (ii) the anisotropy of the radiation field \citep{2015ApJ...809...78K, 2016MNRAS.456.2794Y}. When the slab is too optically thin, the radiation field is too weak, and Stokes $Q$ remains small. When the slab becomes too optically thick, the radiation field approaches isotropy, and Stokes $Q$ decreases. Around $\tau_\mathrm{max} \simeq 1 - 2$, the radiation field is not weak and anisotropic. As a result, Stokes $Q$ attains its maximum around $\tau_\mathrm{max} \simeq 1 - 2$. As a function of $\omega$, no clear monotonic trend is seen. This is because scattering polarization (Stokes $Q$) increases with both the radiation field (Stokes $I$) and scattering efficiency $\omega$, while Stokes $I$ decreases with increasing $\omega$ in the upper panel of Fig. \ref{fig:IQPF_inc45}, leading to a complex dependence of Stokes $Q$ on $\omega$.

In the bottom panel of Fig. \ref{fig:IQPF_inc45}, the polarization fraction peaks at $\tau_\mathrm{max} \simeq 1$ and then decreases toward an asymptotic value for $\tau_\mathrm{max} \gg 1$. The value of $\tau_\mathrm{max}$ at which the polarization fraction peaks is provided in Section \ref{subsec:fitting_PF}. This $\tau_\mathrm{max}$-trend is consistent with \citet{2017MNRAS.472..373Y}. As a function of $\omega$, for $\tau_\mathrm{max} < 10$, the polarization fraction increases with $\omega$. In this regime, since Stokes $Q$ generally increases with both Stokes $I$ and $\omega$ and the polarization fraction is defined as $Q/I$, the polarization fraction tends to increase with $\omega$. However, for $\tau_\mathrm{max} > 10$, the dependence on $\omega$ is not monotonic. This deviation arises because, in the very optically thick regime, scattering not only produces polarization, but also attenuates scattered light.

We have verified that the qualitative trends persist at other inclinations. Additional results for $i=0\tcdegree, 15\tcdegree, 30\tcdegree, 60\tcdegree$ and $75\tcdegree$ are provided in Appendix \ref{Appendix:AddInc}.

\subsubsection{Dependence on inclination and albedo}
\label{subsub:Dependence_on_inclination}

\begin{figure}[htbp]
\resizebox{\hsize}{!}{\includegraphics{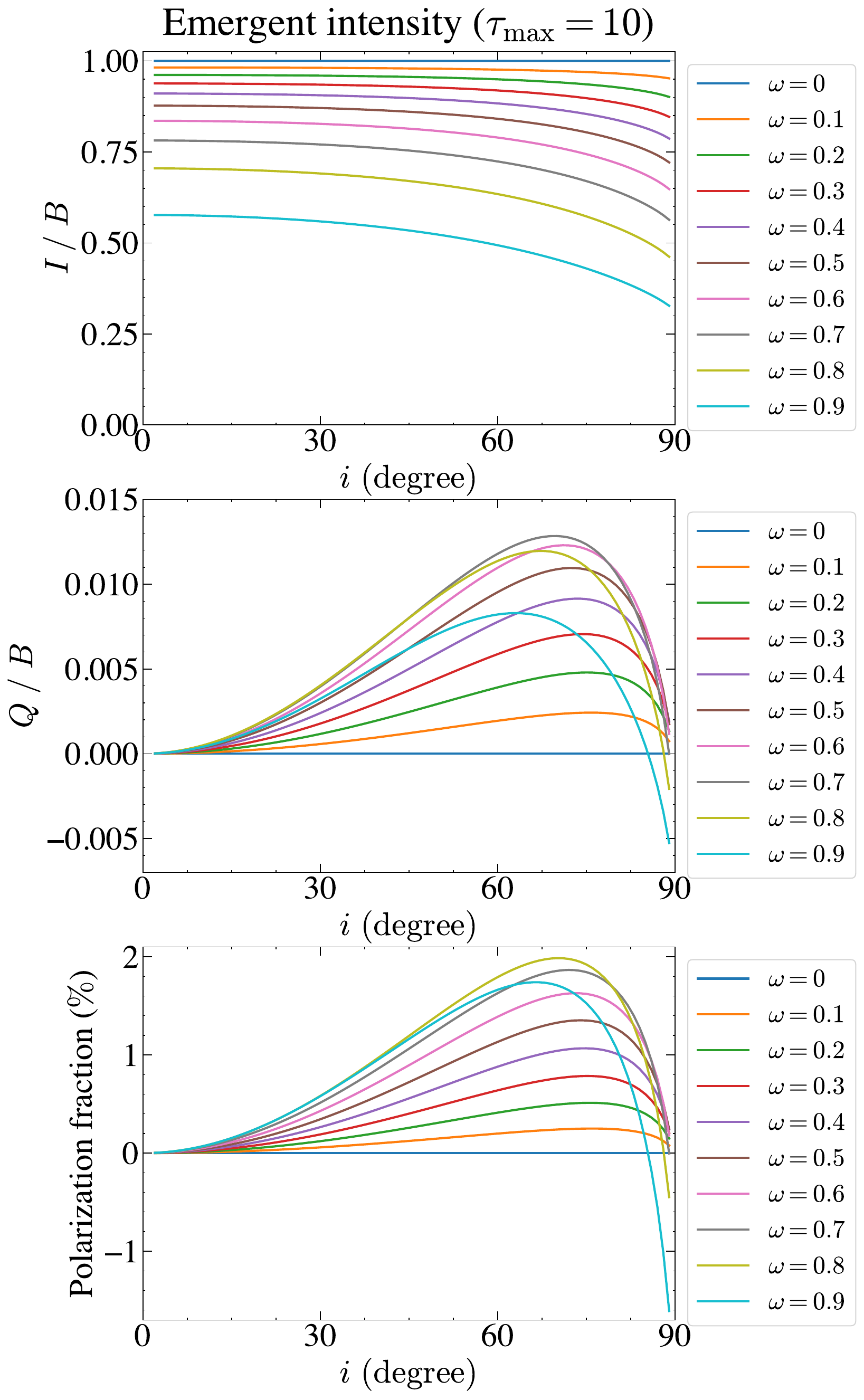}}
\caption{Stokes $I$, Stokes $Q$, and polarization fraction ($\equiv Q/I$) of the emergent intensity as functions of disk inclination for $\tau_\mathrm{max}=10$, obtained by numerically solving Eq. (\ref{eq:RT_fullStokes}). Stokes $I$ and $Q$ are normalized by the Planck function, $B$. Here, the polarization fraction is defined as $100 \times Q/I$ (in percent), without taking the absolute value.
Each curve represents a different albedo value, $\omega$. While the numerical results are discrete in $\mu$, the number of data points is large, so we display them as lines rather than dots here.}
\label{fig:Inclination_dependence}
\end{figure}

In Section \ref{subsub:Dependence_on_inclination}, we investigate how Stokes $I$, $Q$, and the polarization fraction of the emergent intensity depend on inclination $i$ and dust albedo $\omega$. Fig. \ref{fig:Inclination_dependence} shows these results for $\tau_\mathrm{max}=10$. In the top panel of Fig. \ref{fig:Inclination_dependence}, for $\omega=0$, the emergent Stokes $I$ equals the Planck function regardless of $i$. This is because the slab remains optically thick for any inclination. Once scattering is included ($\omega > 0$), the emergent Stokes $I$ decreases with increasing inclination for all $\omega$. This trend is consistent with \citet{2019ApJ...877L..18Z}. The detailed mechanism by which Stokes $I$ decreases with increasing inclination in the optically thick regime is described in Section \ref{subsubsec:surface-layer}. As a function of $\omega$, the emergent Stokes $I$ decreases with increasing $\omega$. This mechanism is also described in Section \ref{subsubsec:surface-layer}.

In the middle panel of Fig. \ref{fig:Inclination_dependence}, Stokes $Q$ increases with inclination, except near $90 \tcdegree$. This increase can also be understood in terms of two factors, as in Section \ref{subsubsec:Dependence_on_albedo}: the strength of the radiation field and its anisotropy. Stokes $I$ decreases with increasing inclination $i$, but the variation is small, as shown in the top panel of Fig. \ref{fig:Inclination_dependence}. Instead, a higher $i$ produces a more anisotropic radiation field, which enhances Stokes $Q$ \citep{2016MNRAS.456.2794Y, 2017MNRAS.472..373Y} (see Appendix \ref{subsec:simple_app}). However, for inclinations close to $90 \tcdegree$, Stokes $Q$ decreases with increasing inclination. This is because the path length becomes long enough that attenuation of the scattered light overwhelms the anisotropy effect, causing Stokes $Q$ to decline \citep{2017MNRAS.472..373Y}. As a function of $\omega$, Stokes $Q$ remains non-monotonic, similar to  the complex behavior seen in the middle panel of Fig. \ref{fig:IQPF_inc45}.

\begin{figure}[thbp]
\resizebox{\hsize}{!}{\includegraphics{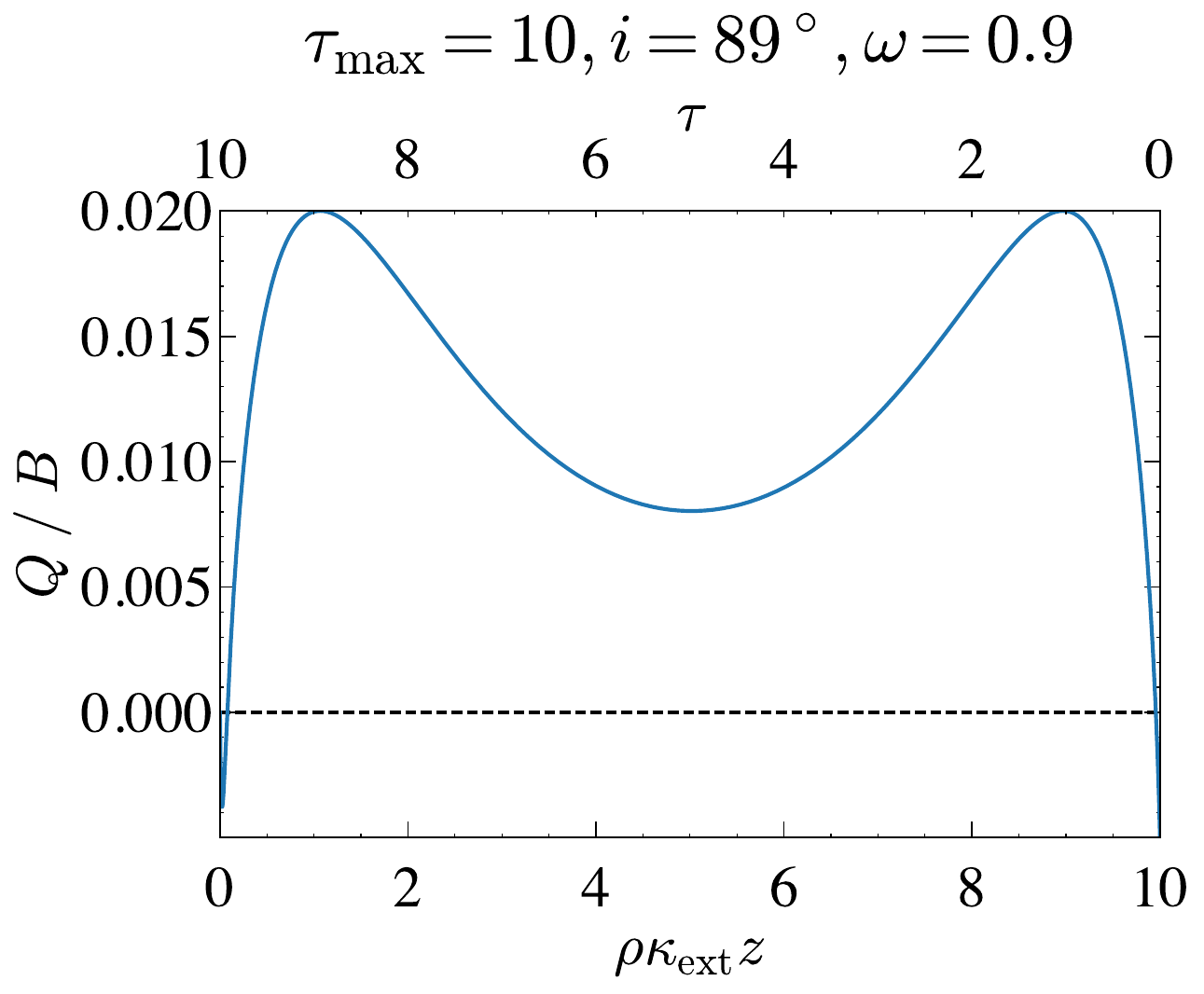}}
\caption{Stokes $Q$ as a function of $z$ and the optical depth from the surface $\tau$, obtained by numerically solving Eq. (\ref{eq:RT_fullStokes}). The slab is inclined at $89\tcdegree$ with $\omega=0.9$ and $\tau_\mathrm{max}=10$. Stokes $Q$ is normalized by the Planck function, $B$. The lower axis shows $\rho \kappa_\mathrm{ext} z$, while the upper axis shows $\tau$. $\tau=0$ corresponds to the slab surface facing the observer.}
\label{fig:Q_i89_omega0_9}
\end{figure}

However, near $i \sim 90 \tcdegree$ and for $\omega \geq 0.8$, Stokes $Q$ becomes negative; this behavior cannot be explained by attenuation alone. The negative Stokes $Q$ indicates that an unusual anisotropy of the radiation field dominates the polarized emission in this regime.
By our convention, positive Stokes $Q$ is along $\boldsymbol{e}_\mathrm{inc}$, so that scattering of radiation from a direction perpendicular to the sheet (along $\mathbf{e}_\mathrm{prep}$) yields positive Stokes $Q$, whereas scattering of radiation from a direction parallel to the sheet produces negative Stokes $Q$. Near the surface ($\tau \sim 0$ and $\tau \sim \tau_\mathrm{max}$), a high albedo strongly attenuates the radiation perpendicular to the sheet through scattering, while the radiation parallel to the sheet remains relatively strong; consequently, the polarization induced by irradiation parallel to the sheet dominates, and Stokes $Q$ becomes negative. Moreover, at larger inclination angles, the path length through this surface layer, where negative Stokes $Q$ is generated, increases. Fig. \ref{fig:Q_i89_omega0_9} shows the $\tau$-dependence of Stokes $Q$ within the slab for $i=89\tcdegree$, $\omega=0.9$ and $\tau_\mathrm{max}=10$, demonstrating that Stokes $Q$ becomes negative near $\tau\sim0$ and $\tau \sim 10$. Thus, for nearly edge-on viewing angles and high albedo, the negative contribution near the surface is amplified, resulting in a negative emergent Stokes $Q$.

In the bottom panel of Fig. \ref{fig:Inclination_dependence}, the $i$-dependence of the polarization fraction shows a similar trend to Stokes $Q$. As for $\omega$, the polarization fraction does not increase monotonically with $\omega$, similar to the very optically thick regime in the fiducial run at $45\tcdegree$ shown in the bottom panel of Fig. \ref{fig:IQPF_inc45}.

\begin{figure}[htbp]
\resizebox{\hsize}{!}{\includegraphics{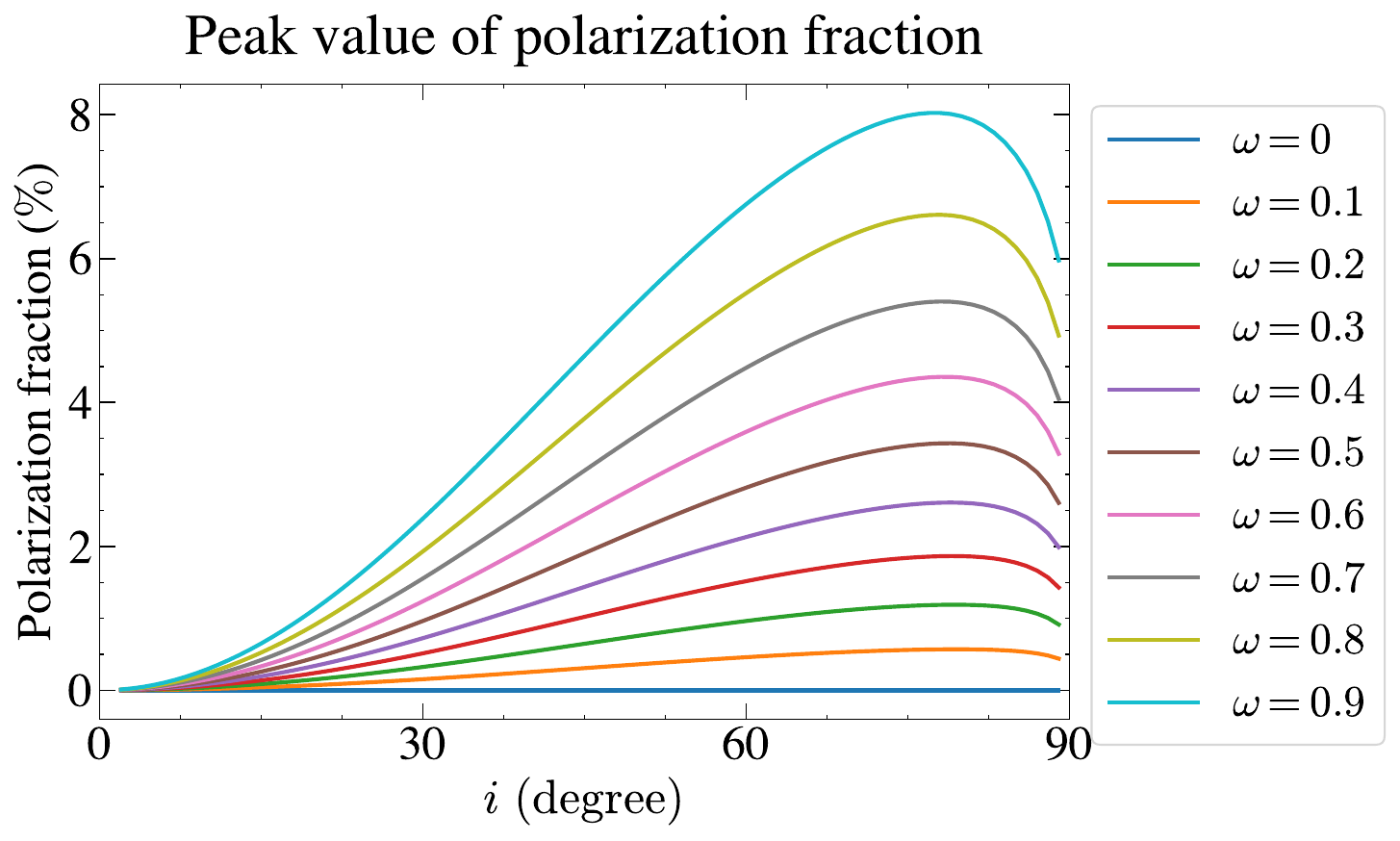}}
\caption{Peak emergent polarization fraction as a function of disk inclination $i$ and dust albedo $\omega$ for slabs of varying total optical depth. Each curve represents a different $\omega$.}
\label{fig:PF_peak}
\end{figure}

Additionally, Fig. \ref{fig:PF_peak} illustrates how the peak polarization fraction of the emergent intensity depends on $i$ and $\omega$ for slabs of varying total optical depth. 
The peak polarization fraction increases with inclination, except near $90\tcdegree$, and also increases with $\omega$. As shown in the bottom panel of Fig. \ref{fig:IQPF_inc45}, the polarization fraction peaks at $\tau_\mathrm{max} \sim 1$; thus, this plot can also be interpreted as showing the $i$ and $\omega$-dependence of the polarization fraction at $\tau_\mathrm{max} \sim 1$.

\subsection{Physical origin of scattering effects}
\label{subsec:inseide_slab}
\subsubsection{Surface-layer effect}
\label{subsubsec:surface-layer}

\begin{figure}[htbp]
\resizebox{\hsize}{!}{\includegraphics{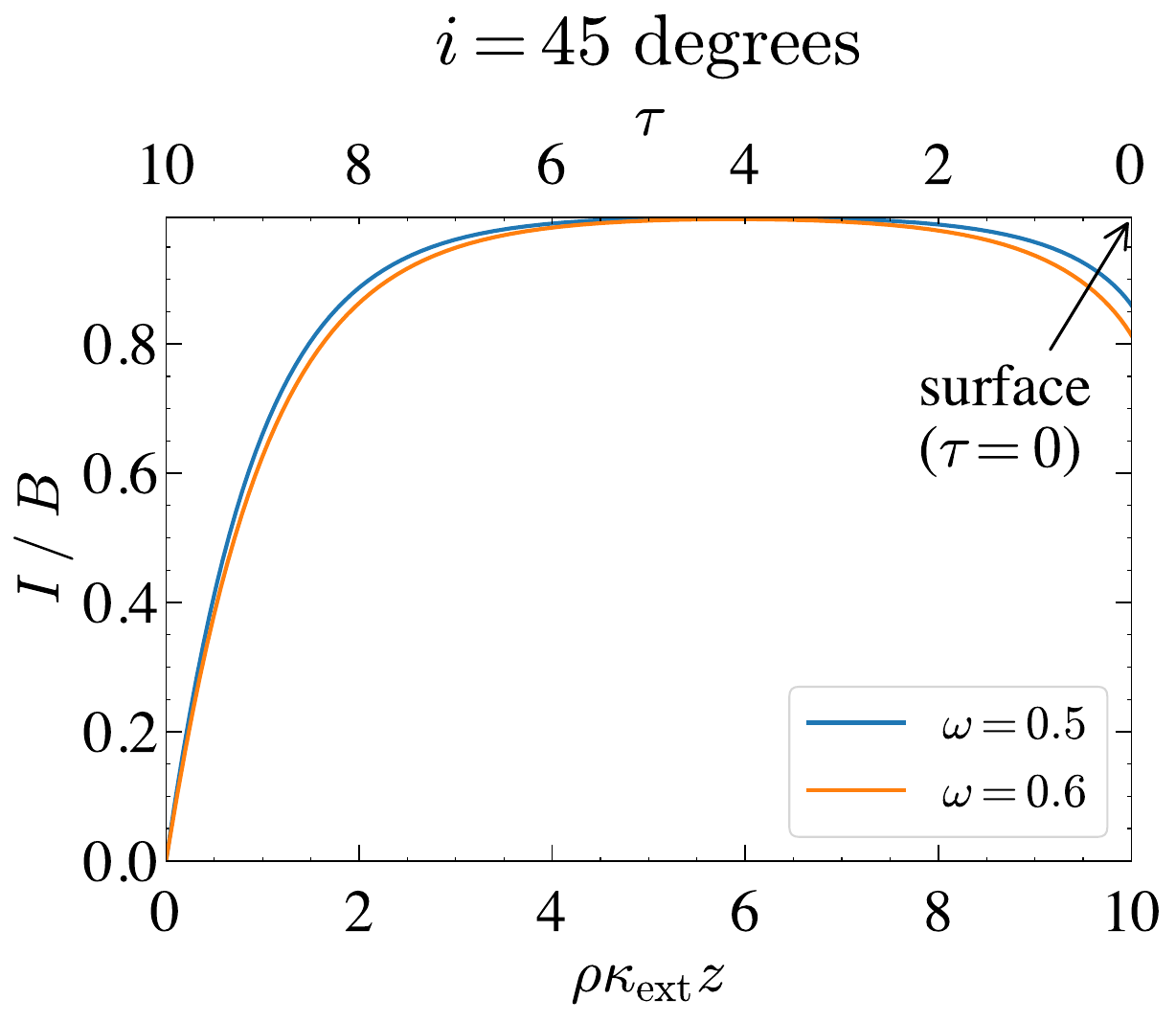}}
\caption{Stokes $I$ as a function of $z$ and the optical depth from the surface $\tau$, obtained by numerically solving Eq. (\ref{eq:RT_fullStokes}). The slab is inclined at 45°; results are shown for $\omega=0.5$ and 0.6. Stokes $I$ is normalized by the Planck function, $B$. The lower axis shows $\rho \kappa_\mathrm{ext} z$, while the upper axis shows $\tau$. $\tau=0$ corresponds to the slab surface facing the observer.}
\label{fig:I_inc45_omega0_5_0_6_tau10}
\end{figure}

The “surface-layer effect,” which is the attenuation of Stokes $I$ in the slab’s surface layer facing the observer, explains the scattering-induced reduction of the emergent Stokes $I$, as seen in Section \ref{subsec:Emergent_Stokes}. In section \ref{subsubsec:surface-layer}, we define the “surface” as the observer-facing boundary at $\tau=0$.
Fig. \ref{fig:I_inc45_omega0_5_0_6_tau10} shows the $\tau$-dependence of Stokes $I$ within the slab for $\tau_\mathrm{max}=10$. Stokes $I$ decreases from its midplane value toward the surface layer ($\tau\sim0$), which is the surface-layer effect. The mechanism of this effect is explained as follows. Around $\tau \sim 4$, Stokes $I$ along rays directed toward the surface is $\sim B$, whereas the intensity incident from the surface side (i.e., toward increasing $\tau$) is $< B$, as can be seen from the fact that Stokes $I$ already falls below $B$ around $\tau \sim 6$ in Fig. \ref{fig:I_inc45_omega0_5_0_6_tau10}.
We then consider the radiative transfer equation for Stokes $I$
\begin{equation}
\begin{split}
\label{eq:RTeq_I}
\mu\frac{dI}{d\tau} = I - S
\end{split}
\end{equation}
where, using the Rayleigh scattering matrix, the source function is given by
\begin{equation} \label{eq:SourceFunction}
S \equiv (1-\omega)B +\frac{3}{8\pi} \omega \int \frac{1}{2}(1+\cos^2{\theta_\mathrm{s}})I d\Omega,
\end{equation}
neglecting the contribution of the polarized component of the incoming intensity (the Stokes $Q$, $U$, and $V$ components in the scattering term of Eq. (\ref{eq:RT_fullStokes})). This approximation is justified in Section \ref{subsub:Scattering_polarized_light}. 
Because the incoming intensity from the surface is smaller than  $B$, the scattering term is smaller than the value obtained by replacing $I$ with $B$, $3/(8\pi)\times\omega \int \frac{1}{2}(1+\cos^2{\theta_\mathrm{s}})B d\Omega = \omega B$, and thus Eq. (\ref{eq:SourceFunction}) yields $S < (1-\omega)B+\omega B=B$. Using $I \sim B$ in Eq. (\ref{eq:RTeq_I}), we obtain $dI/d\tau > 0$. Since radiation propagates toward decreasing $\tau$, $dI/d\tau > 0$ means that Stokes $I$ decreases toward the surface and approaches $S$. Moreover, since $S$ decreases as $\tau$ decreases from $\sim 4$ toward the surface, Stokes $I$ continues to decrease. This is the physical origin of the surface-layer effect.

The surface-layer effect accounts for the $\omega$-dependence of the emergent Stokes $I$ shown in Fig. \ref{fig:IQPF_inc45}. This is because the surface-layer effect becomes stronger with increasing $\omega$, as seen in Fig. \ref{fig:I_inc45_omega0_5_0_6_tau10}, which compares the $\tau$-dependence of Stokes $I$ for $\omega=0.5$ and $\omega=0.6$. This trend can be understood as follows: as $\omega$ increases, the source function $S$ near the surface tends to decrease and the positive gradient $dI/d\tau$ increases (Eq. (\ref{eq:RTeq_I})), causing Stokes $I$ to fall further below $B$ toward the surface. Consequently, the emergent Stokes $I$ decreases with increasing $\omega$.

\begin{figure}[thbp]
\resizebox{\hsize}{!}{\includegraphics{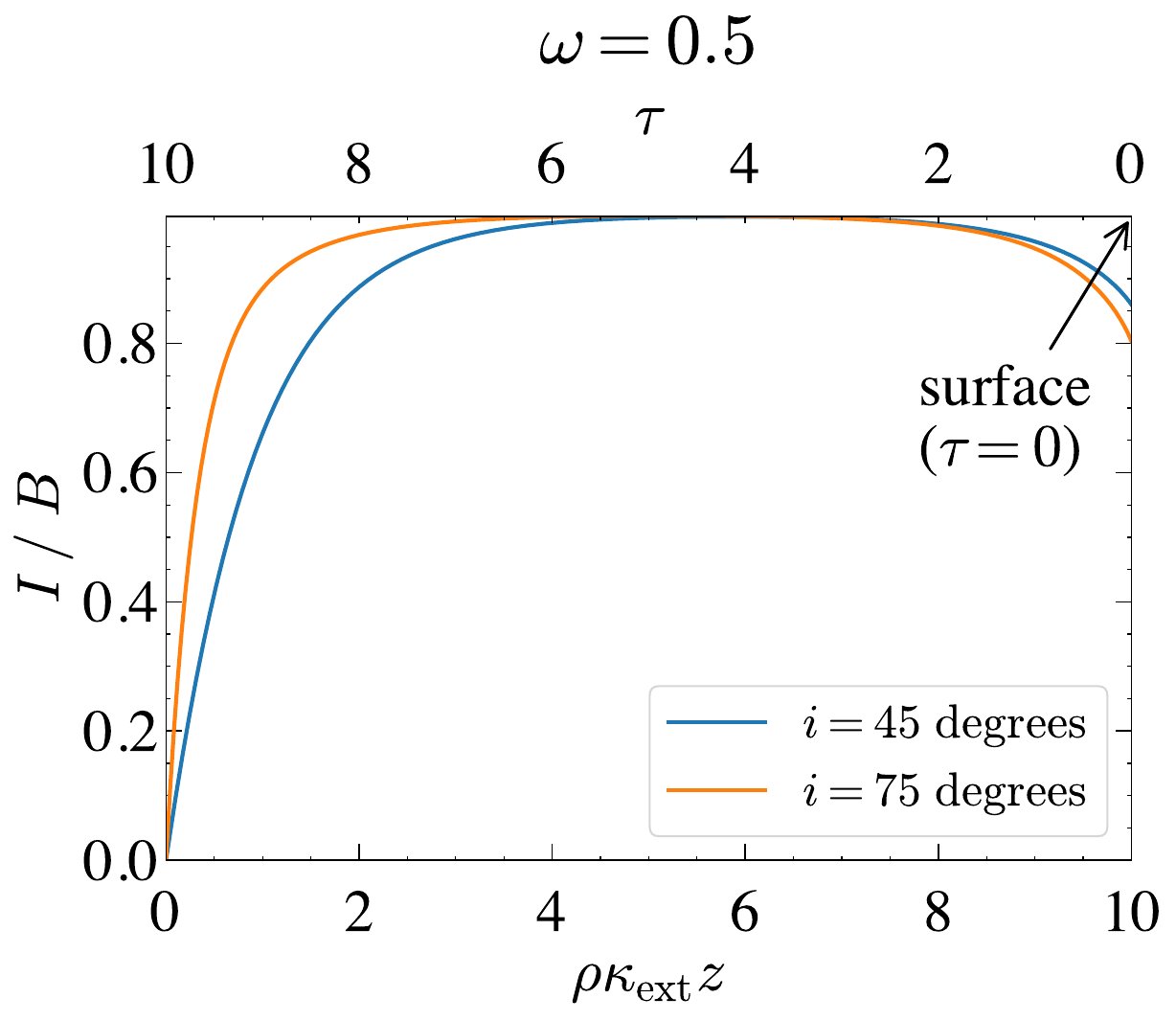}}
\caption{Stokes $I$ as a function of $z$ and the optical depth from the surface $\tau$, obtained by numerically solving Eq. (\ref{eq:RT_fullStokes}). Results are shown $i=45\tcdegree$ and $75 \tcdegree$, with $\omega=0.5$. Stokes $I$ is normalized by the Planck function, $B$. The lower axis shows $\rho \kappa_\mathrm{ext} z$, while the upper axis shows $\tau$.}
\label{fig:I_inc45_75_omega0_5_tau10}
\end{figure}

The surface-layer effect also explains the inclination dependence seen in Fig. \ref{fig:Inclination_dependence}. This is because the surface-layer effect becomes stronger with increasing $i$, as seen in Fig. \ref{fig:I_inc45_75_omega0_5_tau10}, which compares the $\tau$-dependence of Stokes $I$ for $i=45\tcdegree$ and $i=75\tcdegree$.
This trend is expected from Eq. (\ref{eq:RTeq_I}), which gives $dI/d\tau=(I-S)/\mu$ with $\mu\equiv\cos{i}$; thus, a larger inclination (smaller $\mu$) leads to a steeper variation of $I$ with $\tau$ along the ray. As a result, the emergent Stokes $I$ decreases with increasing $i$. 
However, the above discussions on the $\omega-$ and $i-$dependence are qualitative. In practice, because the integral in the scattering term depends on $\omega$ and $i$ in a non-trivial way, $dI /d\tau$ does not necessarily increase monotonically with increasing $\omega$ or $i$.

\begin{figure}[thbp]
\resizebox{\hsize}{!}{\includegraphics{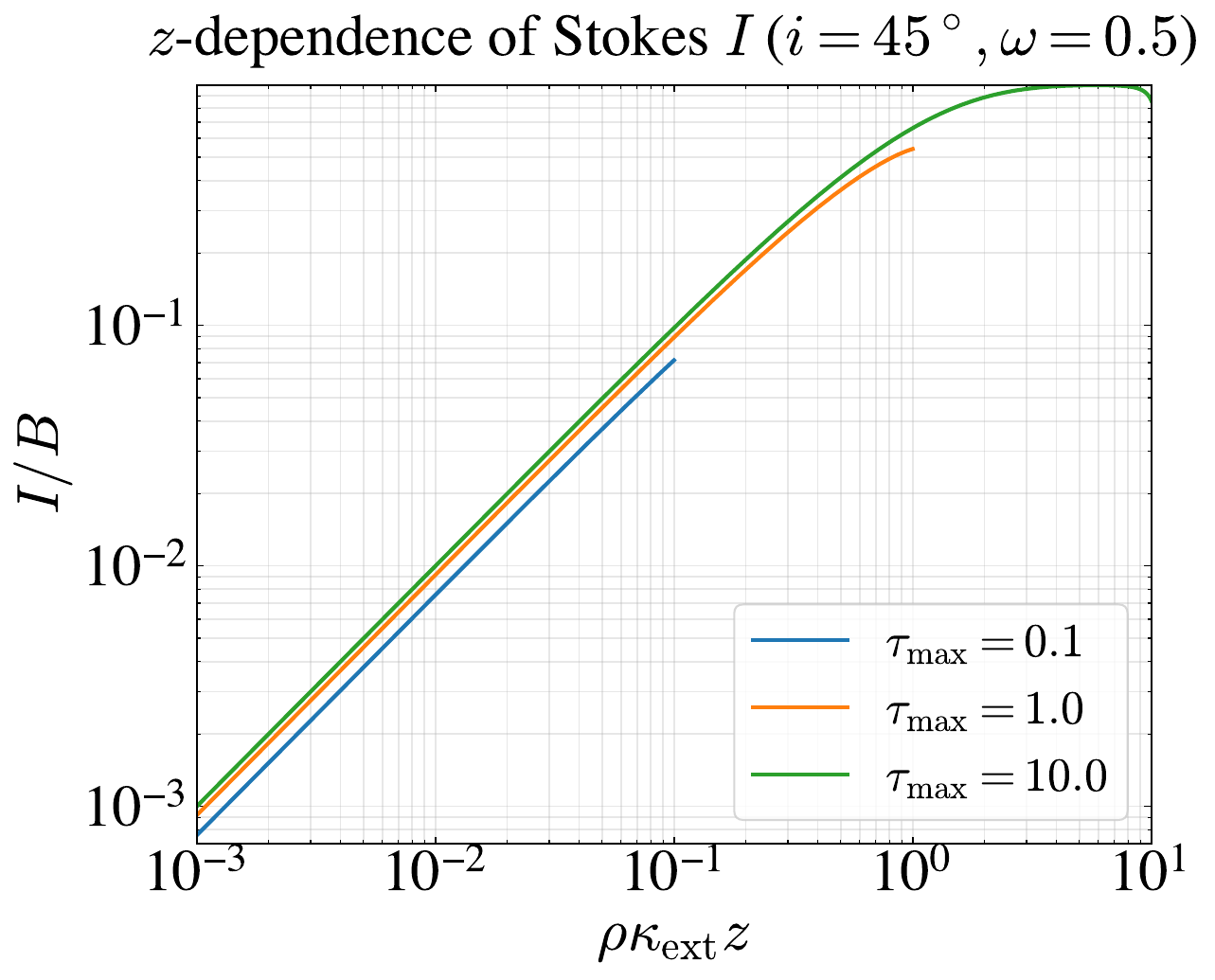}}
\caption{Emergent Stokes $I$ as a function of $z$, obtained by numerically solving Eq. (\ref{eq:RT_fullStokes}). The slab is inclined at $45\tcdegree$ with $\omega=0.5$. Results are shown for $\tau_\mathrm{max}$ = 0.1, 1.0, and 10.0. Stokes $I$ is normalized by the Planck function, $B$.}
\label{fig:I_inc45_omega0_5}
\end{figure}

The surface-layer effect is significant primarily in optically thick slabs. Fig. \ref{fig:I_inc45_omega0_5} presents the $z$-dependence of Stokes $I$ within the slab for $\tau_\mathrm{max} = 0.1, 1,$ and $10$. For $\tau_\mathrm{max}=0.1$, Stokes $I$ varies almost linearly with $z$. For $\tau_\mathrm{max}=1$, a slight deviation from linearity appears near the surface. In contrast, a clear attenuation near the surface is seen only in the optically thick case, $\tau_\mathrm{max}=10$. We also note that the three cases yield different values of Stokes $I$ even at the same $z$. This difference arises because the strength of the radiation field incident from the surface side varies substantially with $\tau_\mathrm{max}$.

\subsubsection{Scattering of the polarized intensity}
\label{subsub:Scattering_polarized_light}

\begin{figure}[htbp]
\resizebox{\hsize}{!}{\includegraphics{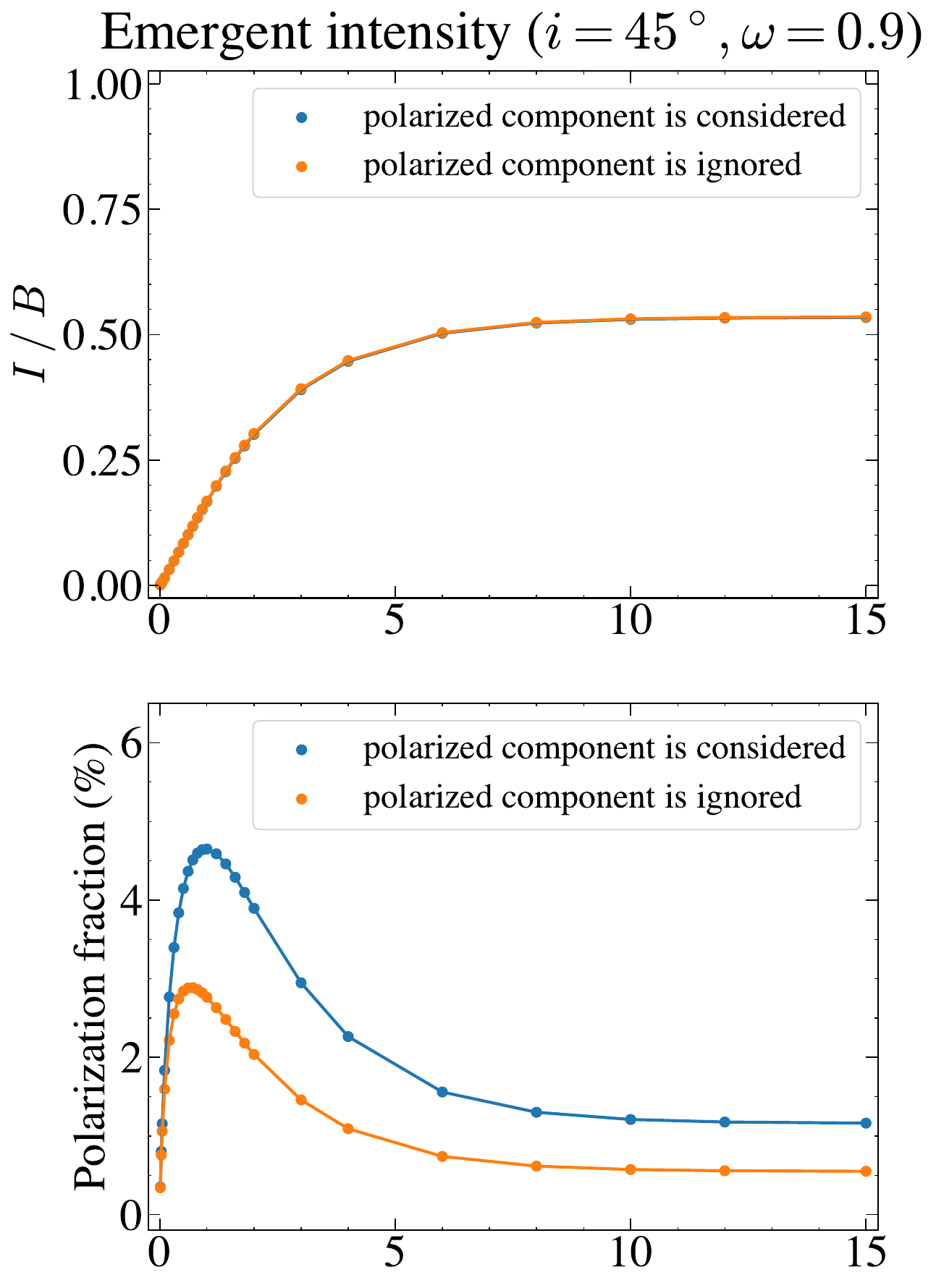}}
\caption{Comparison of the emergent Stokes $I$ and polarization fraction as functions of total optical depth ($\tau_\mathrm{max}$) for a slab inclined at $45\tcdegree$ with $\omega=0.9$, computed both including (blue curves, reproducing Fig. \ref{fig:IQPF_inc45}) and neglecting (orange curves, calculated via Eq. (\ref{eq:RT_fullStokes_Iin})) the polarization of the incoming intensity at all scattering events. This incoming intensity does not represent the background light but the radiation from dust grains around the scatterer. In the upper panel, the two curves nearly overlap.}
\label{fig:Pol_nonPol}
\end{figure}

Scattering of the incoming polarized component can make a non-negligible contribution to the emergent polarization. This is notable because the polarization fraction of the incoming intensity is only a few percent (Figs. \ref{fig:IQPF_inc45} and \ref{fig:Inclination_dependence}). Fig. \ref{fig:Pol_nonPol} compares Stokes $I$ and the polarization fraction of the emergent intensity computed with and without accounting for the polarized component of the incoming intensity at all scattering events. In both calculations, we use the same Stokes $I$ profile of the incoming intensity obtained from the full multiple-scattering solution, thereby isolating the effect of the polarized components of the incoming intensity. For the case of the unpolarized incoming intensity, we set the $Q$, $U$, and $V$ components of the incoming intensity to zero and solve
\begin{equation}
\begin{split}
    \label{eq:RT_fullStokes_Iin}
    \mu
    \frac{d}{d\tau} 
    \begin{pmatrix}
        I(\mu,\tau)\\
        Q(\mu,\tau)\\
        U(\mu,\tau)\\
        V(\mu,\tau)
    \end{pmatrix}
    &=
    \begin{pmatrix}
        I(\mu,\tau)\\
        Q(\mu,\tau)\\
        U(\mu,\tau)\\
        V(\mu,\tau)
    \end{pmatrix} 
    -
    (1 - \omega) 
    \begin{pmatrix}
        B(T)\\
        0\\
        0\\
        0
    \end{pmatrix} \\
    &- \frac{3}{8\pi} \omega
    \int 
    \bm{M} \bm{Z}^\mathrm{R} \bm{M'}
    \begin{pmatrix}
        I_{\mathrm{in}}(\mu',\tau)\\
        0\\
        0\\
        0
    \end{pmatrix}
    d\Omega
    .
\end{split}
\end{equation}
We adopt an albedo of 0.9 and an inclination of $45 \tcdegree$. The lower panel of Fig. \ref{fig:Pol_nonPol} displays the polarization fraction of the emergent intensity. Here, including the polarization of the incoming intensity approximately doubles the polarization fraction of the emergent intensity, compared with the case of ignoring the polarized component of the incoming intensity. This increase indicates that the contribution from the polarized component of the incoming intensity is comparable to that from Stokes $I$.

\begin{figure}[thbp]
\resizebox{\hsize}{!}{\includegraphics{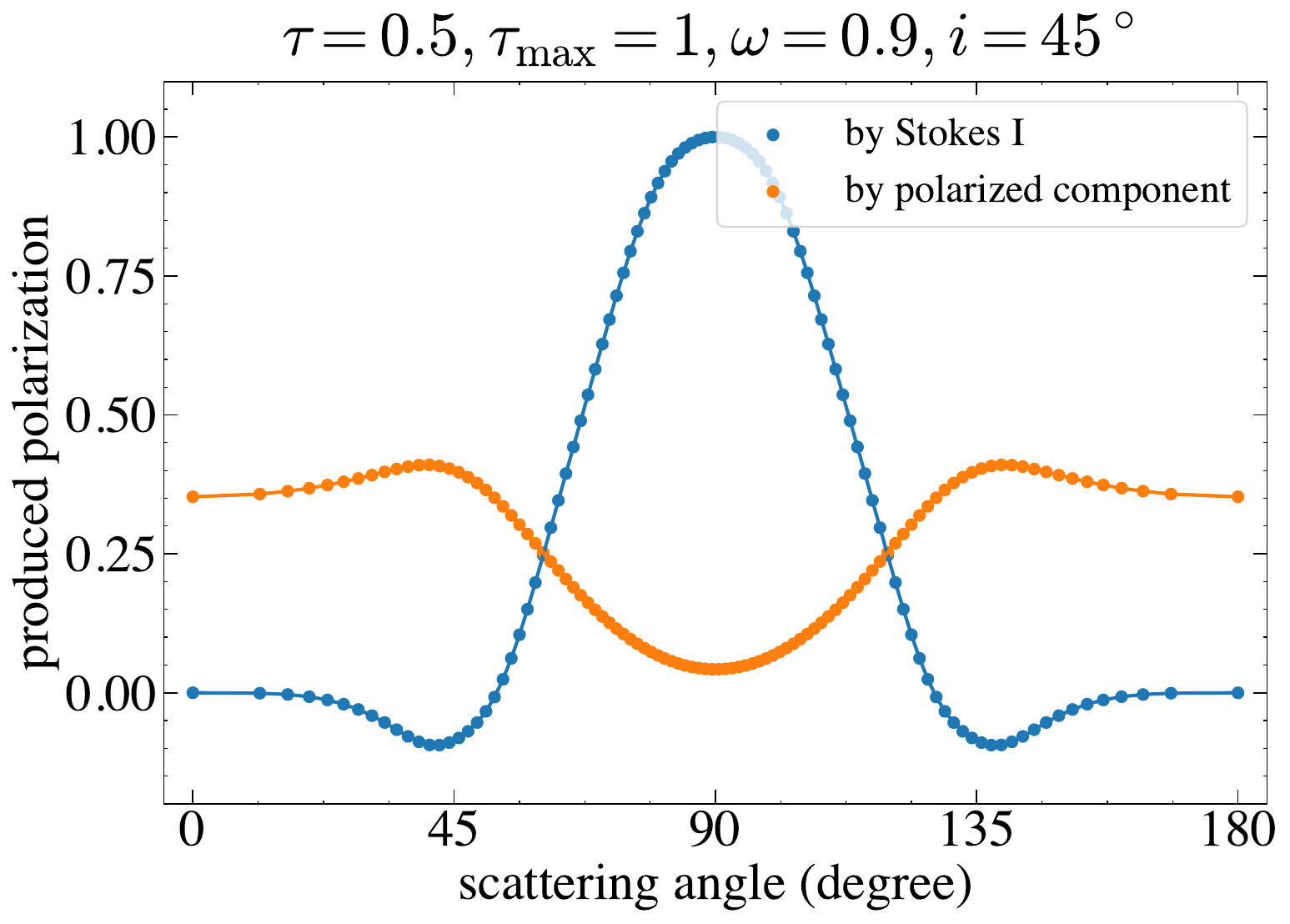}}
\caption{Comparison of the scattering‐angle dependence of the polarization produced by  Stokes $I$ of incoming intensity and by the polarized component of incoming intensity at $\tau = 0.5$ in a plane‐parallel slab with $\tau_\mathrm{max}=1.0$, $\omega = 0.9$, and $i = 45\tcdegree$. The produced polarization on the vertical axis is normalized. Positive values correspond to polarization aligned with $\boldsymbol{e}_\mathrm{inc}$, whereas negative values correspond to polarization aligned with $\boldsymbol{e}_\mathrm{perp}$.}
\label{fig:Pol_scaangle_depend}
\end{figure}

Forward and backward scattering of the incoming polarized component preferentially amplifies the polarization. Fig. \ref{fig:Pol_scaangle_depend} illustrates this effect by showing the scattering-angle dependence of the polarized intensity sourced by incoming Stokes $I$ and by incoming polarized component at an optical depth of $\tau=0.5$ from the slab surface, for a slab with $\tau_\mathrm{max}=1$, $\omega=0.9$, and $i=45\tcdegree$. The curves are normalized to the peak polarized intensity produced by incoming Stokes $I$. The $I$-sourced polarization peaks at a scattering angle of $90\tcdegree$, whereas the polarization sourced by the incoming polarized component is strong around $0\tcdegree$ and $180\tcdegree$. This angular behavior supports the view that forward/backward scattering preferentially amplifies pre-existing polarization.

\begin{figure}[htbp]
\resizebox{\hsize}{!}{\includegraphics{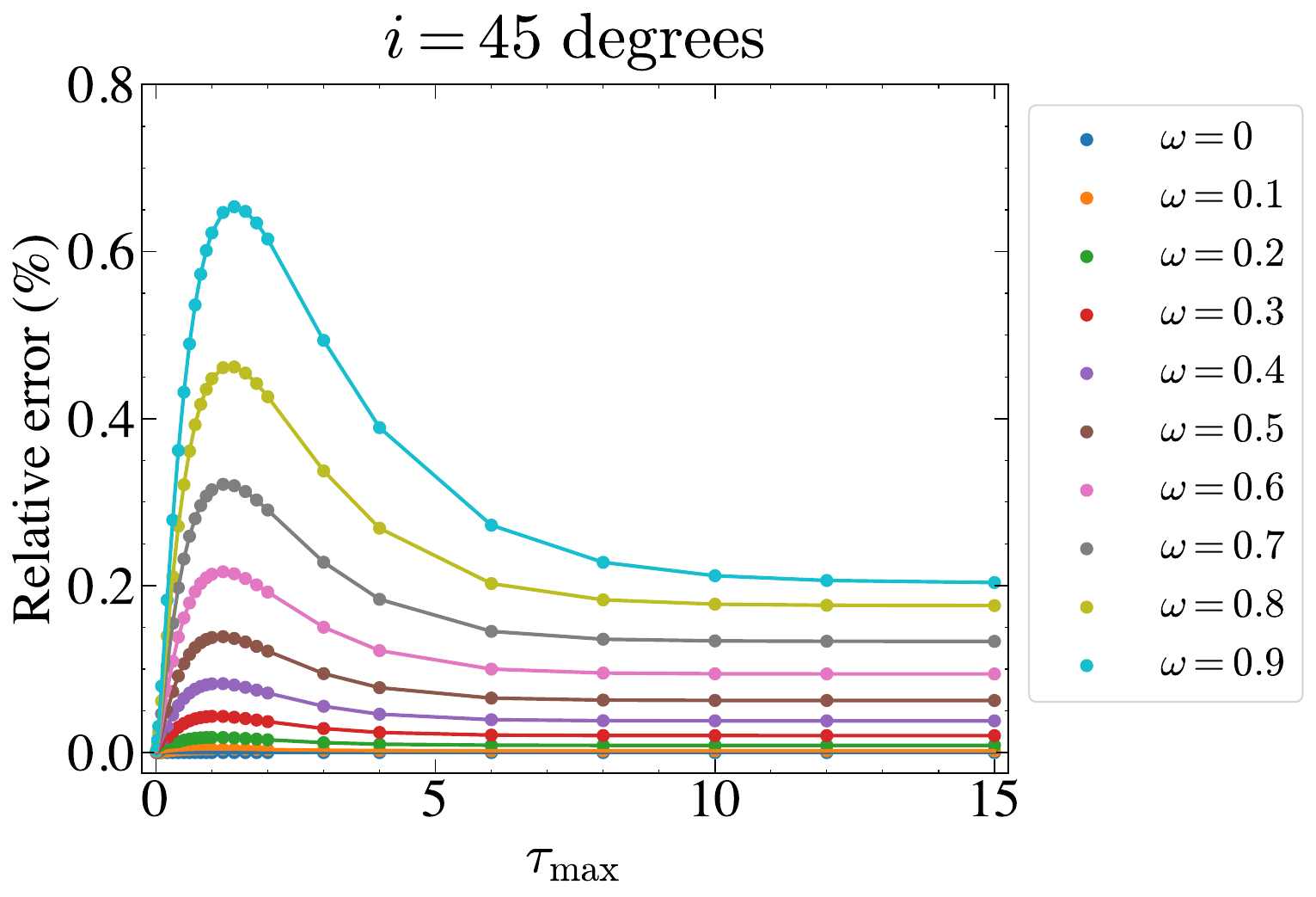}}
\caption{Relative error in the emergent Stokes $I$ between calculations neglecting and including the polarization of the incoming intensity, defined as $({I_\mathrm{unpol} - I_\mathrm{pol}})/{I_\mathrm{pol}}$.
}
\label{fig:Pol_nonPol_Relative_error}
\end{figure}

By contrast, the polarized component of the incoming intensity has a negligible impact on the emergent Stokes $I$. In the upper panel of Fig. \ref{fig:Pol_nonPol}, the emergent Stokes $I$ is plotted as a function of the total optical depth $\tau_\mathrm{max}$ for both polarized and unpolarized incoming intensity. The two curves nearly overlap, with the unpolarized case yielding only slightly higher Stokes $I$. Fig. \ref{fig:Pol_nonPol_Relative_error} displays the relative error $(I_\mathrm{unpol} - I_\mathrm{pol})/I_\mathrm{pol}$, where $I_\mathrm{unpol}$ and $I_\mathrm{pol}$ denote the emergent Stokes $I$ for unpolarized and polarized incoming intensity, respectively. The error remains below $1\%$ over the explored range of $\tau_\mathrm{max}$. This indicates that treating the incoming intensity as unpolarized at all scattering events introduces a negligible bias in current ALMA Stokes $I$ measurements, because ALMA’s typical absolute flux calibration uncertainty is on the order of $5-10\%$ \citep{ALMA_technical_handbook}.

\section{Accuracy of approximate formulae} \label{sec:Comparison}
In this section, we show that widely used approximate formulae for the emergent Stokes $I$ \citep{Carrasco-Gonzalez_2019, 2018ApJ...869L..45B, 2019ApJ...877L..18Z} are systematically lower than our numerical results. This deviation implies that previous SED fitting based on those approximations suffers from non-negligible errors. We begin by summarizing how each approximation was derived.

We first describe \citet{MIYAKE199320}, which provides the basis for widely used approximations. \citet{MIYAKE199320} used the Eddington approximation, the two-stream approximation, and the assumption of isotropic scattering to solve the radiative transfer equation including scattering for an isothermal, constant–density, plane-parallel slab. Under the isotropic-scattering assumption, the radiative transfer equation reduces to Eq. (\ref{eq:RT_iso}), where $J=\int I \Omega$. To determine $J$, they employed the Eddington approximation, in which the radiation field inside the slab is assumed to be nearly isotropic. In this approximation, the specific intensity is expanded to first order in $\mu$ as $I(\mu) = c_1 + c_2 \mu$, where $c_1$ and $c_2$ are constants. Using this Eddington approximation and Eq.~(\ref{eq:RT_iso}), they obtained
\begin{equation} 
\frac{1}{3}\frac{\partial^2 J}{\partial \tau^2} = (1 - \omega)(J - B).
\end{equation}
They then solved this equation by imposing the two-stream approximation to set the boundary conditions, which represent the boundary radiation field by two discrete beams propagating in opposite directions. This procedure yields Eq. (\ref{eq:meanIntensity}). Finally, using Eq.~(\ref{eq:meanIntensity}) and the two-stream boundary conditions, they derived an analytic expression for the emergent intensity evaluated at $\mu = 1/\sqrt3$,
\begin{equation} \label{eq:MN93}
I = B\Big \{ \frac{2\sqrt{1-\omega}(e^{-\sqrt{3(1-\omega)}\tau_\mathrm{max}} - 1)}{e^{-\sqrt{3(1-\omega)}\tau_\mathrm{max}}(\sqrt{1-\omega} - 1)-(\sqrt{1-\omega } + 1)} \Big \}.
\end{equation}

\citet{Carrasco-Gonzalez_2019} generalized the emergent-intensity expression of \citet{MIYAKE199320} to arbitrary $\mu$. They substituted Eq. (\ref{eq:meanIntensity}) into Eq.~(\ref{eq:RT_iso}), and directly solved for the emergent intensity as
\begin{equation} \label{eq:CG19}
I = B\{(1-e^{-\frac{\tau_\mathrm{max}}{\mu}})+\omega F(\tau_\mathrm{max}, \omega, \mu)\},
\end{equation}
where
\begin{equation}
\begin{split}
F(\tau_\mathrm{max}, \omega, \mu) &= \frac{1}{e^{-\sqrt{3}\epsilon \tau_\mathrm{max}}(\epsilon-1) - (\epsilon+1)} \\
&\times
\left(\frac{1-e^{-(\sqrt{3}\epsilon + \frac{1}{\mu})\tau_\mathrm{max}}}{\sqrt{3}\epsilon \mu + 1}  
+
\frac{e^{-\frac{\tau_\mathrm{max}}{\mu}} - e^{-\sqrt{3}\epsilon \tau_\mathrm{max}}}{\sqrt{3}\epsilon \mu - 1}\right)
\end{split}
\end{equation}
and $\epsilon = \sqrt{1-\omega}$. Substituting $\mu = {1}/{\sqrt{3}}$ into Eq.~(\ref{eq:CG19}) exactly matches Eq.~(\ref{eq:MN93}).

\citet{2018ApJ...869L..45B} used a modified Eddington-Barbier approximation to derive the emergent intensity.
In this approximation, the emergent intensity is determined by the source function at $\tau \sim 2/3\mu$, yielding
\begin{equation} \label{eq:Till18}
I = (1-e^{-\frac{\tau_\mathrm{max}}{\mu}})S ( \tau=2\mu/3 ),
\end{equation}
where $S=(1-\omega)B + \omega J$. They adopted $J$ given by Eq.~(\ref{eq:meanIntensity}). If $\tau_\mathrm{max}<4\mu/3$, $\tau$ in $S$ was chosen as $\tau=\tau_\mathrm{max}/2$.

\citet{2019ApJ...877L..18Z} adopted the same Eddington–Barbier approximation but eliminated the piecewise optical‐depth dependence introduced by \citet{2018ApJ...869L..45B}, yielding 
\begin{equation} \label{eq:Zhu19}
I = (1-e^{-\frac{\tau_\mathrm{max}}{\mu}})S\left(\tau=\frac{2\mu\tau_\mathrm{max}}{3\tau_\mathrm{max}+1}\right).
\end{equation}

We summarize the key features of these approximate formulae. \citet{MIYAKE199320} and \citet{Carrasco-Gonzalez_2019} adopted the Eddington approximation, the two‐stream approximation, and the assumption of isotropic scattering. They differ in the range of $\mu$ covered: \citet{MIYAKE199320} evaluated the emergent intensity at $\mu=1/\sqrt{3}$, whereas \citet{Carrasco-Gonzalez_2019} derived an expression valid for arbitrary $\mu$.
\citet{2018ApJ...869L..45B} and \citet{2019ApJ...877L..18Z} likewise adopted the Eddington approximation, the two‐stream approximation, and the assumption of isotropic scattering, but additionally invoked the Eddington–Barbier approximation. The \citet{2018ApJ...869L..45B} formula requires piecewise cases in $\tau_\mathrm{max}$, whereas \citet{2019ApJ...877L..18Z} provides a unified expression without case distinctions.

\begin{figure*}[htbp]
\resizebox{\hsize}{!}{\includegraphics{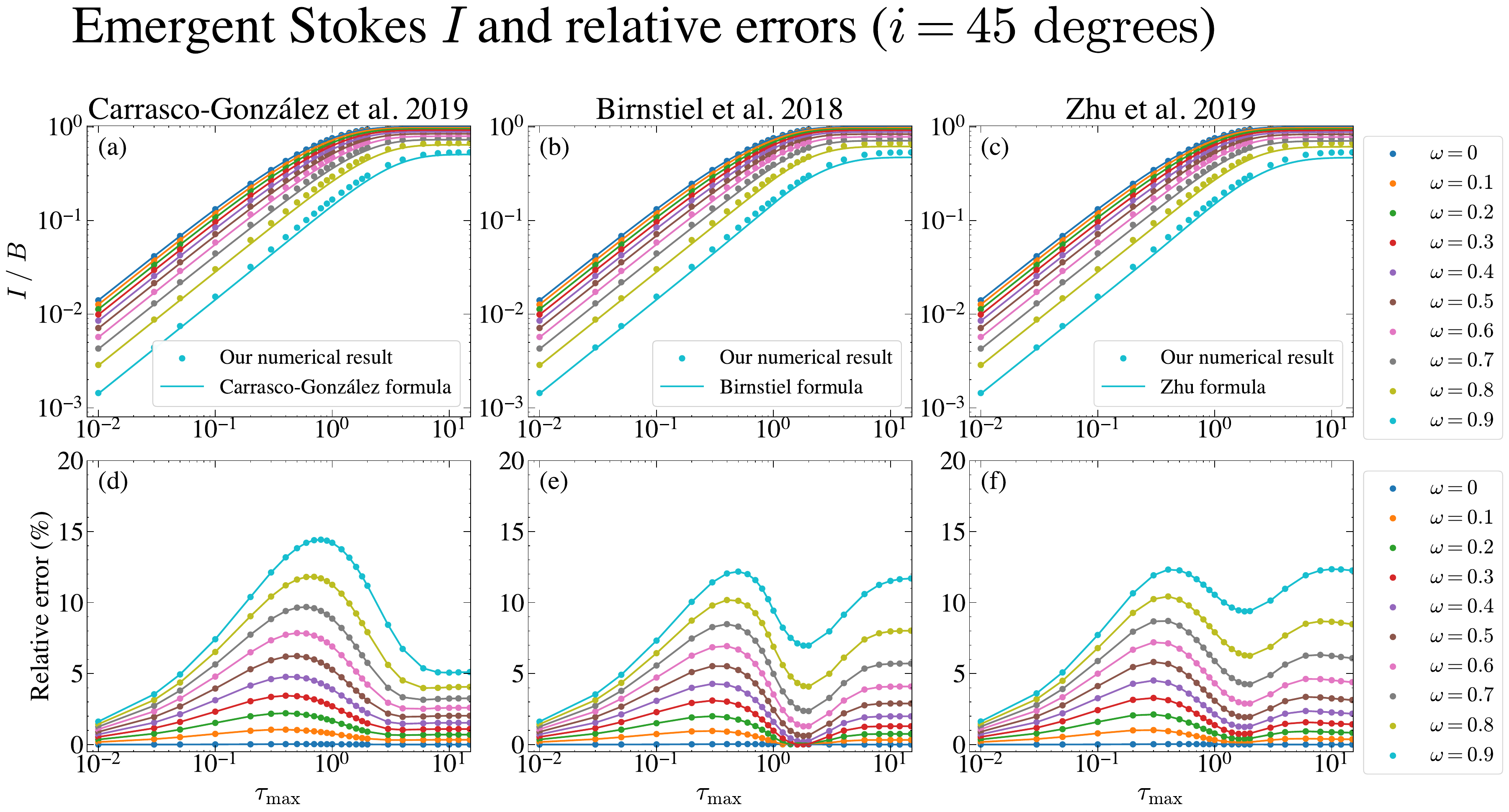}}
\caption{The upper panels compare the approximate formulae for the emergent Stokes $I$ given by \citet{Carrasco-Gonzalez_2019}, \citet{2018ApJ...869L..45B}, and \citet{2019ApJ...877L..18Z} with our numerical results, as a function of the total optical depth and the albedo. Solid lines denote the approximations, and dots indicate the numerical results. The slab is inclined at $45\tcdegree$.
The lower panels plot the corresponding relative error between each approximation and the numerical results. From left to right, the panels show comparisons with \citet{Carrasco-Gonzalez_2019}, \citet{2018ApJ...869L..45B}, and \citet{2019ApJ...877L..18Z}, respectively.}
\label{fig:Relative_error}
\end{figure*}

Next, we show that these widely used approximate formulae for the emergent Stokes $I$ deviate from our numerical results. The upper panels of Fig. \ref{fig:Relative_error} compare our numerical results for the emergent Stokes $I$ with the approximations of \citet{Carrasco-Gonzalez_2019}, \citet{2018ApJ...869L..45B}, and \citet{2019ApJ...877L..18Z} as functions of $\tau_\mathrm{max}$ and $\omega$. We adopt $i=45\tcdegree$ as the fiducial inclination. Results for other inclinations are provided in Appendix \ref{Appendix:AddInc_comparison}. The lower panels of Fig. \ref{fig:Relative_error} show the relative error, e.g., $100 \times |(I_\mathrm{num} - I_\mathrm{app})/I_\mathrm{num}|$, where $I_\mathrm{num}$ is our numerical result and $I_\mathrm{app}$ is the value from the approximations. All three approximations tend to underestimate the emergent Stokes $I$ compared with our numerical results. They yield small errors when $\tau_\mathrm{max} \ll 1$, but the error peaks at about 10 to $15\%$ around $\tau_\mathrm{max} \sim 1$. The error for \citet{Carrasco-Gonzalez_2019} then decreases as $\tau_\mathrm{max}$ increases, whereas the errors for \citet{2018ApJ...869L..45B} and \citet{2019ApJ...877L..18Z} remain above $10 \%$ even for $\tau_\mathrm{max} \gg 1$. Because these errors exceed ALMA’s typical absolute flux calibration uncertainty of about 5 to 10\% \citep{ALMA_technical_handbook}, these errors are non-negligible for ALMA data analysis. For more accurate data analysis, we therefore recommend using our numerical results or our new fitting formula (see Section \ref{sec:FittingFormulae}) rather than these approximations. Our full numerical dataset is available on the following website\footnote{https://github.com/naoyakitade-astro/emergentintensity\label{fn:url}}.

These errors arise because the underlying assumptions do not necessarily hold in our numerical calculations. For example, the Eddington approximation assumes a nearly isotropic radiation field, which is generally valid only in optically thick regions. Therefore, it becomes invalid in an optically thin slab and also near the surface layers of an optically thick disk. The assumption of isotropic scattering is also not satisfied in our calculations, because we use the Rayleigh scattering matrix. In addition, other assumptions, such as the two-stream approximation and the Eddington–Barbier approximation, are not strictly valid in general.

These errors imply that SED fitting based on these approximations tends to overestimate the optical depth and underestimate the dust albedo, because approximations generally yield a lower emergent Stokes $I$ than our calculations for the same optical depth and albedo. This means that the disk mass can be overestimated and the dust size can be inferred incorrectly. The dust temperature may also be overestimated.

\section{New fitting formulae} \label{sec:FittingFormulae}
In this section, we introduce new fitting formulae that reproduce the numerical results presented in Section \ref{sec:NumericalResults}. This new fitting formula for the emergent Stokes $I$ will be more useful for data analysis than widely used approximations because, as shown in Section \ref{sec:Comparison}, those approximations exhibit non-negligible errors. This new fitting formula for the emergent polarization will also be useful because no such formula has been developed yet. In Section \ref{susbsec:fitting_I}, we present the fitting formula for emergent Stokes $I$, and in Section \ref{subsec:fitting_PF}, we provide the fitting formula for the polarization fraction of emergent intensity.

\subsection{Emergent Stokes $I$} \label{susbsec:fitting_I}

We empirically develop a fitting formula for the emergent Stokes $I$, expressed as 
\begin{equation} \label{eq:fitting_I}
\left\{
\begin{aligned}
I &= \left\{1-\omega_\mathrm{I}(\omega, \mu)\right\} \frac{\tau_\mathrm{max}} {\mu} B  \qquad \qquad \quad \quad(\tau_\mathrm{max} < 0.04) \\
I &=  \Big[ I_\mathrm{conv}(\omega, \mu) \left\{1-\exp\left(-A_\mathrm{I}(\omega, \mu)\tau_\mathrm{max}^{B_\mathrm{I}(\omega, \mu)}\right)\right\}  \\
      &\quad- I_\mathrm{conv}(\omega, \mu) \left\{1-\exp\left(-A_\mathrm{I}(\omega, \mu)0.04^{B_\mathrm{I}(\omega, \mu)}\right)\right\} \\
      &\quad+ \left\{1-\omega_\mathrm{I}(\omega, \mu)\right\} \frac{0.04}{\mu}  \Big]B
      \qquad \qquad \quad (\tau_\mathrm{max} > 0.04),
\end{aligned}
\right.
\end{equation} 
based on our numerical results. The coefficients $A_\mathrm{I}, B_\mathrm{I}, I_\mathrm{conv}$ and $\omega_\mathrm{I}$ are fitting parameters that depend on the dust albedo $\omega$ and the cosine of the inclination $\mu$. This formula is not motivated by the radiative transfer equation, but is chosen to reproduce the systematic trends seen in our numerical results. This formula is continuous at $\tau_\mathrm{max} = 0.04$ but not differentiable there.

\begin{figure}[htbp]
\resizebox{\hsize}{!}{\includegraphics{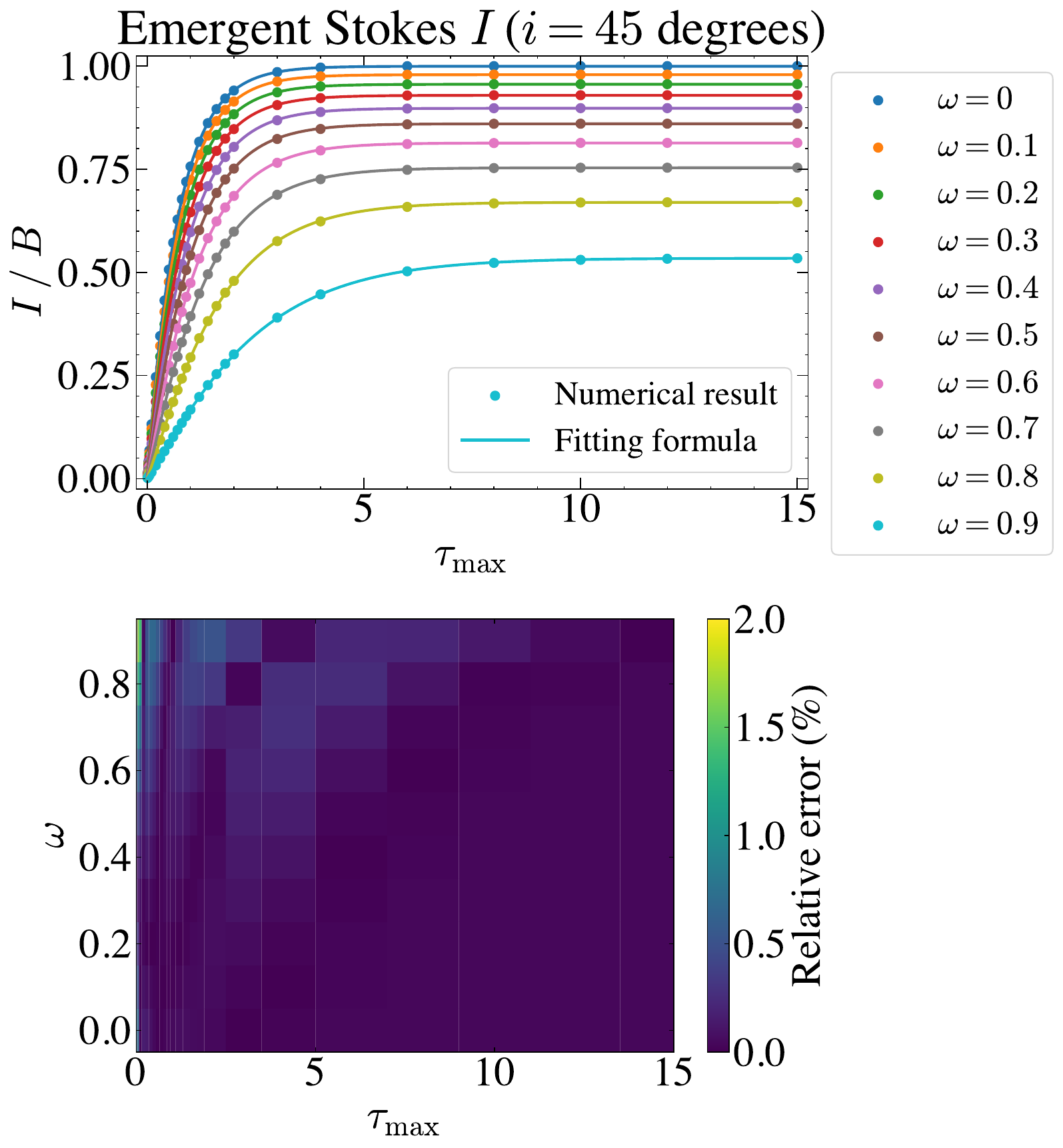}}
\caption{Upper panel: Comparison between our numerical results for the emergent Stokes $I$ and the proposed fitting formula, Eq. (\ref{eq:fitting_I}), for $i=45 \tcdegree$.
Lower panel: Relative error of the fitting formula with respect to the numerical results, defined as $|(I_\mathrm{num} - I_\mathrm{fit})/I_\mathrm{num}|$.}
\label{fig:Fitting_intensity}
\end{figure}

This fitting formula can accurately reproduce our numerical results. The upper panel of Fig. \ref{fig:Fitting_intensity} compares the fitting formula (Eq. (\ref{eq:fitting_I})) with our numerical results for $i=45\tcdegree$. The fitting parameters are determined by minimizing the relative errors between the fitting formula and the numerical results. Specifically, we minimize $\sum_\mathrm{n} \{ (I_\mathrm{num, n} - I_\mathrm{fit, n}) / I_\mathrm{num, n} \}^2$, where $I_\mathrm{fit}$ is the emergent Stokes $I$ given by Eq. (\ref{eq:fitting_I}), and $I_\mathrm{num}$ is the emergent Stokes $I$ from our numerical results. The lower panel of Fig. \ref{fig:Fitting_intensity} shows the relative errors. The relative errors are below $1\%$ over most of the parameter space and reach at most $\sim 2\%$. This indicates that the fitting formula is useful for the analysis of ALMA data because these errors are below ALMA's typical observational uncertainties ($\sim 5-10 \%$)\citep{ALMA_technical_handbook}. Although we show the $i=45\tcdegree$ case here, we obtain comparably small relative errors at other angles. The best-fit parameters ($A_\mathrm{I}, B_\mathrm{I}, I_\mathrm{conv}$, $\omega_\mathrm{I}$) for $i=0\tcdegree, 15\tcdegree, 30\tcdegree, 45\tcdegree, 60\tcdegree, 75\tcdegree$ are provided in Appendix \ref{Appendix:AddInc_fitting}. These fitting parameters are available on the website\footnotemark[\getrefnumber{fn:url}].

\subsection{Polarization fraction of emergent intensity} \label{subsec:fitting_PF}

We empirically develop a fitting formula for the polarization fraction of the emergent intensity, expressed as
\begin{equation}
\begin{split} \label{eq:fitting_PF}
PF \equiv \frac{Q}{I} &= A_\mathrm{PF}(\omega, \mu) \tau_\mathrm{max}^{B_\mathrm{PF}(\omega, \mu)} \exp\left(-(\frac{\tau_\mathrm{max}}{\tau_\mathrm{PF}(\omega, \mu)})^{0.8} \right)  \\
&\quad+ PF_\mathrm{conv}(\omega, \mu)\left \{ 1 - \exp \left( -(\frac{\tau_\mathrm{max}}{\tau_\mathrm{PF}(\omega, \mu)})^{0.8} \right)\right \},
\end{split}
\end{equation}
based on our numerical results. The coefficients $A_\mathrm{PF}, B_\mathrm{PF}, PF_\mathrm{conv}$ and $\tau_\mathrm{PF}$ are fitting parameters that depend on the dust albedo $\omega$ and the cosine of the inclination $\mu$. This formula is also not motivated by the radiative transfer equation, but is chosen to reproduce the systematic trends seen in our numerical results.

\begin{figure}[htbp]
\resizebox{\hsize}{!}{\includegraphics{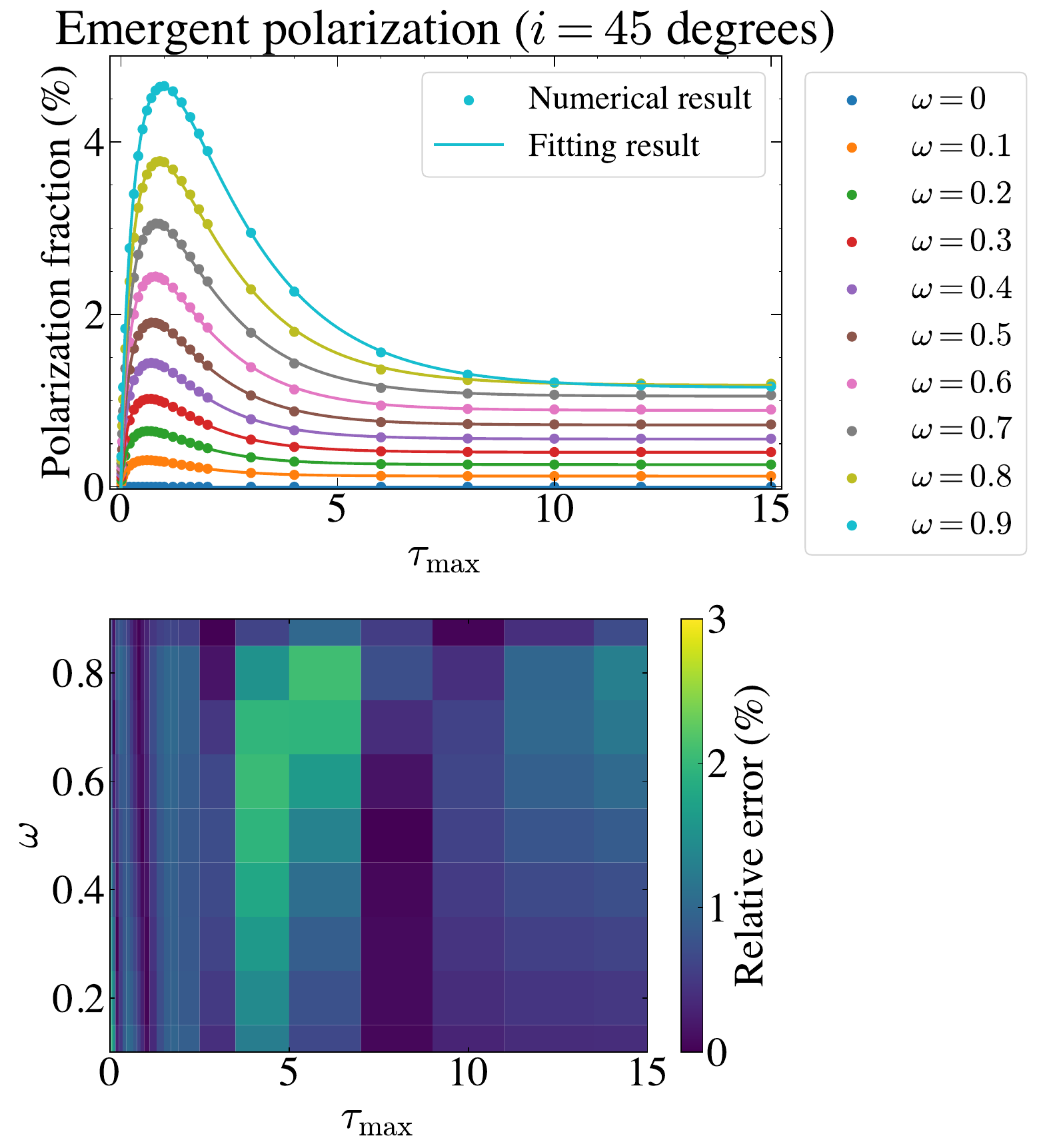}}
\caption{Upper panel: Comparison between our numerical results for the polarization fraction of the emergent intensity, $PF_\mathrm{num}$, and the proposed fitting formula, $PF_\mathrm{fit}$ (Eq. (\ref{eq:fitting_PF})), for $i=45 \tcdegree$.
Lower panel: Relative error of the fitting formula with respect to the numerical results, defined as $|(PF_\mathrm{num} - PF_\mathrm{fit})/PF_\mathrm{num}|$.}
\label{fig:Fitting_polarization}
\end{figure}

This fitting formula can accurately reproduce our numerical results. The upper panel of Fig. \ref{fig:Fitting_polarization} compares the fitting formula (Eq. (\ref{eq:fitting_PF})) with our numerical results for $i=45\tcdegree$, our fiducial inclination. The fitting parameters are determined in the same manner as for Stokes $I$. The lower panel of Fig. \ref{fig:Fitting_polarization} shows the relative errors. The relative errors reach at most $\sim 3\%$. This indicates that the fitting formula is useful for the analysis of ALMA data because these errors are below ALMA's typical observational uncertainties ($\sim 5-10 \%$)\citep{ALMA_technical_handbook}. Although we show the $i=45\tcdegree$ case here, we obtain comparably small relative errors at other angles. The best-fit parameters ($A_\mathrm{PF}, B_\mathrm{PF}, PF_\mathrm{conv}$, $\tau_\mathrm{PF}$) for $i= 15\tcdegree, 30\tcdegree, 45\tcdegree, 60\tcdegree, 75\tcdegree$ are provided in Appendix \ref{Appendix:AddInc_fitting}. These fitting parameters are available on the website\footnotemark[\getrefnumber{fn:url}].

\begin{figure}
\resizebox{\hsize}{!}{\includegraphics{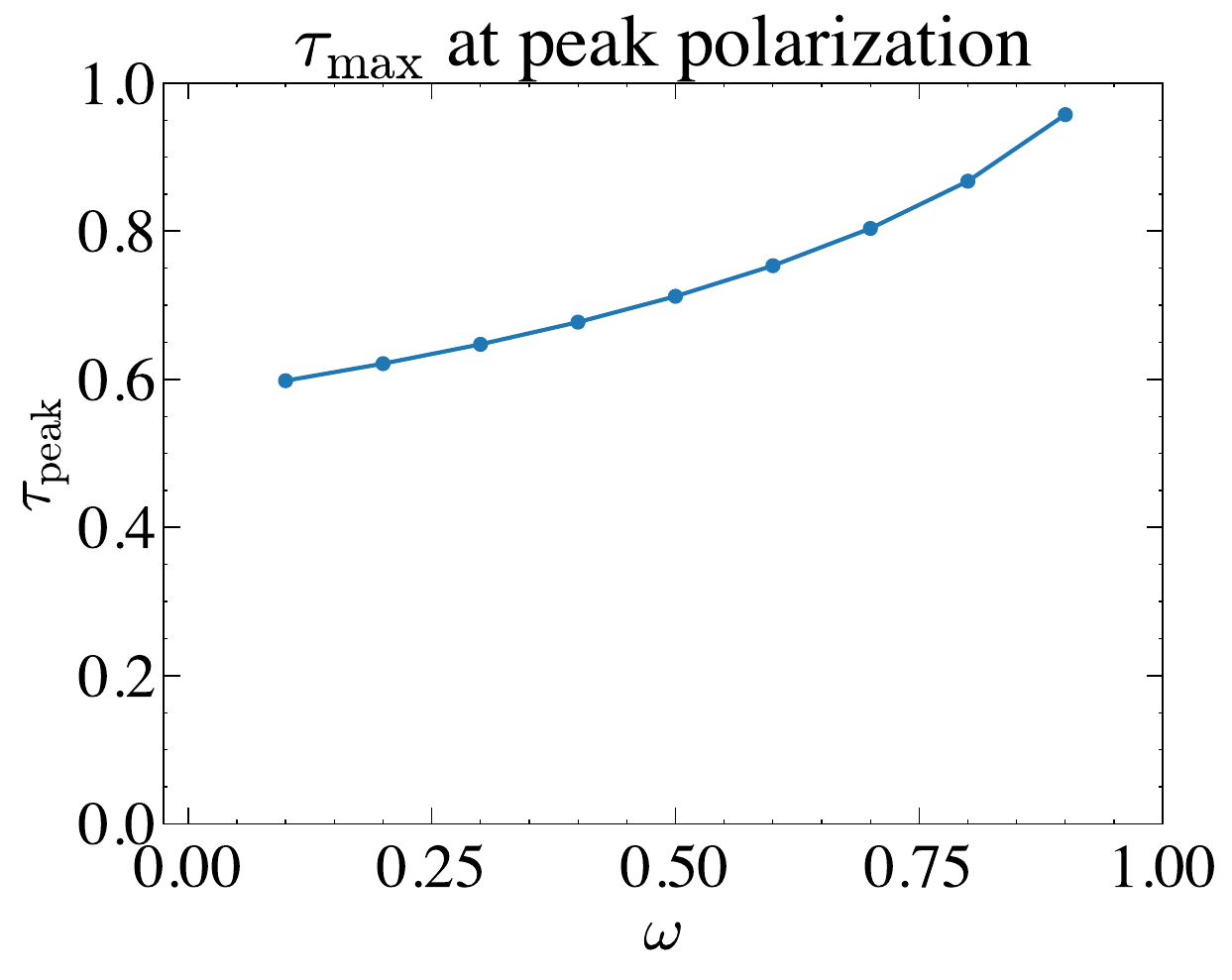}}
\caption{$\omega$-dependence of $\tau_\mathrm{peak}$ for $i=45\tcdegree$, where $\tau_\mathrm{peak}$ is the total optical depth at which the polarization fraction of the emergent intensity peaks.}
\label{fig:Polarization_peak}
\end{figure}

From Eq. (\ref{eq:fitting_PF}), we can derive $\tau_\mathrm{peak}$, the total optical depth at which the polarization fraction of the emergent intensity peaks. Specifically, $\tau_\mathrm{peak}$ is obtained by solving $\partial PF / \partial \tau_\mathrm{max}=0$ using Eq. (\ref{eq:fitting_PF}). Fig. \ref{fig:Polarization_peak} shows the dependence of $\tau_\mathrm{peak}$ on $\omega$ for our fiducial inclination $i=45\tcdegree$. We find that $\tau_\mathrm{peak}$ increases with $\omega$, although the physical origin of this trend remains unclear. $\tau_\mathrm{peak}$ for inclinations other than $i=45\tcdegree$ are provided in Appendix \ref{Appendix:AddInc_taupeak}.

\section{Estimating Mie-Scattering disk emission from Rayleigh-scattering results} \label{subsec:Mie}

In this section, we estimate Mie-scattering disk emission from our Rayleigh-scattering results. Mie scattering is crucial for millimeter observations because dust grains in protoplanetary disks are expected to grow to sizes comparable to the observing wavelength. However, accounting for Mie scattering over a wide range of dust models when computing disk emission is computationally expensive. Therefore, estimating Mie-scattering disk emission from our Rayleigh-scattering results would provide a useful and efficient approach for analyzing millimeter observational data.

Previous studies have estimated Mie-scattering disk emission using some approximations. To estimate the emergent Stokes $I$, Mie scattering has often been treated as effectively isotropic by introducing an effective scattering opacity \citep[e.g.,][]{Macias2021, 2025ApJ...990..183U}. For the emergent polarization, it has been commonly assumed that, as in Rayleigh scattering, the polarization is dominated by $\sim 90 \tcdegree$ scattering even for Mie scattering \citep{2015ApJ...809...78K, 2019ApJ...885...52T}. However, these approximations have not been quantitatively tested. 

We assess under what conditions Mie-scattering disk emission can be estimated based on these approximations. To this end, we compare their estimations for several dust models with numerical calculations that account for Mie scattering. If the estimations agree with the numerical results, Mie-scattering disk emission can be efficiently estimated using these approximations without performing numerical calculations.

This section is organized as follows. First, we describe the numerical method used to compute emission including Mie scattering, and we summarize the dust models. We then compare our Rayleigh-based estimations for Mie-scattering disk emission with numerical results that account for Mie scattering, separately for the emergent Stokes $I$ and the emergent polarization.
 
\subsection{Numerical method and dust model} \label{subsubsec:Mie_method_model}
We numerically solve the radiative transfer equation, Eq. \ref{eq:RT_fullStokes}, following the same procedure as in Section \ref{sec:method}. For the scattering matrix, we use the Mie scattering matrices $\bm{Z}^\mathrm{Mie}$. In calculating $\bm{Z}^\mathrm{Mie}$, we consider compact ($f=1$) or porous ($f=0.1$) spherical grains, where $f$ denotes the volume filling factor. The composition is taken to be the DSHARP dust composition: a mixture of $20 \%$ water ice, $33\%$ astronomical silicates, $7\%$ troilite and $40\%$ refractory organics by mass \citep{2018ApJ...869L..45B}. The refractive indices of water ice are taken from \citet{2008JGRD..11314220W}, astronomical silicate from \citet{2003ApJ...598.1017D}, and troilite and refractory organics from \citet{1996A&A...311..291H}. We use \texttt{dsharp\_opac} to calculate $\bm{Z}^\mathrm{Mie}$ \citep{2018ApJ...869L..45B}. Dust grains are assumed to have a size distribution of $n(a) \propto a^{-3.5}$. The minimum size is taken to be $0.1~\mathrm{\mu m}$. The observing wavelength is $870~\mathrm{\mu m}$, which corresponds to ALMA Band 7. The fiducial inclination is set to $45\tcdegree$. Results for other inclinations are provided in Appendix \ref{Appendix:AddInc_Mie}.

For illustration, we choose four representative cases for the maximum grain size $a_\mathrm{max}$ and the volume filling factor $f$: $a_\mathrm{max} f = 100~\mathrm{\mu m}$ and $200~\mathrm{\mu m}$ with $f=1$ (compact) and $f=0.1$ (porous). We choose these grain sizes because grains that are either significantly smaller or larger do not produce sufficient scattering polarization when the dust is compact. The scattering parameters of four dust models used in the following discussion are summarized in Table \ref{tab:Mie_models}. In this table, the asymmetry parameter $g$ is defined as $g \equiv \int_{-1}^{1} \cos{\theta_\mathrm{s}} Z_{11}d(\cos{\theta_\mathrm{s}}) / \int_{-1}^1 Z_{11}d(\cos{\theta_\mathrm{s}})$, and the effective albedo $\omega_\mathrm{eff}$ is defined as $\omega_\mathrm{eff} \equiv \kappa_\mathrm{sca}^\mathrm{eff}/(\kappa_\mathrm{abs} + \kappa_\mathrm{sca}^\mathrm{eff})$, where $\kappa_\mathrm{sca}^\mathrm{eff} \equiv (1-g) \kappa_\mathrm{sca}$.

\begin{table*}
\caption{}
\label{tab:Mie_models}
\centering
\resizebox{\textwidth}{!}{%
\begin{tabular}{cccccc}
 & $a_\mathrm{max}f = 100 \mathrm{~\mu m}$ $(f=1.0)$ & $a_\mathrm{max}f = 200\mathrm{~\mu m}$ $(f=1.0)$ & $a_\mathrm{max} f = 100\mathrm{~\mu m}$ $(f=0.1)$ & $a_\mathrm{max} f = 200\mathrm{~\mu m}$ $(f=0.1)$  \\
\hline \hline
$\omega$ & 0.68 & 0.91 & 0.93 & 0.96 \\
$g$ & 0.10 & 0.40 & 0.84 & 0.92 \\
$\omega_\mathrm{eff} $ & 0.66 & 0.86 & 0.67 & 0.66 \\
$|Z_{12}(\theta_\mathrm{s}=90\tcdegree)/Z_{11}(\theta_\mathrm{s}=90\tcdegree)|$ & 0.99 & 0.26 & 0.95 & 0.91 \\
\hline
\end{tabular}}
\tablefoot{Scattering properties of the four DSHARP dust models. Here $f$ is the volume filling factor, $g$ is the asymmetry parameter defined by $g\equiv \frac{\int_{-1}^1 \cos{\theta_\mathrm{s}} Z_{11}d(\cos{\theta_\mathrm{s}})}{\int_{-1}^1 Z_{11}d(\cos{\theta_\mathrm{s}})}$, $\omega_\mathrm{eff}$ is the effective albedo defined by
$\omega_\mathrm{eff} \equiv \frac{\kappa_\mathrm{sca}^\mathrm{eff}}{\kappa_\mathrm{abs}+\kappa_\mathrm{sca}^\mathrm{eff}}$, and $Z_{11}$ and $Z_{12}$ are elements of the scattering matrix.}
\end{table*}

\subsection{Emergent Stokes $I$} \label{subsubsec:Mie_intensity}

\begin{figure*}[htbp]
\resizebox{\hsize}{!}{\includegraphics{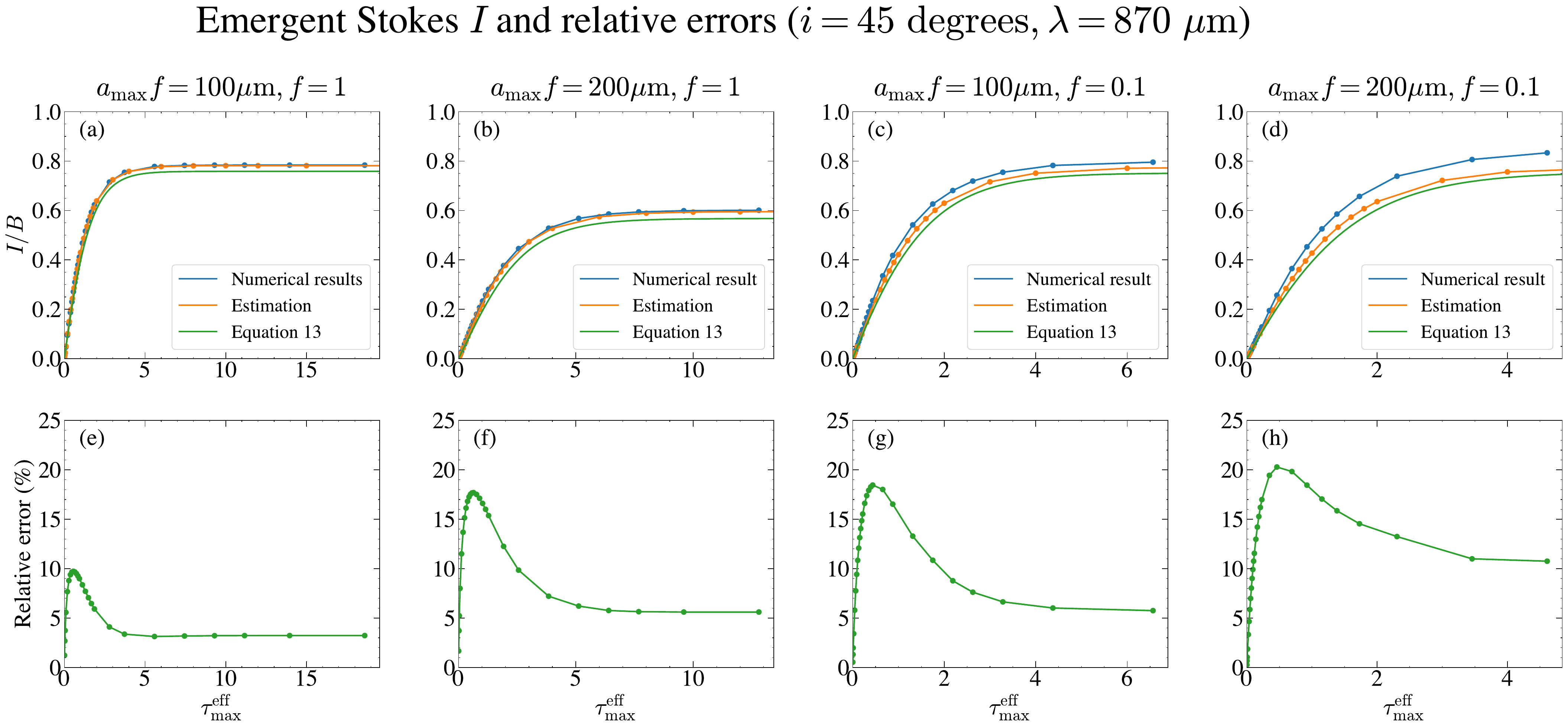}}
\caption{Upper panels: Comparison, for the four dust models, between the numerically computed emergent Stokes $I$ including Mie scattering and two estimates based on the effective scattering opacity, for $i=45\tcdegree$ and $\lambda=870~\mathrm{\mu m}$. The first estimate (orange lines) is obtained from our Rayleigh-scattering calculations using $\omega=\omega_\mathrm{eff}$ and $\tau_\mathrm{max}=\tau_\mathrm{max}^\mathrm{eff}$. The second estimate (green lines) is obtained from Eq. \ref{eq:CG19} using $\omega=\omega_\mathrm{eff}$ and $\tau_\mathrm{max}=\tau_\mathrm{max}^\mathrm{eff}$. Lower panels: Relative errors, for the four dust models, of the estimate based on Eq. \ref{eq:CG19} ($I_{\mathrm{Eq.} \ref{eq:CG19}}^\mathrm{eff}$) with respect to the numerically computed emergent Stokes $I$ including Mie scattering ($I_\mathrm{Mie}$), defined as $|(I_\mathrm{Mie} - I_{\mathrm{Eq.} \ref{eq:CG19}}^\mathrm{eff})/I_\mathrm{Mie}|$.}
\label{fig:DSHARP_vs_Rayleigh_I}
\end{figure*}

We find that the effective-scattering-opacity approximation estimates the Mie-scattering emergent Stokes $I$ reasonably well for most of our dust models, except for the case of $a_\mathrm{max}f=200~\mathrm{\mu m}$ and $f=0.1$. The upper panels of Fig. \ref{fig:DSHARP_vs_Rayleigh_I} compares the numerically computed emergent Stokes $I$ that accounts for Mie scattering (blue lines) with the estimations based on the effective scattering opacity $\kappa_\mathrm{sca}^\mathrm{eff}$ (orange lines) \citep[e.g.,][]{1978wpsr.book.....I, 2018ApJ...869L..45B}. To compute the estimations, we use our numerical results for Rayleigh scattering evaluated with $\omega=\omega_\mathrm{eff}$ and $\tau_\mathrm{max}=\tau_\mathrm{max}^\mathrm{eff}$, which is defined as $\tau_\mathrm{max}^\mathrm{eff} = (\kappa_\mathrm{abs} + \kappa_\mathrm{sca}^\mathrm{eff})\Sigma_\mathrm{dust}$ where $\Sigma_\mathrm{dust}$ is the dust surface density. For compact grains (panels (a) and (b) of Fig. \ref{fig:DSHARP_vs_Rayleigh_I}), the estimations agree well with the numerical results.
For porous grains with $a_\mathrm{max}f = 100 ~ \mathrm{\mu m}$ and $f=0.1$ (panel (c) of Fig. \ref{fig:DSHARP_vs_Rayleigh_I}), the agreement is somewhat worse than in the compact-grain cases. However, the relative errors remain within roughly $5\%$, comparable to ALMA’s typical observational uncertainties ($\sim 5 - 10 \%$; \citealt{ALMA_technical_handbook}). Therefore, this estimation is sufficient for ALMA data analysis. In contrast, for porous grains with $a_\mathrm{max}f = 200~\mathrm{\mu m}$ and $f=0.1$ (panel (d) of Fig. \ref{fig:DSHARP_vs_Rayleigh_I}), the estimations do not agree with the numerical results, with the relative errors exceeding $5\%$. Thus, the effective-scattering-opacity approximation is not always valid. For other inclinations, the estimates can become even less accurate in some cases (see Appendix \ref{Appendix:AddInc_Mie}). Since this assessment is based on only four dust models, further validation over a wider parameter space is required.

\begin{figure*}[htbp]
\resizebox{\hsize}{!}{\includegraphics{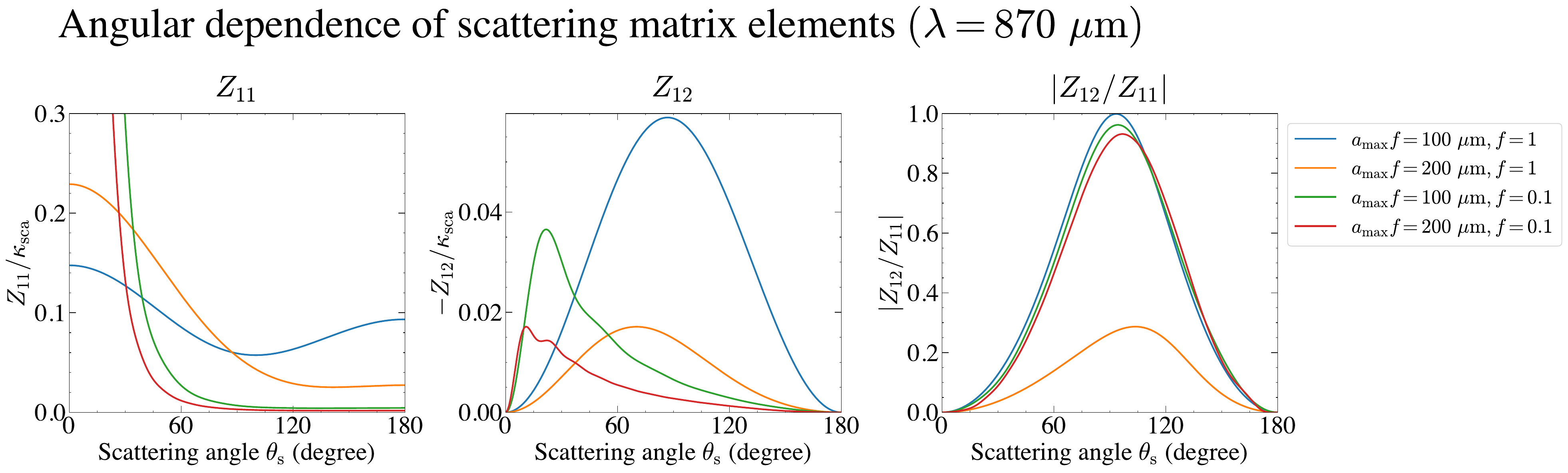}}
\caption{Angular dependence of the scattering matrix elements $Z_{11}$, $Z_{12}$, and $|Z_{12}/Z_{11}|$ at $\lambda = 870 \mathrm{\mu m}$ computed with the DSHARP dust model. The scattering matrices are normalized such that $\int_{4\pi} Z_{11} d\Omega = \kappa_\mathrm{sca}$.}
\label{fig:scattering_matrix}
\end{figure*}

One reason why the estimation is less accurate for porous grains than for compact grains may be the more strongly forward-peaked $Z_{11}$ of porous grains. The left panel of Fig. \ref{fig:scattering_matrix} shows that $Z_{11}$ for porous grains is more strongly forward-peaked than that for compact grains. This suggests that, even for compact grains, the Mie-scattering emergent Stokes $I$ may not be accurately estimated when $Z_{11}$ becomes sufficiently forward-peaked. Other scattering-matrix elements, such as $Z_{12}$ (which equals $Z_{21}$ for spherical grains), have a much smaller impact on the emergent Stokes $I$ than $Z_{11}$ and are therefore neglected here (see Section \ref{subsub:Scattering_polarized_light}).

The estimate based on our Rayleigh-scattering results is more accurate than that obtained by applying the effective-scattering-opacity approximation to conventional approximate formulae for disk emission (e.g., Eq. \ref{eq:CG19}). The upper panels of Fig. \ref{fig:DSHARP_vs_Rayleigh_I} compare the Mie-scattering results with the estimates obtained from the approximation of \citet{Carrasco-Gonzalez_2019} (Eq. \ref{eq:CG19}) using the effective scattering opacity (green lines). These panels demonstrate that the estimates based on our Rayleigh-scattering calculations agree better with the Mie-scattering results than those based on Eq. \ref{eq:CG19}. The lower panels of Fig. \ref{fig:DSHARP_vs_Rayleigh_I} show the relative errors between the estimates based on Eq. \ref{eq:CG19} and the Mie-scattering results. Except for the dust model with $a_\mathrm{max}f = 100~\mathrm{\mu m}$ and $f=1$, the relative errors are much larger than ALMA’s typical observational uncertainties ($\sim 5$--$10\%$; \citealt{ALMA_technical_handbook}). This indicates that estimates based on Eq. \ref{eq:CG19} are not suitable for the analysis of ALMA data. Except for a few cases, the same trends are found for other inclinations as well (see Appendix \ref{Appendix:AddInc_Mie}).

\subsection{Emergent polarization} \label{subsubsec:Mie_pol}

\begin{figure*}[htbp]
\resizebox{\hsize}{!}{\includegraphics{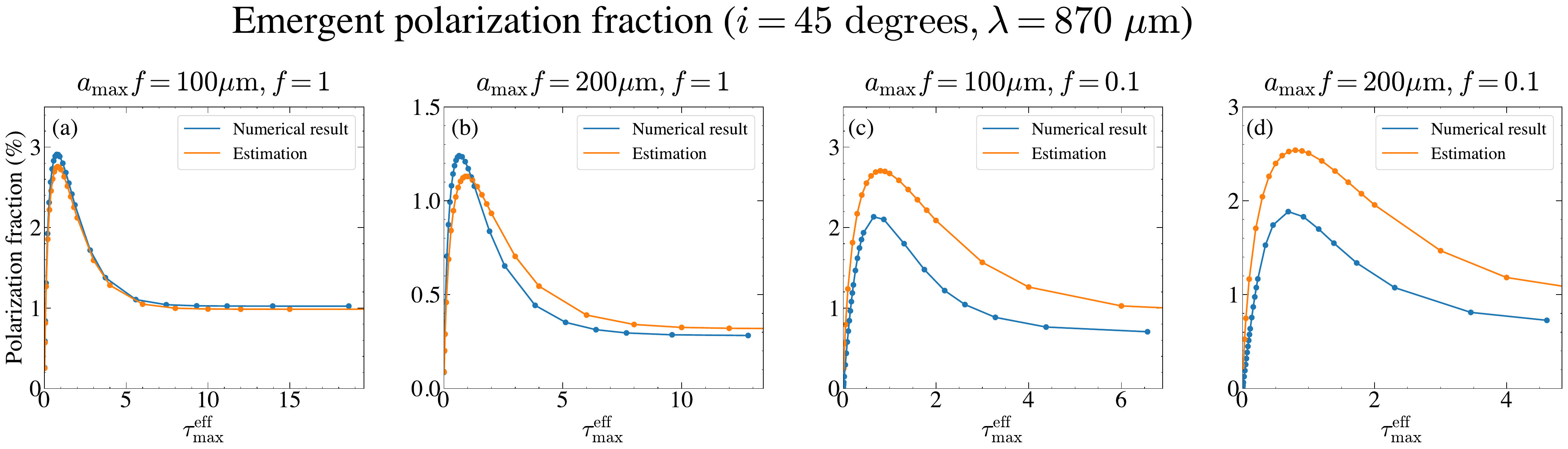}}
\caption{Comparison, for the four dust models, between the numerically computed emergent polarization including Mie scattering and the estimations based on a $90\tcdegree$-scattering rescaling, for $i=45\tcdegree$ and $\lambda=870~\mathrm{\mu m}$. To compute the estimations, we use our numerical results for Rayleigh scattering evaluated with $\omega=\omega_\mathrm{eff}$ and $\tau_\mathrm{max} = \tau_\mathrm{max}^\mathrm{eff}$, and then rescale the resulting polarization using the $90\tcdegree$-scattering polarization efficiency according to Eq. \ref{eq:PF_rescale}. }
\label{fig:DSHARP_vs_Rayleigh_PF}
\end{figure*}

We find that a $90\tcdegree$-scattering–based rescaling of the Rayleigh results reproduces the Mie-scattering polarization fraction reasonably well, but only for compact grains.
Fig. \ref{fig:DSHARP_vs_Rayleigh_PF} compares the numerically computed emergent polarization that accounts for Mie scattering with the estimations using a $90\tcdegree$-scattering–based rescaling \citep{2019ApJ...885...52T}.
To compute the estimations, we use our numerical results for Rayleigh scattering evaluated with $\omega=\omega_\mathrm{eff}$ and $\tau_\mathrm{max} = \tau_\mathrm{max}^\mathrm{eff}$. We then rescale the resulting polarization using the $90\tcdegree$-scattering polarization efficiency. Specifically, we define the estimated polarization fraction $P F_\mathrm{est}$ as
\begin{equation} \label{eq:PF_rescale}
\begin{split}
PF_\mathrm{est} &= PF_\mathrm{Rayleigh}(\omega_\mathrm{eff}, \tau_\mathrm{max}^\mathrm{eff}) \frac{|Z_{12}(\theta_\mathrm{s}=90\tcdegree)/Z_{11}(\theta_\mathrm{s}=90\tcdegree)|_\mathrm{Mie}}{|Z_{12}(\theta_\mathrm{s}=90\tcdegree)/Z_{11}(\theta_\mathrm{s}=90\tcdegree)|_\mathrm{Rayleigh}} \\
&= PF_\mathrm{Rayleigh}(\omega_\mathrm{eff}, \tau_\mathrm{max}^\mathrm{eff}) |Z_{12}(\theta_\mathrm{s}=90\tcdegree)/Z_{11}(\theta_\mathrm{s}=90\tcdegree)|_\mathrm{Mie},
\end{split}
\end{equation}
where $PF_\mathrm{Rayleigh}$ is the numerically computed polarization fraction for Rayleigh scattering.
For compact grains with $a_\mathrm{max}f=100~\mathrm{\mu m}$ and $f=1$ (panel (a) of Fig. \ref{fig:DSHARP_vs_Rayleigh_PF}), the estimation agrees well with the numerical result. For compact grains with $a_\mathrm{max}f = 200~\mathrm{\mu m}$ and $f=1$ (panel (b) of Fig. \ref{fig:DSHARP_vs_Rayleigh_PF}), the agreement is somewhat worse than in the case of $a_\mathrm{max}f = 100~\mathrm{\mu m}$ and $f=1$. However, the deviation remains within roughly $0.1 \%$, comparable to ALMA's typical observational uncertainties in the polarization fraction ($\sim 0.1\%$; \citealt{ALMA_technical_handbook}). Therefore, this estimation is sufficient for ALMA data analysis. In contrast, for porous grains (panels (c) and (d) of Fig. \ref{fig:DSHARP_vs_Rayleigh_PF}), the estimations do not agree with the numerical results. Thus, a $90\tcdegree$-scattering–based rescaling of the Rayleigh results is valid only in limited situations. Since this assessment is based on only four dust models, further validation over a wider parameter space is required.

One reason why the estimation is less accurate for porous grains than for compact grains may be the more strongly forward-peaked $Z_{11}$ and $Z_{12}$ of porous grains. Both $Z_{11}$ and $Z_{12}$ can strongly affect the estimated polarization: $Z_{12}$ converts incoming Stokes $I$ into polarization, whereas $Z_{11}$ (which equals $Z_{22}$ for spherical grains) converts the polarized component of the incoming intensity into scattered polarization. 
Therefore, differences in $Z_{11}$ and $Z_{12}$ between the compact- and porous-grain models may be responsible for the reduced estimation accuracy for porous grains. To assess these differences, the left and center panels of Fig. \ref{fig:scattering_matrix} compare $Z_{11}$ and $Z_{12}$ for four dust models as functions of the scattering angle. A key difference between compact and porous dust models is the forward-peakedness of $Z_{11}$ and $Z_{12}$: for porous grains, both $Z_{11}$ and $Z_{12}$ are more strongly forward-peaked than for compact grains. This suggests that, even for compact grains, the Mie-scattering polarization may not be accurately estimated when $Z_{11}$ or $Z_{12}$ becomes sufficiently forward-peaked.

However, $|Z_{12}/Z_{11}|$ is unlikely to be the primary cause of the failure of the estimation for porous grains. This is because, as shown in the right panel of Fig. \ref{fig:scattering_matrix}, $|Z_{12}/Z_{11}|$ for porous grains is similar to that for $a_\mathrm{max}f = 100~\mathrm{\mu m}$ and $f=1$, for which the estimation agrees well with the numerical results. This provides additional evidence that, contrary to the conventional expectation \citep{2015ApJ...809...78K, 2019ApJ...885...52T}, the polarization level is not determined solely by $|Z_{12}/Z_{11}|$.

\section{Discussion} \label{sec:Discussion}

\subsection{Comparison with Monte Carlo radiative transfer simulations}

\begin{figure*}[htbp]
\resizebox{\hsize}{!}{\includegraphics{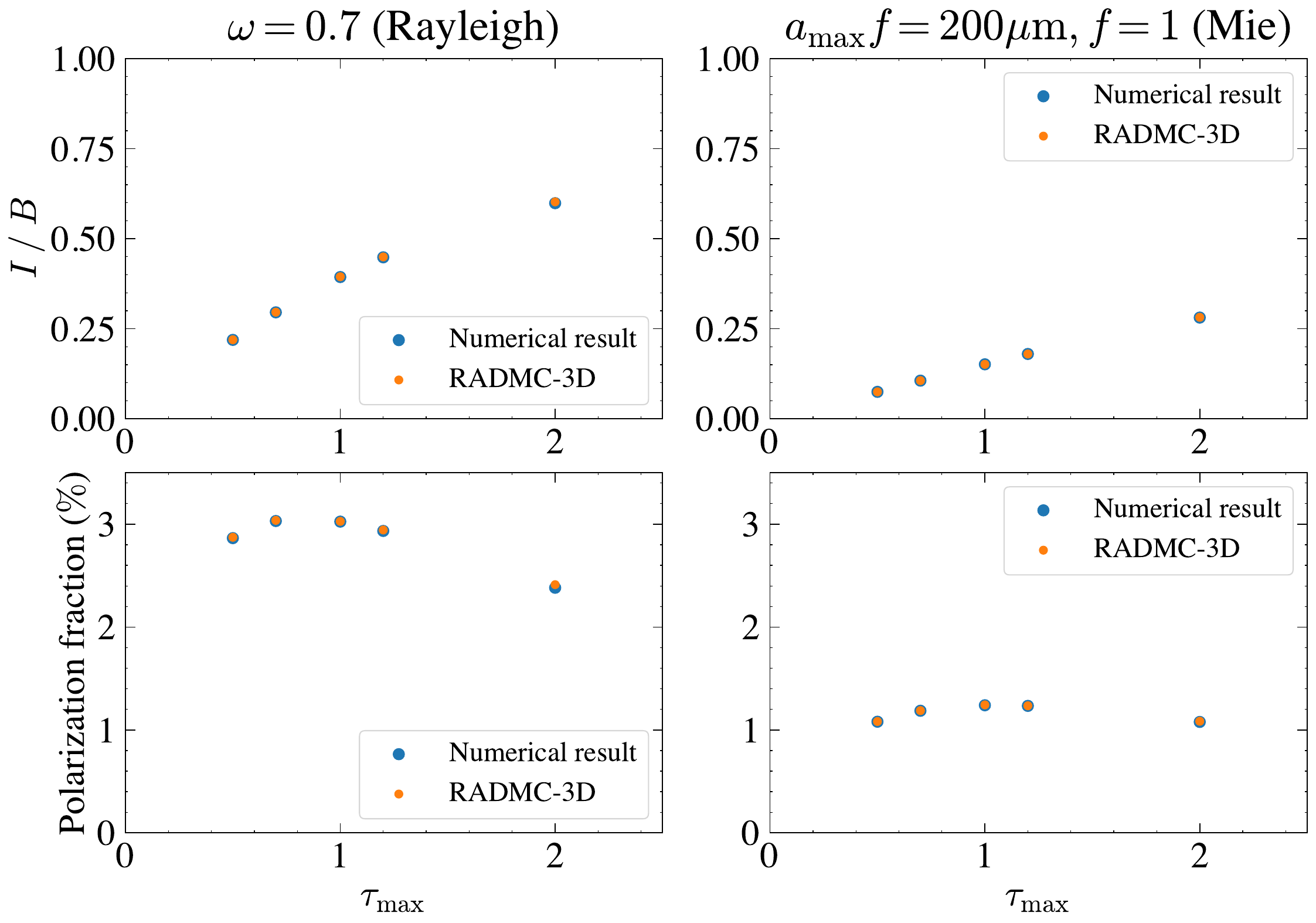}}
\caption{Comparison of the emergent Stokes $I$ and polarization fraction from a plane‐parallel slab inclined at $45 \tcdegree$ between RADMC‐3D \citep{2012ascl.soft02015D} simulations and our numerical solutions. Two cases are considered in the simulations: a Rayleigh scattering matrix with $\omega = 0.7$, and a Mie scattering matrix corresponding to the case with $a_\mathrm{max}f = 200~\mathrm{\mu m}$, $f = 1$, and a wavelength of $870~\mathrm{\mu m}$ adopted in Section \ref{subsec:Mie}. The slab optical depth is taken to be 0.5, 0.7, 1.0, 1.2, and 2.0.}
\label{fig:Comparsion_RADMC}
\end{figure*}

We validate our numerical results by comparing them with Monte Carlo radiative transfer simulations performed using RADMC-3D \citep{2012ascl.soft02015D}. In the Monte Carlo simulations, we adopt an isothermal, uniform-density dust slab with a surface area much larger than its vertical thickness. We evaluate the emergent Stokes $I$ and polarization fraction at the slab center to approximate the radiation from an infinite plane-parallel slab. Each Monte Carlo simulation uses $10^{10}$ photon packets. We perform the validation for two cases: a Rayleigh scattering matrix with $\omega=0.7$, and a Mie scattering matrix corresponding to the case with $a_\mathrm{max}f=200~\mathrm{\mu m}$, $f=1$, and a wavelength of $870~\mathrm{\mu m}$ used in Section \ref{subsec:Mie}. We run simulations for vertical optical depths $\tau_\mathrm{max}$ of 0.5, 0.7, 1.0, 1.2, and 2.0. The slab inclination is set to $45^\tcdegree$. Fig. \ref{fig:Comparsion_RADMC} shows the comparison between RADMC-3D results and our numerical calculations for Rayleigh scattering. For both cases, the close agreement in both emergent Stokes $I$ and polarization fraction confirms the validity of our numerical results. 

\subsection{Validity of the plane-parallel slab approximation} \label{subsec:Validity_slab}
While our analysis adopts an isothermal, constant-density plane-parallel slab approximation, these assumptions may not hold in real protoplanetary disks. If the disk is not vertically isothermal or constant-density, the applicability of our results may be limited. In addition, the plane-parallel approximation assumes that the disk is radially optically thick; this condition may be violated in disks with ring-like substructures or radially optically thin gaps, where caution is warranted. Moreover, the vertical dust distribution in real disks is generally considered to be Gaussian-like, whereas the plane-parallel slab approximation imposes a step-function–like truncation. This structural difference may modify the intensity of the radiation incident at grazing angles that produces the scattered light, and thus can affect the predicted polarization.

\subsection{Contribution of intrinsic polarization from aligned dust grains} \label{subsec:aligned_polarization}
To apply our polarized results to disk polarization observations, polarization contributions other than self-scattering should be accounted for. Our models consider only self-scattering polarization, whereas the polarization observed in real disks is a superposition of intrinsic polarization from aligned grains and scattering-induced polarization \citep[e.g.,][]{10.1093/mnras/stac753, 2024MNRAS.528..843L}. Accordingly, one should first decompose the measured polarization into these two components, and then compare the self-scattering component with our model predictions. This decomposition can be achieved by exploiting the distinct polarization-vector patterns produced by the two mechanisms \citep{2024MNRAS.528..843L}.

\subsection{Application to Observations}
In this subsection, we summarize how to estimate the disk emission based on our results, for use in the analysis of observational data, separately for continuum and polarization.

For the continuum, if one wishes to estimate the emergent Stokes $I$ accounting for Mie scattering, one can use the expression
\begin{equation}\label{eq:fitting_I_Mie}
\left\{
\begin{aligned}
I &= \left\{1-\omega_\mathrm{I}(\omega_\mathrm{eff}, \mu)\right\}
      \frac{\tau_\mathrm{max}^\mathrm{eff}}{\mu}\,B
      \qquad \qquad \qquad(\tau_\mathrm{max}^\mathrm{eff} < 0.04)\\[1mm]
I &= \Big[
      I_\mathrm{conv}(\omega_\mathrm{eff}, \mu)
      \left\{1-\exp\!\left(-A_\mathrm{I}(\omega_\mathrm{eff}, \mu)
      (\tau_\mathrm{max}^\mathrm{eff})^{B_\mathrm{I}(\omega_\mathrm{eff}, \mu)}\right)\right\} \\
  &\quad
      - I_\mathrm{conv}(\omega_\mathrm{eff}, \mu)
      \left\{1-\exp\!\left(-A_\mathrm{I}(\omega_\mathrm{eff}, \mu)
      0.04^{B_\mathrm{I}(\omega_\mathrm{eff}, \mu)}\right)\right\} \\
  &\quad
      + \left\{1-\omega_\mathrm{I}(\omega_\mathrm{eff}, \mu)\right\}
      \frac{0.04}{\mu}
      \Big]B
      \qquad \qquad \quad (\tau_\mathrm{max}^\mathrm{eff} > 0.04)
\end{aligned}
\right.
\end{equation}
obtained by applying the approximation in Section \ref{subsubsec:Mie_intensity} to Eq. (\ref{eq:fitting_I}).
Note, however, that this approximation tends to perform less well for dust models with strongly forward-peaked scattering, such as porous and large grains, as shown in Section \ref{subsubsec:Mie_intensity}.
The fitting coefficients for Eq. (\ref{eq:fitting_I_Mie}) are available on the website\footnotemark[\getrefnumber{fn:url}].

For the polarization, if one wishes to estimate the polarization fraction of the emergent intensity accounting for Mie scattering, one can use the expression
\begin{equation}
\begin{split} \label{eq:fitting_PF_Mie}
PF &= \Biggr[ A_\mathrm{PF}(\omega_\mathrm{eff}, \mu) (\tau_\mathrm{max}^\mathrm{eff})^{B_\mathrm{PF}(\omega_\mathrm{eff}, \mu)} \exp\left(-\left(\frac{\tau_\mathrm{max}^\mathrm{eff}}{\tau_\mathrm{PF}(\omega_\mathrm{eff}, \mu)}\right)^{0.8} \right)  \\
&\quad+ PF_\mathrm{conv}(\omega_\mathrm{eff}, \mu)\left\{ 1 - \exp \left( -\left(\frac{\tau_\mathrm{max}^\mathrm{eff}}{\tau_\mathrm{PF}(\omega_\mathrm{eff}, \mu)}\right)^{0.8} \right)\right\} \Biggr] \\
&\quad \times \left|\frac{Z_{12}(\theta_\mathrm{s}=90\tcdegree)}{Z_{11}(\theta_\mathrm{s}=90\tcdegree)}\right|
\end{split}
\end{equation}
obtained by applying the approximation in Section \ref{subsubsec:Mie_pol} to Eq. (\ref{eq:fitting_PF}).
Note, however, that this approximation tends to perform less well for dust models with strongly forward-peaked scattering, such as porous and large grains, as shown in Section \ref{subsubsec:Mie_pol}.
The fitting coefficients for Eq. (\ref{eq:fitting_PF_Mie}) are available on the website\footnotemark[\getrefnumber{fn:url}].

The same website\footnotemark[\getrefnumber{fn:url}] also provides a code that returns the disk emission by interpolating our numerical results. Readers may use either the fitting formulae or the interpolation code, depending on their purpose.

\section{Conclusion} \label{sec:conclusion}
We studied millimeter continuum emission and self-scattering polarization from protoplanetary disks by solving the radiative transfer equation in a one-dimensional plane-parallel slab, including dust absorption, emission, and self‐scattering with full Stokes parameters. Using the Rayleigh scattering matrix as the primary case and discussing implications for Mie scattering, we assessed the validity of commonly used analytic approximations and provided practical fitting formulae for rapid observational interpretation. Our main conclusions are as follows.
\begin{enumerate}
\item
We quantified how the emergent Stokes $I$, $Q$, and polarization fraction vary with the total optical depth of the slab $\tau_\mathrm{max}$, the dust albedo $\omega$, and the disk’s inclination $i$ (see Figs. \ref{fig:IQPF_inc45} and \ref{fig:Inclination_dependence}). In the optically thick regime, the $\omega$ and $i$ dependence of the emergent Stokes $I$ can be explained by a surface-layer effect.

\item
Scattering of the incoming Stokes $Q$ can make a non-negligible contribution to the emergent polarization. This is notable because the polarization fraction of the incoming intensity is only a few percent. By contrast, the polarized component of the incoming intensity has a negligible impact on the emergent Stokes $I$.

\item
Commonly used analytic approximations for the emergent Stokes $I$ \citep{2019ApJ...876....7S, 2018ApJ...869L..45B, 2019ApJ...877L..18Z} are systematically about 10 to $15\%$ lower than our numerical solutions. These errors are non-negligible because these errors exceed ALMA’s typical absolute flux calibration uncertainty of about 5 to 10\%. SED analyses of (sub)millimeter observations that adopt these formulae \citep[e.g.,][]{Carrasco-Gonzalez_2019, Macias2021, Sierra_2021, Guerra-Alvarado2024} are likely to overestimate the optical depth (and thus the disk mass) and the dust temperature, and underestimate the albedo (and thus altering the inferred constraints on grain size).

\item
We provide empirical fitting formulae that reproduce our numerical solutions for both the emergent Stokes $I$ and the polarization fraction (Eqs. (\ref{eq:fitting_I}) and (\ref{eq:fitting_PF})). These formulae enable efficient observational analyses, such as SED fitting. Specifically, our new fitting formula for the emergent Stokes $I$ will be more useful for observational data analysis than widely used approximations, which exhibit non-negligible errors. In addition, our new fitting formula for the emergent polarization fraction enables much faster polarization analyses than conventional approaches that rely on Monte Carlo radiative transfer simulations.

\item
For the case of Mie scattering, we also propose fitting formulae for the emergent Stokes $I$ and polarization fraction (Eqs. (\ref{eq:fitting_I_Mie}) and (\ref{eq:fitting_PF_Mie})). However, these formulae are not universally applicable to all dust models.
They tend to perform less well for dust models with strongly forward-peaked scattering, such as porous grains and large grains.

\end{enumerate}

\begin{acknowledgements}
The authors are grateful to Kiyoaki Doi, Tomohiro Yoshida, Kei Tanaka for useful discussions. This work was supported by JSPS KAKENHI Grant Number JP22K03680. This work was partially supported by Research Fund for Students (2024) of the Astronomical Science Program, The Graduate University for Advanced Studies, SOKENDAI.
\end{acknowledgements}

%
   \bibliographystyle{aa} 
   \bibliography{Kitade_citation} 

@ARTICLE{MIYAKE199320,
       author = {{Miyake}, Kotaro and {Nakagawa}, Yoshitsugu},
        title = "{Effects of Particle Size Distribution on Opacity Curves of Protoplanetary Disks around T Tauri Stars}",
      journal = {\icarus},
         year = 1993,
        month = nov,
       volume = {106},
       number = {1},
        pages = {20-41},
          doi = {10.1006/icar.1993.1156},
       adsurl = {https://ui.adsabs.harvard.edu/abs/1993Icar..106...20M}
}

@ARTICLE{Ricci2010,
       author = {{Ricci}, L. and {Testi}, L. and {Natta}, A. and {Neri}, R. and {Cabrit}, S. and {Herczeg}, G.~J.},
        title = "{Dust properties of protoplanetary disks in the Taurus-Auriga star forming region from millimeter wavelengths}",
      journal = {\aap},
         year = 2010,
        month = mar,
       volume = {512},
          eid = {A15},
        pages = {A15},
          doi = {10.1051/0004-6361/200913403},
archivePrefix = {arXiv},
       eprint = {0912.3356},
 primaryClass = {astro-ph.EP},
       adsurl = {https://ui.adsabs.harvard.edu/abs/2010A&A...512A..15R}
}

@ARTICLE{Carrasco-Gonzalez_2019,
       author = {{Carrasco-Gonz{\'a}lez}, Carlos and {Sierra}, Anibal and {Flock}, Mario and {Zhu}, Zhaohuan and {Henning}, Thomas and {Chandler}, Claire and {Galv{\'a}n-Madrid}, Roberto and {Mac{\'\i}as}, Enrique and {Anglada}, Guillem and {Linz}, Hendrik and {Osorio}, Mayra and {Rodr{\'\i}guez}, Luis F. and {Testi}, Leonardo and {Torrelles}, Jos{\'e} M. and {P{\'e}rez}, Laura and {Liu}, Yao},
        title = "{The Radial Distribution of Dust Particles in the HL Tau Disk from ALMA and VLA Observations}",
      journal = {\apj},
         year = 2019,
        month = sep,
       volume = {883},
       number = {1},
          eid = {71},
        pages = {71},
          doi = {10.3847/1538-4357/ab3d33},
archivePrefix = {arXiv},
       eprint = {1908.07140},
 primaryClass = {astro-ph.EP},
       adsurl = {https://ui.adsabs.harvard.edu/abs/2019ApJ...883...71C}
}

@ARTICLE{Macias2021,
       author = {{Mac{\'\i}as}, E. and {Guerra-Alvarado}, O. and {Carrasco-Gonz{\'a}lez}, C. and {Ribas}, {\'A}. and {Espaillat}, C.~C. and {Huang}, J. and {Andrews}, S.~M.},
        title = "{Characterizing the dust content of disk substructures in TW Hydrae}",
      journal = {\aap},
         year = 2021,
        month = apr,
       volume = {648},
          eid = {A33},
        pages = {A33},
          doi = {10.1051/0004-6361/202039812},
archivePrefix = {arXiv},
       eprint = {2102.04648},
 primaryClass = {astro-ph.EP},
       adsurl = {https://ui.adsabs.harvard.edu/abs/2021A&A...648A..33M}
}

@ARTICLE{Ueda_2020,
       author = {{Ueda}, Takahiro and {Kataoka}, Akimasa and {Tsukagoshi}, Takashi},
        title = "{Scattering-induced Intensity Reduction: Large Mass Content with Small Grains in the Inner Region of the TW Hya disk}",
      journal = {\apj},
         year = 2020,
        month = apr,
       volume = {893},
       number = {2},
          eid = {125},
        pages = {125},
          doi = {10.3847/1538-4357/ab8223},
archivePrefix = {arXiv},
       eprint = {2003.09353},
 primaryClass = {astro-ph.EP},
       adsurl = {https://ui.adsabs.harvard.edu/abs/2020ApJ...893..125U}
}

@ARTICLE{Ueda_2021,
       author = {{Ueda}, Takahiro and {Kataoka}, Akimasa and {Zhang}, Shangjia and {Zhu}, Zhaohuan and {Carrasco-Gonz{\'a}lez}, Carlos and {Sierra}, Anibal},
        title = "{Impact of Differential Dust Settling on the SED and Polarization: Application to the Inner Region of the HL Tau Disk}",
      journal = {\apj},
         year = 2021,
        month = jun,
       volume = {913},
       number = {2},
          eid = {117},
        pages = {117},
          doi = {10.3847/1538-4357/abf7b8},
archivePrefix = {arXiv},
       eprint = {2104.05927},
 primaryClass = {astro-ph.EP},
       adsurl = {https://ui.adsabs.harvard.edu/abs/2021ApJ...913..117U}
}

@ARTICLE{Ueda_2022,
       author = {{Ueda}, Takahiro and {Kataoka}, Akimasa and {Tsukagoshi}, Takashi},
        title = "{Massive Compact Dust Disk with a Gap around CW Tau Revealed by ALMA Multiband Observations}",
      journal = {\apj},
         year = 2022,
        month = may,
       volume = {930},
       number = {1},
          eid = {56},
        pages = {56},
          doi = {10.3847/1538-4357/ac634d},
archivePrefix = {arXiv},
       eprint = {2203.16236},
 primaryClass = {astro-ph.EP},
       adsurl = {https://ui.adsabs.harvard.edu/abs/2022ApJ...930...56U}
}

@ARTICLE{Sierra_2021,
       author = {{Sierra}, Anibal and {P{\'e}rez}, Laura M. and {Zhang}, Ke and {Law}, Charles J. and {Guzm{\'a}n}, Viviana V. and {Qi}, Chunhua and {Bosman}, Arthur D. and {{\"O}berg}, Karin I. and {Andrews}, Sean M. and {Long}, Feng and {Teague}, Richard and {Booth}, Alice S. and {Walsh}, Catherine and {Wilner}, David J. and {M{\'e}nard}, Fran{\c{c}}ois and {Cataldi}, Gianni and {Czekala}, Ian and {Bae}, Jaehan and {Huang}, Jane and {Bergner}, Jennifer B. and {Ilee}, John D. and {Benisty}, Myriam and {Le Gal}, Romane and {Loomis}, Ryan A. and {Tsukagoshi}, Takashi and {Liu}, Yao and {Yamato}, Yoshihide and {Aikawa}, Yuri},
        title = "{Molecules with ALMA at Planet-forming Scales (MAPS). XIV. Revealing Disk Substructures in Multiwavelength Continuum Emission}",
      journal = {\apjs},
         year = 2021,
        month = nov,
       volume = {257},
       number = {1},
          eid = {14},
        pages = {14},
          doi = {10.3847/1538-4365/ac1431},
archivePrefix = {arXiv},
       eprint = {2109.06433},
 primaryClass = {astro-ph.EP},
       adsurl = {https://ui.adsabs.harvard.edu/abs/2021ApJS..257...14S}
}

@ARTICLE{Guerra-Alvarado2024,
       author = {{Guerra-Alvarado}, Osmar M. and {Carrasco-Gonz{\'a}lez}, Carlos and {Mac{\'\i}as}, Enrique and {van der Marel}, Nienke and {Houge}, Adrien and {Maud}, Luke T. and {Pinilla}, Paola and {Villenave}, Marion and {Asaki}, Yoshiharu and {Humphreys}, Elizabeth},
        title = "{Into the thick of it: ALMA 0.45 mm observations of HL Tau at a resolution of 2 au}",
      journal = {\aap},
         year = 2024,
        month = jun,
       volume = {686},
          eid = {A298},
        pages = {A298},
          doi = {10.1051/0004-6361/202349046},
archivePrefix = {arXiv},
       eprint = {2404.04164},
 primaryClass = {astro-ph.EP},
       adsurl = {https://ui.adsabs.harvard.edu/abs/2024A&A...686A.298G}
}

@BOOK{1979rpa..book.....R,
       author = {{Rybicki}, George B. and {Lightman}, Alan P.},
        title = "{Radiative processes in astrophysics}",
        publisher = {New York: Wiley-Interscience},
         year = 1979,
       adsurl = {https://ui.adsabs.harvard.edu/abs/1979rpa..book.....R}
}

@ARTICLE{2019ApJ...876....7S,
       author = {{Sierra}, Anibal and {Lizano}, Susana and {Mac{\'\i}as}, Enrique and {Carrasco-Gonz{\'a}lez}, Carlos and {Osorio}, Mayra and {Flock}, Mario},
        title = "{An Analytical Model of Radial Dust Trapping in Protoplanetary Disks}",
      journal = {\apj},
         year = 2019,
        month = may,
       volume = {876},
       number = {1},
          eid = {7},
        pages = {7},
          doi = {10.3847/1538-4357/ab1265},
archivePrefix = {arXiv},
       eprint = {1903.08769},
 primaryClass = {astro-ph.EP},
       adsurl = {https://ui.adsabs.harvard.edu/abs/2019ApJ...876....7S}
}

@ARTICLE{2019ApJ...877L..18Z,
       author = {{Zhu}, Zhaohuan and {Zhang}, Shangjia and {Jiang}, Yan-Fei and {Kataoka}, Akimasa and {Birnstiel}, Tilman and {Dullemond}, Cornelis P. and {Andrews}, Sean M. and {Huang}, Jane and {P{\'e}rez}, Laura M. and {Carpenter}, John M. and {Bai}, Xue-Ning and {Wilner}, David J. and {Ricci}, Luca},
        title = "{One Solution to the Mass Budget Problem for Planet Formation: Optically Thick Disks with Dust Scattering}",
      journal = {\apjl},
         year = 2019,
        month = jun,
       volume = {877},
       number = {2},
          eid = {L18},
        pages = {L18},
          doi = {10.3847/2041-8213/ab1f8c},
archivePrefix = {arXiv},
       eprint = {1904.02127},
 primaryClass = {astro-ph.EP},
       adsurl = {https://ui.adsabs.harvard.edu/abs/2019ApJ...877L..18Z}
}

@ARTICLE{2014A&A...568A..42K,
       author = {{Kataoka}, Akimasa and {Okuzumi}, Satoshi and {Tanaka}, Hidekazu and {Nomura}, Hideko},
        title = "{Opacity of fluffy dust aggregates}",
      journal = {\aap},
         year = 2014,
        month = aug,
       volume = {568},
          eid = {A42},
        pages = {A42},
          doi = {10.1051/0004-6361/201323199},
archivePrefix = {arXiv},
       eprint = {1312.1459},
 primaryClass = {astro-ph.EP},
       adsurl = {https://ui.adsabs.harvard.edu/abs/2014A&A...568A..42K}
}

@ARTICLE{2015ApJ...809...78K,
       author = {{Kataoka}, Akimasa and {Muto}, Takayuki and {Momose}, Munetake and {Tsukagoshi}, Takashi and {Fukagawa}, Misato and {Shibai}, Hiroshi and {Hanawa}, Tomoyuki and {Murakawa}, Koji and {Dullemond}, Cornelis P.},
        title = "{Millimeter-wave Polarization of Protoplanetary Disks due to Dust Scattering}",
      journal = {\apj},
         year = 2015,
        month = aug,
       volume = {809},
       number = {1},
          eid = {78},
        pages = {78},
          doi = {10.1088/0004-637X/809/1/78},
archivePrefix = {arXiv},
       eprint = {1504.04812},
 primaryClass = {astro-ph.EP},
       adsurl = {https://ui.adsabs.harvard.edu/abs/2015ApJ...809...78K}
}

@ARTICLE{2016ApJ...820...54K,
       author = {{Kataoka}, Akimasa and {Muto}, Takayuki and {Momose}, Munetake and {Tsukagoshi}, Takashi and {Dullemond}, Cornelis P.},
        title = "{Grain Size Constraints on HL Tau with Polarization Signature}",
      journal = {\apj},
         year = 2016,
        month = mar,
       volume = {820},
       number = {1},
          eid = {54},
        pages = {54},
          doi = {10.3847/0004-637X/820/1/54},
archivePrefix = {arXiv},
       eprint = {1507.08902},
 primaryClass = {astro-ph.EP},
       adsurl = {https://ui.adsabs.harvard.edu/abs/2016ApJ...820...54K}
}

@ARTICLE{2017ApJ...844L...5K,
       author = {{Kataoka}, Akimasa and {Tsukagoshi}, Takashi and {Pohl}, Adriana and {Muto}, Takayuki and {Nagai}, Hiroshi and {Stephens}, Ian W. and {Tomisaka}, Kohji and {Momose}, Munetake},
        title = "{The Evidence of Radio Polarization Induced by the Radiative Grain Alignment and Self-scattering of Dust Grains in a Protoplanetary Disk}",
      journal = {\apjl},
         year = 2017,
        month = jul,
       volume = {844},
       number = {1},
          eid = {L5},
        pages = {L5},
          doi = {10.3847/2041-8213/aa7e33},
archivePrefix = {arXiv},
       eprint = {1707.01612},
 primaryClass = {astro-ph.EP},
       adsurl = {https://ui.adsabs.harvard.edu/abs/2017ApJ...844L...5K}
}

@ARTICLE{2016MNRAS.456.2794Y,
       author = {{Yang}, Haifeng and {Li}, Zhi-Yun and {Looney}, Leslie and {Stephens}, Ian},
        title = "{Inclination-induced polarization of scattered millimetre radiation from protoplanetary discs: the case of HL Tau}",
      journal = {\mnras},
         year = 2016,
        month = mar,
       volume = {456},
       number = {3},
        pages = {2794-2805},
          doi = {10.1093/mnras/stv2633},
archivePrefix = {arXiv},
       eprint = {1507.08353},
 primaryClass = {astro-ph.SR},
       adsurl = {https://ui.adsabs.harvard.edu/abs/2016MNRAS.456.2794Y}
}

@ARTICLE{2017MNRAS.472..373Y,
       author = {{Yang}, Haifeng and {Li}, Zhi-Yun and {Looney}, Leslie W. and {Girart}, Josep M. and {Stephens}, Ian W.},
        title = "{Scattering-produced (sub)millimetre polarization in inclined discs: optical depth effects, near-far side asymmetry and dust settling}",
      journal = {\mnras},
         year = 2017,
        month = nov,
       volume = {472},
       number = {1},
        pages = {373-388},
          doi = {10.1093/mnras/stx1951},
archivePrefix = {arXiv},
       eprint = {1705.05432},
 primaryClass = {astro-ph.SR},
       adsurl = {https://ui.adsabs.harvard.edu/abs/2017MNRAS.472..373Y}
}

@ARTICLE{2019ApJ...885...52T,
       author = {{Tazaki}, Ryo and {Tanaka}, Hidekazu and {Kataoka}, Akimasa and {Okuzumi}, Satoshi and {Muto}, Takayuki},
        title = "{Unveiling Dust Aggregate Structure in Protoplanetary Disks by Millimeter-wave Scattering Polarization}",
      journal = {\apj},
         year = 2019,
        month = nov,
       volume = {885},
       number = {1},
          eid = {52},
        pages = {52},
          doi = {10.3847/1538-4357/ab45f0},
archivePrefix = {arXiv},
       eprint = {1907.00189},
 primaryClass = {astro-ph.EP},
       adsurl = {https://ui.adsabs.harvard.edu/abs/2019ApJ...885...52T}
}

@ARTICLE{2023ApJ...953...96Z,
       author = {{Zhang}, Shangjia and {Zhu}, Zhaohuan and {Ueda}, Takahiro and {Kataoka}, Akimasa and {Sierra}, Anibal and {Carrasco-Gonz{\'a}lez}, Carlos and {Mac{\'\i}as}, Enrique},
        title = "{Porous Dust Particles in Protoplanetary Disks: Application to the HL Tau Disk}",
      journal = {\apj},
         year = 2023,
        month = aug,
       volume = {953},
       number = {1},
          eid = {96},
        pages = {96},
          doi = {10.3847/1538-4357/acdb4e},
archivePrefix = {arXiv},
       eprint = {2306.00158},
 primaryClass = {astro-ph.EP},
       adsurl = {https://ui.adsabs.harvard.edu/abs/2023ApJ...953...96Z}
}

@ARTICLE{2024NatAs...8.1148U,
       author = {{Ueda}, Takahiro and {Tazaki}, Ryo and {Okuzumi}, Satoshi and {Flock}, Mario and {Sudarshan}, Prakruti},
        title = "{Support for fragile porous dust in a gravitationally self-regulated disk around IM Lup}",
      journal = {Nature Astronomy},
         year = 2024,
        month = sep,
       volume = {8},
        pages = {1148-1158},
          doi = {10.1038/s41550-024-02308-6},
archivePrefix = {arXiv},
       eprint = {2406.07427},
 primaryClass = {astro-ph.EP},
       adsurl = {https://ui.adsabs.harvard.edu/abs/2024NatAs...8.1148U}
}

@ARTICLE{2018ApJ...869L..45B,
       author = {{Birnstiel}, Tilman and {Dullemond}, Cornelis P. and {Zhu}, Zhaohuan and {Andrews}, Sean M. and {Bai}, Xue-Ning and {Wilner}, David J. and {Carpenter}, John M. and {Huang}, Jane and {Isella}, Andrea and {Benisty}, Myriam and {P{\'e}rez}, Laura M. and {Zhang}, Shangjia},
        title = "{The Disk Substructures at High Angular Resolution Project (DSHARP). V. Interpreting ALMA Maps of Protoplanetary Disks in Terms of a Dust Model}",
      journal = {\apjl},
         year = 2018,
        month = dec,
       volume = {869},
       number = {2},
          eid = {L45},
        pages = {L45},
          doi = {10.3847/2041-8213/aaf743},
archivePrefix = {arXiv},
       eprint = {1812.04043},
 primaryClass = {astro-ph.SR},
       adsurl = {https://ui.adsabs.harvard.edu/abs/2018ApJ...869L..45B}
}

@misc{2012ascl.soft02015D,
  author       = {{Dullemond}, C.~P. and {Juh{\'a}sz}, A. and {Pohl}, A. and
                  {Sereshti}, F. and {Shetty}, R. and {Peters}, T. and
                  {Commer{\c c}on}, B. and {Flock}, M.},
  title        = "{RADMC-3D: A multi-purpose radiative transfer tool}",
  year         = {2012},
  note         = {Astrophysics Source Code Library, record ascl:1202.015},
}

@ARTICLE{10.1093/mnras/stac753,
       author = {{Lin}, Zhe-Yu Daniel and {Li}, Zhi-Yun and {Yang}, Haifeng and {Stephens}, Ian and {Looney}, Leslie and {Harrison}, Rachel and {Fern{\'a}ndez-L{\'o}pez}, Manuel},
        title = "{Thermal emission and scattering by aligned grains: Plane-parallel model and application to multiwavelength polarization of the HL Tau disc}",
      journal = {\mnras},
         year = 2022,
        month = may,
       volume = {512},
       number = {3},
        pages = {3922-3947},
          doi = {10.1093/mnras/stac753},
archivePrefix = {arXiv},
       eprint = {2112.10998},
 primaryClass = {astro-ph.SR},
       adsurl = {https://ui.adsabs.harvard.edu/abs/2022MNRAS.512.3922L}
}

@INCOLLECTION{1971430859765671051,
       author = {{Tsang}, L. and {Kong}, J.~A. and {Shin}, R.~T.},
        title = "{Theory of microwave remote sensing}",
        publisher="Wiley",
    booktitle = {Theory of microwave remote sensing by Tsang},
         year = 1985,
         pages = {},
       adsurl = {https://ui.adsabs.harvard.edu/abs/1985ntrs.book58641T}
}

@BOOK{1983asls.book.....B,
       author = {{Bohren}, Craig F. and {Huffman}, Donald R.},
        title = "{Absorption and scattering of light by small particles}",
        publisher = {New York: Wiley},
         year = 1983,
       adsurl = {https://ui.adsabs.harvard.edu/abs/1983asls.book.....B}
}

@BOOK{1978wpsr.book.....I,
       author = {{Ishimaru}, A.},
        title = "{Wave propagation and scattering in random media. Volume 1 - Single scattering and transport theory}",
        publisher = {New York: Academic},
         year = 1978,
       volume = {1},
          doi = {10.1016/B978-0-12-374701-3.X5001-7},
       adsurl = {https://ui.adsabs.harvard.edu/abs/1978wpsr.book.....I}
}

@ARTICLE{2025ApJ...990..183U,
       author = {{Ueda}, Takahiro and {Andrews}, Sean M. and {Carrasco-Gonz{\'a}lez}, Carlos and {Guerra-Alvarado}, Osmar M. and {Okuzumi}, Satoshi and {Tazaki}, Ryo and {Kataoka}, Akimasa},
        title = "{Multiwavelength Dust Characterization of the HL Tau Disk and Implications for Planet Formation}",
      journal = {\apj},
         year = 2025,
        month = sep,
       volume = {990},
       number = {2},
          eid = {183},
        pages = {183},
          doi = {10.3847/1538-4357/adf214},
archivePrefix = {arXiv},
       eprint = {2507.14443},
 primaryClass = {astro-ph.EP},
       adsurl = {https://ui.adsabs.harvard.edu/abs/2025ApJ...990..183U}
}

@techreport{ALMA_technical_handbook,
    author = {{Cortes}, P. C. and {Carpenter}, J. and {Kameno}, S. and {Loomis}, R. and {Vila-Vilaro}, B. and {Immer}, K. and {Vlahakis}, C. and {Law}, J. and
    {Stoehr}, F. and {Saini}, K. and {Hales}, A.},
    title = "{ALMA Technical Handbook, ALMA Doc. 12.3, ver. 1.5}",
    institution = {ALMA},
    year = 2025,
    doi = {10.5281/zenodo.14933753}
}

@BOOK{1978stat.book.....M,
       author = {{Mihalas}, Dimitri},
        title = "{Stellar atmospheres}",
        publisher = {San Francisco: W.H. Freeman},
         year = 1978,
       adsurl = {https://ui.adsabs.harvard.edu/abs/1978stat.book.....M}
}

@ARTICLE{2008JGRD..11314220W,
       author = {{Warren}, Stephen G. and {Brandt}, Richard E.},
        title = "{Optical constants of ice from the ultraviolet to the microwave: A revised compilation}",
      journal = {Journal of Geophysical Research (Atmospheres)},
         year = 2008,
        month = jul,
       volume = {113},
       number = {D14},
          eid = {D14220},
        pages = {D14220},
          doi = {10.1029/2007JD009744},
       adsurl = {https://ui.adsabs.harvard.edu/abs/2008JGRD..11314220W}
}

@ARTICLE{1996A&A...311..291H,
       author = {{Henning}, T. and {Stognienko}, R.},
        title = "{Dust opacities for protoplanetary accretion disks: influence of dust aggregates.}",
      journal = {\aap},
         year = 1996,
        month = jul,
       volume = {311},
        pages = {291-303},
       adsurl = {https://ui.adsabs.harvard.edu/abs/1996A&A...311..291H}
}

@ARTICLE{1991ApJ...381..250B,
       author = {{Beckwith}, Steven V.~W. and {Sargent}, Anneila I.},
        title = "{Particle Emissivity in Circumstellar Disks}",
      journal = {\apj},
         year = 1991,
        month = nov,
       volume = {381},
        pages = {250},
          doi = {10.1086/170646},
       adsurl = {https://ui.adsabs.harvard.edu/abs/1991ApJ...381..250B}
}

@ARTICLE{1983QJRAS..24..267H,
       author = {{Hildebrand}, R.~H.},
        title = "{The determination of cloud masses and dust characteristics from submillimetre thermal emission.}",
      journal = {\qjras},
         year = 1983,
        month = sep,
       volume = {24},
        pages = {267-282},
       adsurl = {https://ui.adsabs.harvard.edu/abs/1983QJRAS..24..267H}
}

@ARTICLE{2005ApJ...631.1134A,
       author = {{Andrews}, Sean M. and {Williams}, Jonathan P.},
        title = "{Circumstellar Dust Disks in Taurus-Auriga: The Submillimeter Perspective}",
      journal = {\apj},
         year = 2005,
        month = oct,
       volume = {631},
       number = {2},
        pages = {1134-1160},
          doi = {10.1086/432712},
archivePrefix = {arXiv},
       eprint = {astro-ph/0506187},
 primaryClass = {astro-ph},
       adsurl = {https://ui.adsabs.harvard.edu/abs/2005ApJ...631.1134A}
}

@ARTICLE{Adachi1976,
       author = {{Adachi}, I. and {Hayashi}, C. and {Nakazawa}, K.},
        title = "{The gas drag effect on the elliptical motion of a solid body in the primordial solar nebula.}",
      journal = {Progress of Theoretical Physics},
         year = 1976,
        month = dec,
       volume = {56},
        pages = {1756-1771},
          doi = {10.1143/PTP.56.1756},
       adsurl = {https://ui.adsabs.harvard.edu/abs/1976PThPh..56.1756A}
}

@ARTICLE{Weidenschilling1977,
       author = {{Weidenschilling}, S.~J.},
        title = "{Aerodynamics of solid bodies in the solar nebula.}",
      journal = {\mnras},
         year = 1977,
        month = jul,
       volume = {180},
        pages = {57-70},
          doi = {10.1093/mnras/180.2.57},
       adsurl = {https://ui.adsabs.harvard.edu/abs/1977MNRAS.180...57W}
}

@ARTICLE{Chokshi1993,
       author = {{Chokshi}, Arati and {Tielens}, A.~G.~G.~M. and {Hollenbach}, D.},
        title = "{Dust Coagulation}",
      journal = {\apj},
         year = 1993,
        month = apr,
       volume = {407},
        pages = {806},
          doi = {10.1086/172562},
       adsurl = {https://ui.adsabs.harvard.edu/abs/1993ApJ...407..806C}
}

@ARTICLE{Brauer2008,
       author = {{Brauer}, F. and {Dullemond}, C.~P. and {Henning}, Th.},
        title = "{Coagulation, fragmentation and radial motion of solid particles in protoplanetary disks}",
      journal = {\aap},
         year = 2008,
        month = mar,
       volume = {480},
       number = {3},
        pages = {859-877},
          doi = {10.1051/0004-6361:20077759},
archivePrefix = {arXiv},
       eprint = {0711.2192},
 primaryClass = {astro-ph},
       adsurl = {https://ui.adsabs.harvard.edu/abs/2008A&A...480..859B}
}

@ARTICLE{2024MNRAS.528..843L,
       author = {{Lin}, Zhe-Yu Daniel and {Li}, Zhi-Yun and {Stephens}, Ian W. and {Fern{\'a}ndez-L{\'o}pez}, Manuel and {Carrasco-Gonz{\'a}lez}, Carlos and {Chandler}, Claire J. and {Pasetto}, Alice and {Looney}, Leslie W. and {Yang}, Haifeng and {Harrison}, Rachel E. and {Sadavoy}, Sarah I. and {Henning}, Thomas and {Hughes}, A. Meredith and {Kataoka}, Akimasa and {Kwon}, Woojin and {Muto}, Takayuki and {Segura-Cox}, Dominique},
        title = "{Panchromatic (Sub)millimeter polarization observations of HL Tau unveil aligned scattering grains}",
      journal = {\mnras},
         year = 2024,
        month = feb,
       volume = {528},
       number = {1},
        pages = {843-862},
          doi = {10.1093/mnras/stae040},
archivePrefix = {arXiv},
       eprint = {2309.10055},
 primaryClass = {astro-ph.EP},
       adsurl = {https://ui.adsabs.harvard.edu/abs/2024MNRAS.528..843L}
}

@ARTICLE{2003ApJ...598.1017D,
       author = {{Draine}, B.~T.},
        title = "{Scattering by Interstellar Dust Grains. I. Optical and Ultraviolet}",
      journal = {\apj},
         year = 2003,
        month = dec,
       volume = {598},
       number = {2},
        pages = {1017-1025},
          doi = {10.1086/379118},
archivePrefix = {arXiv},
       eprint = {astro-ph/0304060},
 primaryClass = {astro-ph},
       adsurl = {https://ui.adsabs.harvard.edu/abs/2003ApJ...598.1017D}
}

\begin{appendix}





\section{Derivation of the radiative transfer equation}
\label{Appendix:derivation_RTeq}

\begin{figure*}[htbp]
\resizebox{\hsize}{!}{\includegraphics{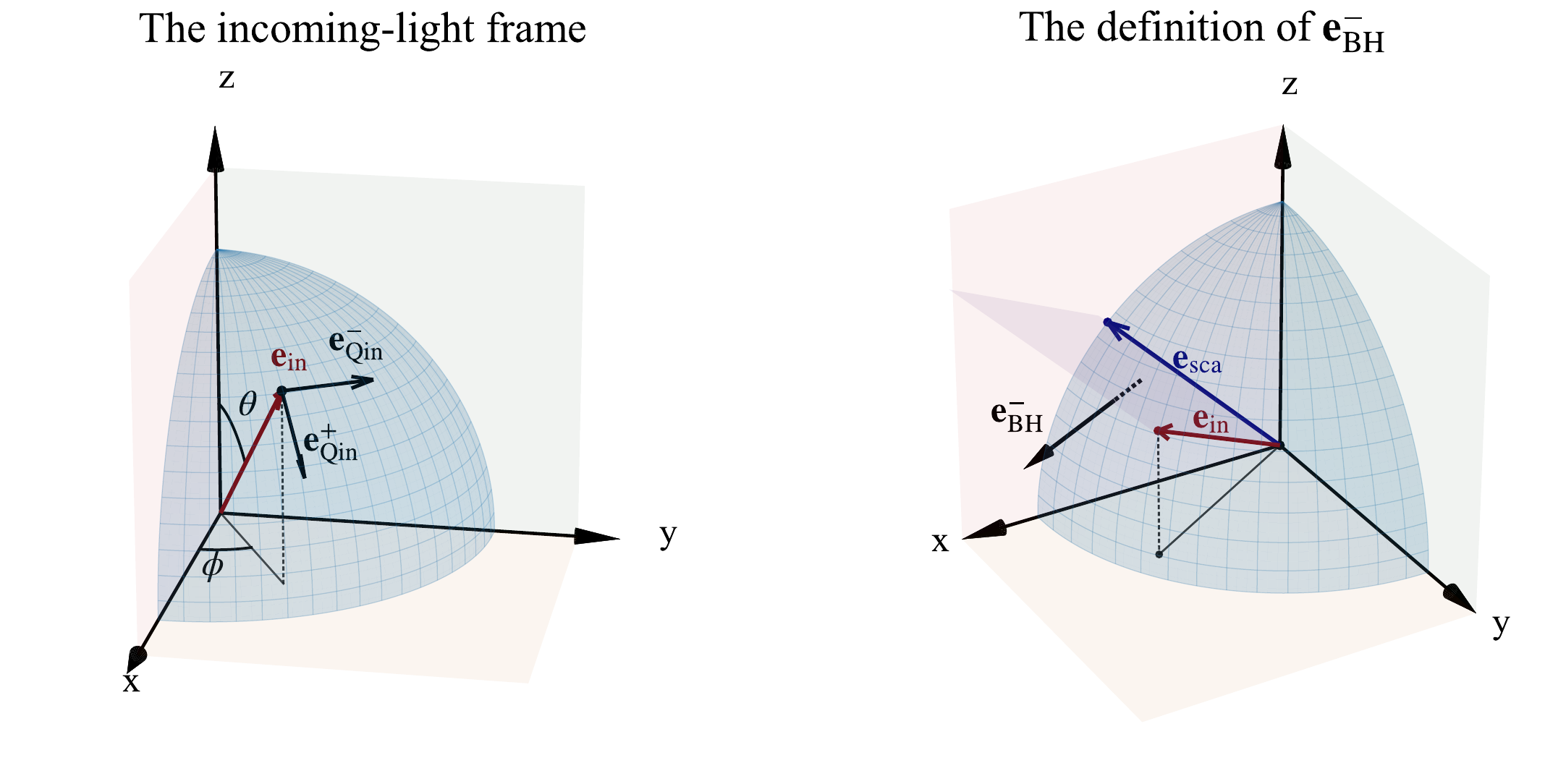}}
\caption{Left panel: Schematic representation of the incoming-light frame.
Right panel: Schematic representation of $\bm{e}_\mathrm{BH}^{-}$.
}
\label{fig:Stokes_reference}
\end{figure*}

We provide the detailed derivation of Eq. (\ref{eq:RT_fullStokes}). 
In previous works \citep[e.g.,][]{1971430859765671051}, the radiative transfer equation with full Stokes parameters, including absorption, emission, and scattering, is written as  
\begin{equation} \label{eq:RT_original}
\frac{d}{ds}
\begin{pmatrix}
I \\ Q \\U \\ V
\end{pmatrix}
= 
-nC_\mathrm{ext}
\begin{pmatrix}
I \\ Q \\U \\ V
\end{pmatrix}
+ 
nC_\mathrm{abs}
\begin{pmatrix}
B \\ 0 \\ 0 \\ 0
\end{pmatrix}
+
n \int  \bm{M}\frac{C_\mathrm{sca} \bm{Z}}{\kappa_\mathrm{sca}} \bm{M}'
\begin{pmatrix}
I_\mathrm{in} \\ Q_\mathrm{in} \\U_\mathrm{in} \\ V_\mathrm{in}
\end{pmatrix} d\Omega,
\end{equation}
where the terms on the right-hand side account for extinction, thermal emission, and scattering.
Here, $s$ is the distance along the light path. $I, Q, U, V$ are the Stokes parameters, and $B$ is the Planck function. $n$ is the number density of the grains. $C_\mathrm{ext}, C_\mathrm{abs}$, and $C_\mathrm{sca}$ are the extinction, absorption and scattering cross-sections. $\kappa_\mathrm{sca}$ is the mass scattering opacity. $\bm{M}$ and $\bm{M}'$ are the coordinate transformation matrices.
$\bm{Z}$ represents the scattering matrix.
To avoid ambiguity due to the definition of normalization, we specify our normalization; $\int_{4\pi} Z_{11} d\Omega = \kappa_\mathrm{sca}$, which we justify later. Recasting Eq. (\ref{eq:RT_original}) as a differential equation in the vertical extinction optical depth $\tau$ yields Eq. (\ref{eq:RT_fullStokes}).

We define four Stokes reference frames: (i) the incoming-light frame, (ii) the incoming-light frame associated with the scattering matrix, (iii) the scattered-light frame associated with the scattering matrix, and (iv) the laboratory frame.
First, we define the incoming-light frame (the left panel of Fig. \ref{fig:Stokes_reference}). When the propagation direction of the incoming light is given by
\begin{equation}
\bm{e}_\mathrm{in} = 
\begin{pmatrix}
\sin{\theta} \cos{\phi} \\
\sin{\theta} \sin{\phi} \\
\cos{\theta}
\end{pmatrix},
\end{equation}
we define the direction in which the Stokes parameter $Q_\mathrm{in}$ is positive as 
\begin{equation}
\bm{e}_{Q\mathrm{in}}^+ = 
\begin{pmatrix}
\cos{\theta} \cos{\phi} \\
\cos{\theta} \sin{\phi} \\
-\sin{\theta}
\end{pmatrix},
\end{equation}
and the direction in which it is negative as 
\begin{equation}
\bm{e}_\mathrm{Qin}^- = 
\begin{pmatrix}
-\sin{\phi} \\
\cos{\phi} \\
0
\end{pmatrix}.
\end{equation}
We adopt this definition so that the Stokes parameter $U$ of the incident radiation ($U_\mathrm{in}$) is always zero.
Next, we define the incoming- and scattered-light frames associated with the scattering matrix $\bm{Z}$. These frames follow the convention of \citet{1983asls.book.....B}. In both frames, the direction corresponding to negative Stokes $Q$, denoted by $\bm{e}_\mathrm{BH}^{-}$ (the right panel of Fig. \ref{fig:Stokes_reference}), is chosen to be perpendicular to both the incoming propagation direction $\bm{e}_\mathrm{in}$ and the scattered propagation direction $\bm{e}_\mathrm{sca}$, which coincides with the light path shown in Fig. \ref{fig:coordinate2}, where
\begin{equation}
\bm{e}_\mathrm{sca} = 
\begin{pmatrix}
\sin{i} \\
0 \\
\cos{i}
\end{pmatrix}.
\end{equation}
That is,
\begin{equation}
\begin{split}
\bm{e}_\mathrm{BH}^- = \frac{\bm{e}_\mathrm{in} \times \bm{e}_\mathrm{sca}}{|\bm{e}_\mathrm{in} \times \bm{e}_\mathrm{sca}|} 
= \frac{1}{t_\mathrm{norm} } \times 
\begin{pmatrix}
\sin{\theta} \sin{\phi} \cos{i} \\ 
\cos{\theta} \sin{i} - \sin{\theta} \cos{\phi} \cos{i} \\
-\sin{\theta} \sin{\phi} \sin{i}
\end{pmatrix},
\end{split}
\end{equation}
where the normalization factor is
\begin{equation}
t_\mathrm{norm} = \sqrt{\sin^2{\theta} \sin^2{\phi} + (\cos{\theta} \sin{i} - \sin{\theta} \cos{\phi} \cos{i})^2}.
\end{equation}
The direction corresponding to positive Stokes $Q$ is then defined as the vector perpendicular to both $\bm{e}_{\mathrm{BH}}^{-}$ and the propagation directions.
Finally, the laboratory frame is defined following Section \ref{sec:method} (Fig. \ref{fig:coordinate2}). In this frame, the direction corresponding to positive Stokes $Q$ is $\bm{e}_\mathrm{inc}$, while that corresponding to negative Stokes $Q$ is $\bm{e}_\mathrm{perp}$, defines as 
\begin{equation}
\bm{e}_\mathrm{inc} = 
\begin{pmatrix}
\cos{i} \\
0 \\
-\sin{i}
\end{pmatrix}
,
\end{equation}
\begin{equation}
\bm{e}_\mathrm{perp} = 
\begin{pmatrix}
0 \\
1 \\
0
\end{pmatrix}
.
\end{equation}

Using the Stokes reference frames defined above, we derive the rotation angles $A_\mathrm{rot}$ and $B_\mathrm{rot}$ that specify the coordinate transformation matrices $\bm{M}'$ and $\bm{M}$. The matrix $\bm{M}'$ transforms the incoming-light frame into the incoming-light frame associated with the scattering matrix $\bm{Z}$. Accordingly, the angle $A_\mathrm{rot}$ is defined as the oriented angle (right-hand rule about $\bm{e}_\mathrm{in}$) from $\bm{e}_\mathrm{Qin}^-$ to $\bm{e}_\mathrm{BH}^-$.
Hence,
\begin{equation}
\begin{split}
|A_\mathrm{rot}| &= \arccos{(\bm{e}_\mathrm{Qin}^- \cdot \bm{e}_\mathrm{BH}^-)} \\
&= \arccos{\left(\frac{-\sin{\theta} \cos{i} + \cos{\theta} \cos{\phi} \sin{i}}{\sqrt{\sin^2{\theta} \sin^2{\phi} + (\cos{\theta} \sin{i} - \sin{\theta} \cos{\phi} \cos{i})^2}}\right)}.
\end{split}
\end{equation}
The sign of $A_\mathrm{rot}$ coincides with the sign of $(\bm{e}_\mathrm{Qin}^- \times \bm{e}_\mathrm{BH}^-) \cdot \bm{e}_\mathrm{in}$.
Similarly, the matrix $\bm{M}$ transforms the scattered-light frame associated with the scattering matrix $\bm{Z}$ into the laboratory frame. Accordingly, the angle $B_\mathrm{rot}$ is defined as the oriented angle (right-hand rule about $\bm{e}_\mathrm{sca}$) from $\bm{e}_\mathrm{BH}^-$ to $\bm{e}_\mathrm{perp}$. Thus, 
\begin{equation} \label{eq:phidash}
\begin{split}
|B_\mathrm{rot}| &= \arccos{(\bm{e}_\mathrm{BH}^- \cdot \bm{e}_\mathrm{perp})} \\
&= \arccos{\left ( \frac{\cos{\theta} \sin{i} - \sin{\theta} \cos{\phi} \cos{i}}{\sqrt{\sin^2{\theta} \sin^2{\phi} + (\cos{\theta} \sin{i} - \sin{\theta} \cos{\phi} \cos{i})^2}}  \right )}.
\end{split}
\end{equation}
The sign of $B_\mathrm{rot}$ coincides with the sign of $(\bm{e}_\mathrm{BH}^- \times \bm{e}_\mathrm{perp}) \cdot \bm{e}_\mathrm{sca}$.

Next, we justify the normalization $\int_{4\pi} Z_{11} d\Omega = \kappa_\mathrm{sca}$ for Rayleigh scattering.
Following \citet{2012ascl.soft02015D}, we obtain the radiative transfer equation neglecting polarization as 
\begin{equation} \label{eq:RADMCRTeq}
\begin{split}
\frac{dI}{ds}&= - (\alpha^\mathrm{abs} + \alpha^\mathrm{scat})I + j^\mathrm{therm} + j^\mathrm{scat} \\
&=-nC_\mathrm{ext} I + nC_\mathrm{abs}B + nC_\mathrm{sca}\frac{1}{4\pi} \int \frac{4\pi}{\kappa_\mathrm{sca}} \frac{S_{11, \mathrm{BohrenH}}}{k^2 m_\mathrm{grain}}Id\Omega.
\end{split}
\end{equation}
Here, $\alpha^\mathrm{abs}$ and $\alpha^\mathrm{scat}$ are the extinction coefficients for absorption and scattering.  $k$ is the wave number, and $m_\mathrm{grain}$ is the mass of one dust particle.
$S_{11,\mathrm{BohrenH}}$ is the scattering matrix element defined by \citet{1983asls.book.....B}.
Using Eqs. (4.77) and (5.4) of \citet{1983asls.book.....B}, Eq. (\ref{eq:RADMCRTeq}) for Rayleigh scattering can be written as
\begin{equation}
\begin{split}
\frac{dI}{ds}=&-nC_\mathrm{ext}I + nC_\mathrm{abs}B \\
&+ \int\frac{4\pi}{\kappa_\mathrm{sca}}\frac{1}{k^2 m_\mathrm{grain}}\frac{1}{2}(|S_{1,\mathrm{BohrenH}}|^2+|S_{2,\mathrm{BohrenH}}|^2)Id\Omega \\
=&-nC_\mathrm{ext}I + nC_\mathrm{abs}B \\
&+ \int \frac{4\pi}{\kappa_\mathrm{sca}}\frac{1}{k^2 m_\mathrm{grain}}\frac{1}{2}\left(\frac{3}{2}\right)^2|a_1|^2(1+\cos^2{\theta_\mathrm{s}})Id\Omega,
\end{split}
\end{equation}
where $S_{1,\mathrm{BohrenH}}$ and $S_{2,\mathrm{BohrenH}}$ are the amplitude scattering matrix elements, and $a_1$ is the scattering coefficient.
In addition, following \citet{2012ascl.soft02015D} and \citet{1983asls.book.....B}, $\kappa_\mathrm{sca}$ for Rayleigh scattering can be written as
\begin{equation}
\begin{split}
\kappa_\mathrm{sca} &= \frac{1}{k^2 m_\mathrm{grain}} \int S_\mathrm{11,\mathrm{BohrenH}} d\Omega  \\
&= \frac{1}{k^2 m_\mathrm{grain}} \int \frac{9}{8}|a_1|^2 (1+\cos^2{\theta_\mathrm{s}})d\Omega \\
&= \frac{6}{k^2 m_\mathrm{grain}}|a_1|^2 \pi.
\end{split}
\end{equation}
Combining these relations with $\kappa_\mathrm{sca} = {nC_\mathrm{sca}}/{\rho}$ yields
\begin{equation}
\frac{dI}{ds} = -nC_\mathrm{ext}I + nC_\mathrm{abs}B + \int \frac{3}{8\pi}C_\mathrm{sca}\frac{1}{2}(1+\cos^2{\theta_\mathrm{s}})I d\Omega,
\end{equation}
and by comparing this with Eq. (\ref{eq:RT_original}) one finds $Z_{11} = 3\kappa_\mathrm{sca}/(16\pi)\times(1+\cos^2{\theta_\mathrm{s}})$.
This result implies that $Z_{11}$ should be normalized such that $\int_{4\pi} Z_{11}d\Omega = \kappa_\mathrm{sca}$ for Rayleigh scattering.
Although a separate proof is required for the Mie scattering case, we adopt the same normalization in this paper.

\section{$\tau$-dependence of the Stokes parameters}
\label{sec:Variation_inside_slab}

\begin{figure*}[htbp]
\resizebox{\hsize}{!}{\includegraphics{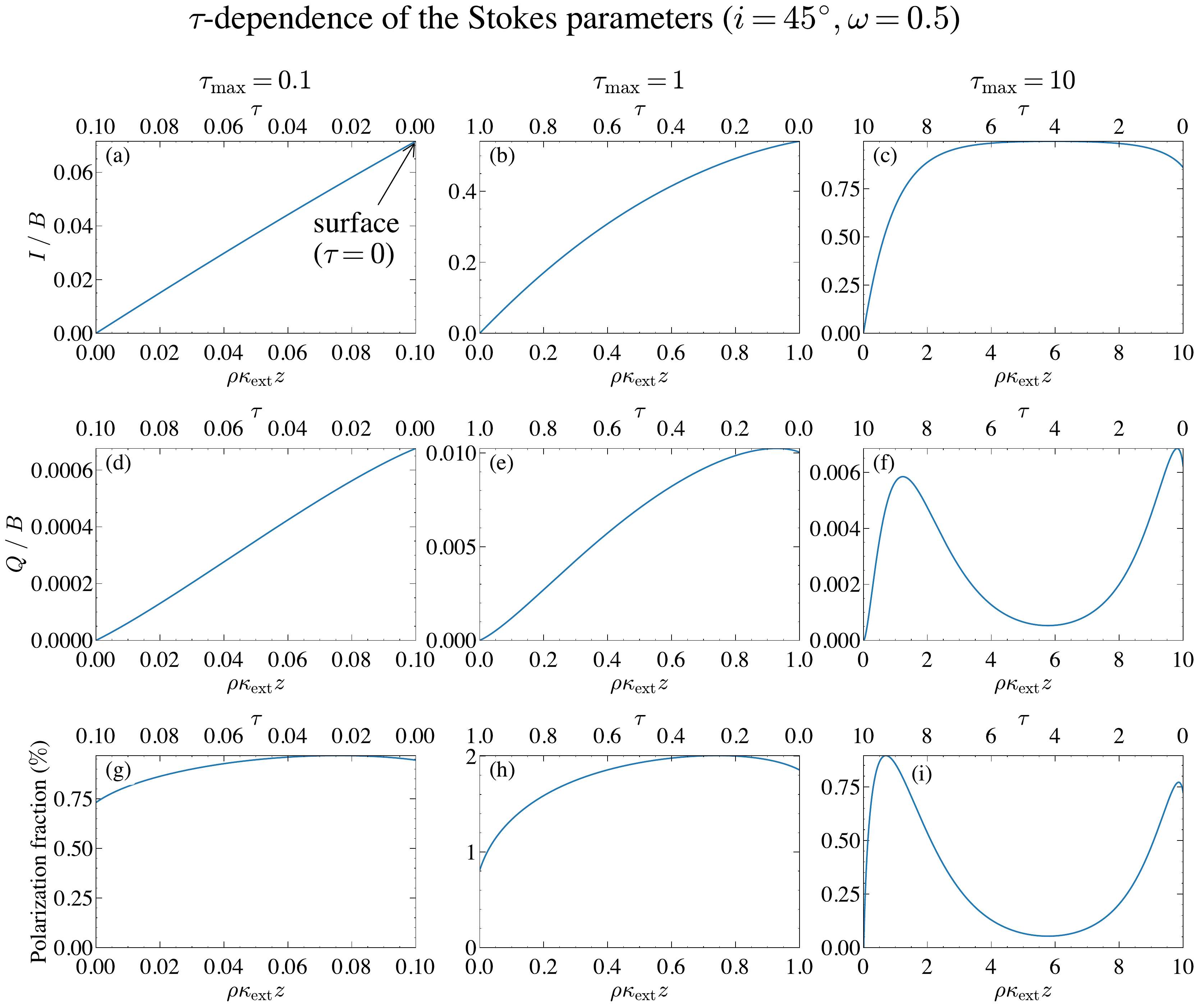}}
\caption{Dependence of the emergent Stokes $I$, Stokes $Q$, and polarization fraction on $z$ and the optical depth from the surface $\tau$, obtained by numerically solving Eq. (\ref{eq:RT_fullStokes}). We set $i=45\tcdegree$ and $\omega=0.5$. The left, middle, and right panels correspond to $\tau_\mathrm{max}$ = 0.1, 1.0, and 10.0, respectively. Stokes $I$ and $Q$ are normalized by the Planck function, $B$. In each panel, the lower axis indicates $\rho \kappa_\mathrm{ext} z$, while the upper axis indicates the vertical optical depth from the surface $\tau$. $\tau=0$ corresponds to the slab surface facing the observer.}
\label{fig:IQPF_inc45_omega0_5}
\end{figure*}

We examine how Stokes $I$, $Q$, and the polarization fraction vary with the vertical optical depth from the surface, $\tau$, within the slab. We adopt the fiducial case, where $i=45\tcdegree$, $\omega=0.5$, and $\tau_\mathrm{max}=0.1, 1.0, 10.0$.

In panel (a) of Fig. \ref{fig:IQPF_inc45_omega0_5}, Stokes $I$ increases as $\tau$ decreases for $\tau_\mathrm{max}=0.1$. This trend is because the slab is optically thin. In panel (d) of Fig. \ref{fig:IQPF_inc45_omega0_5}, Stokes Q also increases as $\tau$ decreases. This trend occurs because, in this optically thin limit, the scattering term’s $Q$ component (the third term on the right-hand side of Eq. (\ref{eq:RT_fullStokes})) contributes an almost constant positive value independent of $\tau$. In contrast, in panel (g) of Fig. \ref{fig:IQPF_inc45_omega0_5}, the polarization fraction $|Q/I|$ remains nearly constant with varying $\tau$. This is because the $\tau$-dependences of $I$ and $Q$ cancel out. Its finite value as $\tau \to 0.1$ reflects genuine physical behavior rather than a numerical artifact due to insufficient grid resolution in $\tau$ (see Eq. (\ref{eq:rough_app_PF})). Analytical expressions in the optically thin limit are discussed in Appendix \ref{Appendix:approximation}.

In panel (b) of Fig. \ref{fig:IQPF_inc45_omega0_5}, Stokes $I$ continues to increase as $\tau$ decreases for $\tau_\mathrm{max}=1.0$. In panels (e) and (h) of Fig. \ref{fig:IQPF_inc45_omega0_5}, Stokes $Q$ and the polarization fraction increase to a peak as $\tau$ decreases, and then decline toward smaller $\tau$.

Panel (c) is discussed in Section \ref{subsubsec:surface-layer}.
In panel (f) of Fig. \ref{fig:IQPF_inc45_omega0_5}, Stokes $Q$ takes very small values near the midplane ($\tau \sim 5$) for $\tau_\mathrm{max}=10$. This is because the polarization strength is determined by the anisotropy of the radiation field, which is nearly isotropic at the midplane. In panel (i) of Fig. \ref{fig:IQPF_inc45_omega0_5}, the polarization fraction follows the same $\tau$‐dependence as Stokes $Q$.

\section{Analytic approximations in the optically thin limit} \label{Appendix:approximation}

\subsection{Rough approximate formula} \label{subsec:simple_app}
In the optically thin limit, we find that the polarization fraction is proportional to $\sin^2{i}$. This result indicates that the radiation-field anisotropy produced by disk inclination, which determines the polarization strength, contributes a $\sin^2 i$ factor to the polarization. We also find that, in the optically thin limit, the polarization fraction is proportional to $\omega$. To demonstrate these scalings, we approximately solve Eq. (\ref{eq:RT_fullStokes}).

In the optically thin limit, multiple scattering is negligible, and thus the polarization of the incoming intensity can be ignored. In this regime, using the Rayleigh scattering matrix,  Eq. (\ref{eq:RT_fullStokes}) for Stokes $Q$ simplifies to:
\begin{equation} \label{eq:RTeq_Q}
\mu \frac{dQ}{d\tau} = Q - \frac{3}{8\pi}\omega \int \cos({2B_\mathrm{rot}}) \times \frac{1}{2}(\cos^2{\theta_\mathrm{s} - 1})I_\mathrm{in}d\Omega.
\end{equation}
To evaluate the scattering term (the second term on the right-hand side of Eq. (\ref{eq:RTeq_Q})), we adopt the following two approximations.

First, we assume that only grazing rays ($\mu'\sim0$) contribute to the scattered polarization. In an optically thin slab, radiation from other directions is very weak, so the polarized emission arises from grazing rays with $\mu' \sim 0$. 

\begin{figure}[tbp]
\resizebox{\hsize}{!}{\includegraphics{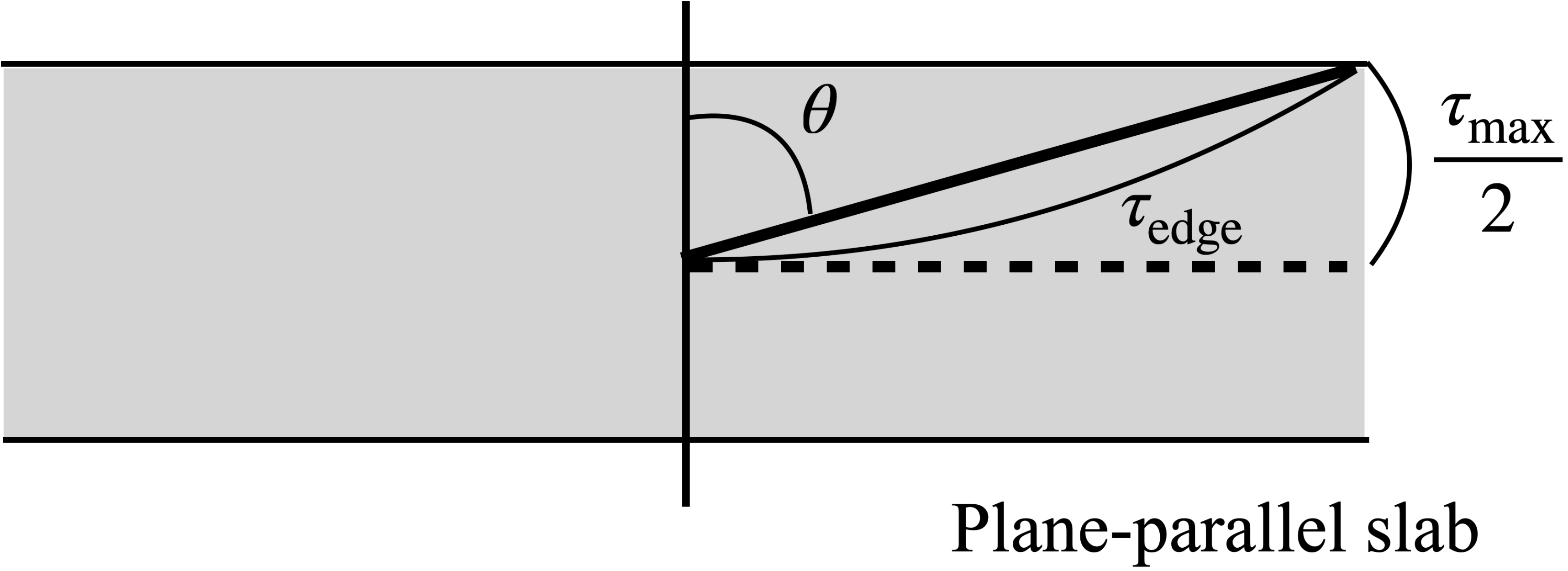}}
\caption{Schematic illustrating the range of incident directions $\mu'$ $(=\cos{\theta})$ that contribute to scattering for rays arriving at the slab midplane ($\tau=\tau_\mathrm{max}/2)$. $\tau_\mathrm{edge}$ is the optical depth from the mid-plane to the slab edge.
}
\label{fig:incidentangle}
\end{figure}

Second, we assume that the angular range of incident directions contributing to the scattering scales in proportion to $\tau_\mathrm{max}$. As noted in the first approximation, the dominant contributors have $\mu' \sim 0$, but in practice radiation with $\mu'$ near zero also contributes.
To estimate the width of this contributing range, we consider incident radiation at the midplane ($\tau=\tau_\mathrm{max}/2$), as shown in Fig. \ref{fig:incidentangle}. The incident directions that yield sufficiently long path lengths through the slab satisfy approximately $\mu' \lesssim \tau_\mathrm{max}/2\tau_\mathrm{edge}$, where $\tau_\mathrm{edge}$ is the optical depth from the midplane to the slab edge. In the optically thin limit, $\tau_\mathrm{edge}$ does not depend on $\tau_\mathrm{max}$; therefore, the width of the contributing $\mu$'-range is proportional to $\tau_\mathrm{max}$. Although this estimate is derived for the midplane ($\tau = \tau_\mathrm{max}/2$), the same approximation holds at other values of $\tau$.

Using the first approximation, Eq. (\ref{eq:phidash}) and the expression for the scattering angle $\theta_{\mathrm{s}}$,
\begin{equation} \label{eq:scattering_angle}
\cos{\theta_\mathrm{s}} = \bm{e}_\mathrm{in} \cdot \bm{e}_\mathrm{sca} = \sin{\theta}\cos{\phi}\sin{i}+\cos{\theta}\cos{i},
\end{equation}
Eq. (\ref{eq:RTeq_Q}) can be written as
\begin{equation} \label{eq:RT_eqQ3}
\begin{split}
&\mu \frac{dQ}{d\tau} = Q - \frac{3}{8\pi} \omega \times \\
                     \iint &\frac{\mu^2\cos^2{\phi} - \sin^2{\phi}}{\mu^2\cos^2{\phi} + \sin^2{\phi}} \times \frac{1}{2}  \left\{(1-\mu^2)\cos^2{\phi} - 1 \right\} I_\mathrm{in}(\mu'=0)d\mu' d\phi \\
                     &\qquad =Q - \frac{3}{8\pi}\omega \iint\ (\sin^2{\phi} - \mu^2\cos^2{\phi})\frac{1}{2}I_\mathrm{in}(\mu'=0) d\mu' d\phi \\
                     &\qquad=Q - \frac{3}{8\pi}\omega\pi (1-\mu^2)\frac{1}{2}\int_{-1}^1 I_\mathrm{in}(\mu'=0)d\mu' 
.
\end{split}
\end{equation}
$I_{\mathrm{in}}(\mu'=0)$ can be derived as follows. Setting $\mu=0$ in Eq. (\ref{eq:RTeq_I}) yields $I=S$. Moreover, the second term on the right-hand side of Eq. (\ref{eq:SourceFunction}) becomes negligible in the optically thin limit. Therefore,
\begin{equation}
I_{\mathrm{in}}(\mu'=0) = (1-\omega)B .
\end{equation}
Using this $I_\mathrm{in}(\mu'=0)$ and the second approximation, Eq. (\ref{eq:RT_eqQ3}) becomes
\begin{equation}
\mu\frac{dQ}{d\tau} = Q - \frac{3C}{16} (1-\mu^2) \omega (1-\omega) \tau_\mathrm{max}B, 
\end{equation}
where $C$ is a dimensionless proportionality constant in the second approximation.
Solving this differential equation for Stokes $Q$ in the optically thin limit yields
\begin{equation}
\begin{split}
Q &= \frac{3C}{16}(1-\mu^2)\omega(1-\omega)\tau_\mathrm{max}\left(1-\exp\left(-\frac{\tau_\mathrm{max} - \tau}{\mu} \right)\right)B \\
&\simeq \frac{3C}{16}\frac{1-\mu^2}{\mu}\omega(1-\omega)\tau_\mathrm{max}(\tau_\mathrm{max} - \tau) B.
\end{split}
\end{equation}
Using the optically thin limit expression for the intensity, $I = (1 - \omega)\{{(\tau_\mathrm{max} - \tau)}/{\mu}\} B$, the polarization fraction becomes:
\begin{equation} \label{eq:rough_app_PF}
PF \equiv\frac{Q}{I} = \frac{3C}{16}(1-\mu^2)\omega\tau_\mathrm{max}.
\end{equation}
From these results, the emergent Stokes $Q$ can be written as
\begin{equation} \label{eq:rough_app_Q}
Q(\tau=0) = \frac{3C}{16}\frac{1-\mu^2}{\mu}\omega(1-\omega)\tau^2_\mathrm{max} B
\end{equation}
and the polarization fraction of the emergent intensity is given by Eq. (\ref{eq:rough_app_PF}).

\begin{figure*}[htbp]
\resizebox{\hsize}{!}{\includegraphics{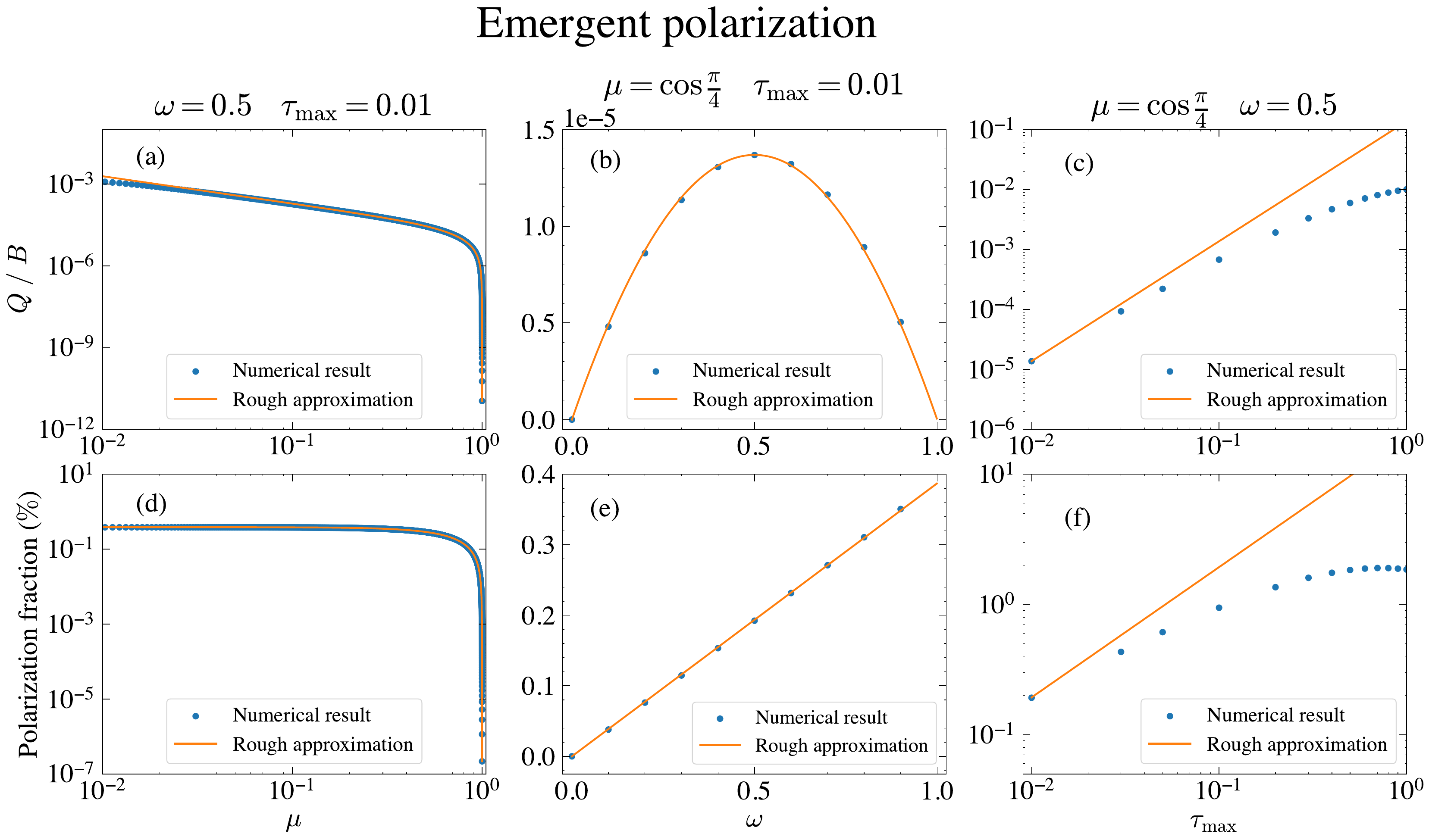}}
\caption{Panels (a)–(c) compare the emergent Stokes $Q$ predicted by the simplified approximation (Eq. (\ref{eq:rough_app_Q})) with the numerical solution, illustrating the dependencies on $\mu$, $\omega$, and $\tau_\mathrm{max}$, respectively. Panels (d)–(f) similarly compare the polarization fraction of the emergent intensity from the simplified formula (Eq. (\ref{eq:rough_app_PF})) with the numerical results for the same parameter dependencies. We choose $C=4.13$ to best match the numerical results.}
\label{fig:rough_analytic}
\end{figure*}

Fig. \ref{fig:rough_analytic} compares our approximations for the emergent Stokes $Q$ (Eq. (\ref{eq:rough_app_Q})) and polarization fraction (Eq. (\ref{eq:rough_app_PF})) with our numerical results. We adopt $\omega=0.5$, $\tau_\mathrm{max}=0.01$, and $\mu = \cos{({\pi}/{4})}$ as our fiducial parameter set. We choose $C=4.13$ to best match the numerical results. In panels (a), (b), (d), and (e) of Fig. \ref{fig:rough_analytic}, the $\mu$- and $\tau$-dependences of our approximations are in good agreement with the numerical results. In contrast, panels (c) and (f) of Fig. \ref{fig:rough_analytic} show that the $\tau_\mathrm{max}$-dependence of our approximations deviates significantly from the numerical results.

Panel (d) of Fig. \ref{fig:rough_analytic} shows that the polarization fraction of the emergent intensity scales with $1-\mu^{2}$, i.e., $\sin^{2} i$.
It is well understood that the polarization strength is controlled by the radiation-field anisotropy induced by disk inclination \citep{2015ApJ...809...78K, 2016MNRAS.456.2794Y}; however, the magnitude of this inclination effect has not been quantified explicitly.
In the optically thin limit, we find this scaling analytically, and we also confirm it with the numerical results: the inclination-induced anisotropy enters the polarization fraction as a $\sin^{2} i$ factor.

Panel (e) of Fig. \ref{fig:rough_analytic} shows that the polarization fraction of the emergent intensity scales with $\omega$.
Scattering-induced polarization is expected to depend sensitively on the scattering efficiency $\omega$, but the form of this dependence has not been quantified explicitly.
In the optically thin limit, we find this dependence analytically, and we also confirm it with the numerical results: the polarization fraction is proportional to $\omega$.

Panels (c) and (f) of Fig.\ref{fig:rough_analytic} show that our approximations do not reproduce the correct $\tau_{\mathrm{max}}$-dependence.
This may be because the range of $\tau_{\mathrm{max}}$ used in the plots is not sufficiently small and may exceed the extremely thin limit in which our approximations are valid.
Capturing the correct $\tau_{\mathrm{max}}$-dependence therefore requires the more rigorous analytical solution developed in Section~\ref{subsec:rigorous_app}.

Finally, we note that the polarization fraction derived above (Eq. (\ref{eq:rough_app_PF})) is independent of $\tau$. This is consistent with the result in Appendix \ref{sec:Variation_inside_slab} that the polarization fraction remains finite even in the limit $\tau \to 0$.

\subsection{Rigorous approximate formula} \label{subsec:rigorous_app}

\begin{figure*}[htbp]
\resizebox{\hsize}{!}{\includegraphics{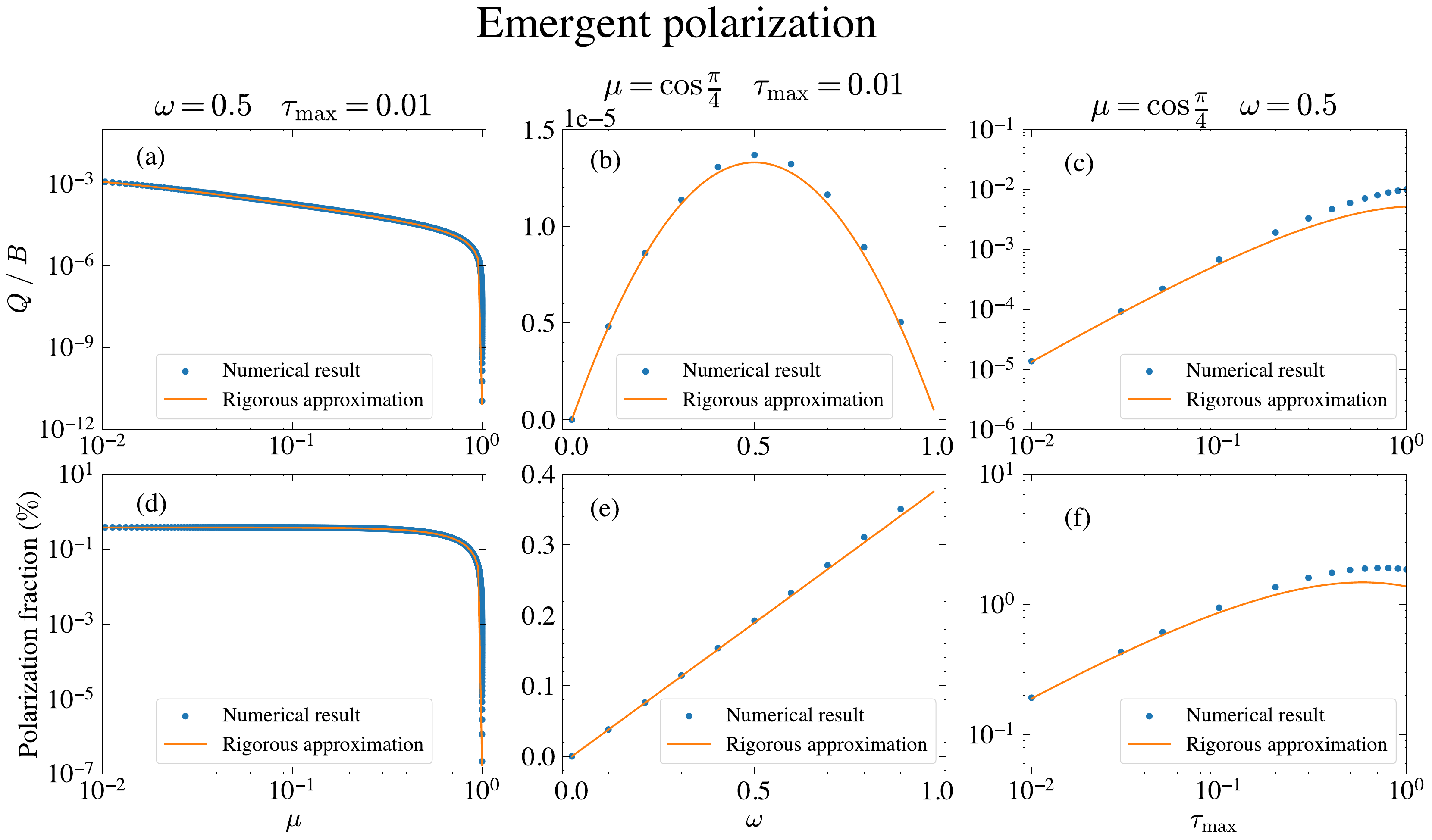}}
\caption{Panels (a)–(c) compare the emergent Stokes $Q$ predicted by the rigorous approximation (Eq. (\ref{eq:rigorous_app_emergent_Q})) with the numerical solution, illustrating the dependencies on $\mu$, $\omega$, and $\tau_\mathrm{max}$, respectively. Panels (d)–(f) similarly compare the emergent polarization fraction from the rigorous formula (Eq. (\ref{eq:rigorou_app_PF})) with the numerical results for the same parameter dependencies.}
\label{fig:rigorous_analytic}
\end{figure*}

In this appendix, we derive approximations that reproduce the numerical results more accurately than those in Appendix \ref{subsec:simple_app}, but require a simple numerical integration. For an extremely optically thin slab, the numerical calculation described in Section \ref{sec:method} becomes computationally expensive because a much finer discretization in the $\mu$ direction is required. The approximations derived here are therefore useful for obtaining the Stokes parameters for such extremely optically thin slabs.

Here we adopt only two simplifications: the assumption of the optically thin limit and the approximation of neglecting the polarization of the incoming intensity.
In the optically thin limit, the second term on the right-hand side of Eq. (\ref{eq:SourceFunction}) vanishes, so Eq. (\ref{eq:RTeq_I}) can be solved analytically. As a result, Stokes $I$ within the slab is given by
\begin{align} \label{eq:rigorous_app_I}
&I=\begin{cases}
(1-\omega)B (1-e^{\tau / \mu}) & \text{($\mu<0$)} \\
(1-\omega)B (1-e^{-(\tau_\mathrm{max} - \tau) / \mu}) & \text{($\mu > 0$)} \\
\end{cases}.
\end{align}
Substituting this form of the Stokes $I$ into Eq. (\ref{eq:RTeq_Q}), the radiative transfer equation for Stokes $Q$ is expressed as
\begin{equation}
\begin{split}
&\mu \frac{dQ}{d\tau} = Q - \frac{3}{8\pi}\omega \times \\
\biggr\{ &\int_0^{2\pi} \int_{-1}^0 \cos{(2B_\mathrm{rot})}\frac{1}{2} (\cos^2{\theta_\mathrm{s}} - 1) (1-\omega) 
B (1 - e^{\frac{\tau}{\mu'}}) d\mu' d\phi \\
+ &\int_0^{2\pi} \int_0^1 \cos{(2B_\mathrm{rot})}\frac{1}{2} (\cos^2{\theta_\mathrm{s}} - 1) (1-\omega) 
B (1 - e^{-\frac{\tau_\mathrm{max} - \tau}{\mu'}}) d\mu' d\phi \biggr\},
\end{split}
\end{equation}
which can be solved analytically to give
\begin{equation} \label{eq:rigotous_app_Q}
\begin{split}
&Q       = \frac{3}{8\pi}\omega \times   \\
&\biggr \{   \int_0^{2\pi} \int_{-1}^0      \cos{(2B_\mathrm{rot})} \frac{1}{2} (\cos^2{\theta_\mathrm{s}}-1) (1-\omega) 
B \times \\
            &                               \left(1- e^{-(\tau_\mathrm{max} - \tau) / \mu} - \frac{\mu'}{\mu' - \mu} e^{\tau / \mu'} + \frac{\mu'}{\mu' - \mu} e^{\tau_\mathrm{max} / \mu' -(\tau_\mathrm{max} - \tau)/\mu} \right) d\mu' d\phi   \\
&+            \int_0^{2\pi} \int_0^1        \cos{(2B_\mathrm{rot})} \frac{1}{2} (\cos^2{\theta_\mathrm{s}}-1) (1-\omega) B \times\\
            &                               \left(1- e^{-(\tau_\mathrm{max} - \tau) / \mu} - \frac{\mu'}{\mu' - \mu} e^{-(\tau_\mathrm{max} - \tau) / \mu'} + \frac{\mu'}{\mu' - \mu} e^{-(\tau_\mathrm{max} - \tau)/\mu} \right) d\mu' d\phi
\biggr \}
\end{split}
\end{equation}
under the boundary condition $Q(\tau = \tau_\mathrm{max}) = 0$.
Setting $\tau=0$ in Eq. (\ref{eq:rigotous_app_Q}), the emergent Stokes $Q$ is expressed as
\begin{equation} \label{eq:rigorous_app_emergent_Q}
\begin{split}
&Q(\tau=0)       = \frac{3}{8\pi}\omega \times   \\
\biggr \{   &\int_0^{2\pi} \int_{-1}^0      \cos{(2B_\mathrm{rot})} \frac{1}{2} (\cos^2{\theta_\mathrm{s}}-1) (1-\omega) 
B \times \\
            &                               \left(1- e^{-\tau_\mathrm{max} / \mu} - \frac{\mu'}{\mu' - \mu} + \frac{\mu'}{\mu' - \mu} e^{\tau_\mathrm{max} / \mu' -\tau_\mathrm{max}/\mu} \right) d\mu' d\phi   \\
+           & \int_0^{2\pi} \int_0^1        \cos{(2B_\mathrm{rot})} \frac{1}{2} (\cos^2{\theta_\mathrm{s}}-1) (1-\omega) B \times\\
            &                               \left(1- e^{-\tau_\mathrm{max} / \mu} - \frac{\mu'}{\mu' - \mu} e^{-\tau_\mathrm{max} / \mu'} + \frac{\mu'}{\mu' - \mu} e^{-\tau_\mathrm{max}/\mu} \right) d\mu' d\phi
\biggr \}.
\end{split}
\end{equation}
Here, $B_\mathrm{rot}$ and $\theta_\mathrm{s}$ are given by Eqs. (\ref{eq:phidash}) and (\ref{eq:scattering_angle}).
Although the resulting integrals over $\mu'$ and $\phi$ cannot be expressed in closed form, they are trivial to evaluate numerically. By numerically computing only these angular integrals in Eq. (\ref{eq:rigorous_app_emergent_Q}), we obtain the rigorous analytic approximation for Stokes $Q$.
Using this result, the polarization fraction of the emergent intensity can likewise be expressed as 
\begin{equation} \label{eq:rigorou_app_PF}
PF(\tau=0) = Q(\tau = 0) / \left  \{(1-\omega) \frac{\tau_\mathrm{max}}{\mu} B \right \}.
\end{equation}

These approximations accurately reproduce our numerical results for $\tau_{\max} \lesssim 0.1$.
Figure \ref{fig:rigorous_analytic} compares our approximations for the emergent Stokes $Q$ and
the polarization fraction with the numerical solutions. As a fiducial parameter set, we
adopt $\omega=0.5$, $\tau_{\max}=0.01$, and $\mu=\cos(\pi/4)$.
Panels (a), (b), (d), and (e) of Fig.~\ref{fig:rigorous_analytic} show that the $\mu$- and
$\omega$-dependences of these approximations are in good agreement with our numerical results.
Panels (c) and (f) further demonstrate that the $\tau_{\max}$ dependence also agrees well
for $\tau_{\max} \lesssim 0.1$. The deviations at larger $\tau_{\max}$ likely arise because
slabs with $\tau_{\max} \gtrsim 0.1$ are no longer sufficiently optically thin, and
Eq. (\ref{eq:rigorous_app_I}) therefore becomes a poor approximation.

\section{Numerical results for other inclinations} \label{Appendix:AddInc}

\begin{figure*}[htbp]
\resizebox{\hsize}{!}{\includegraphics{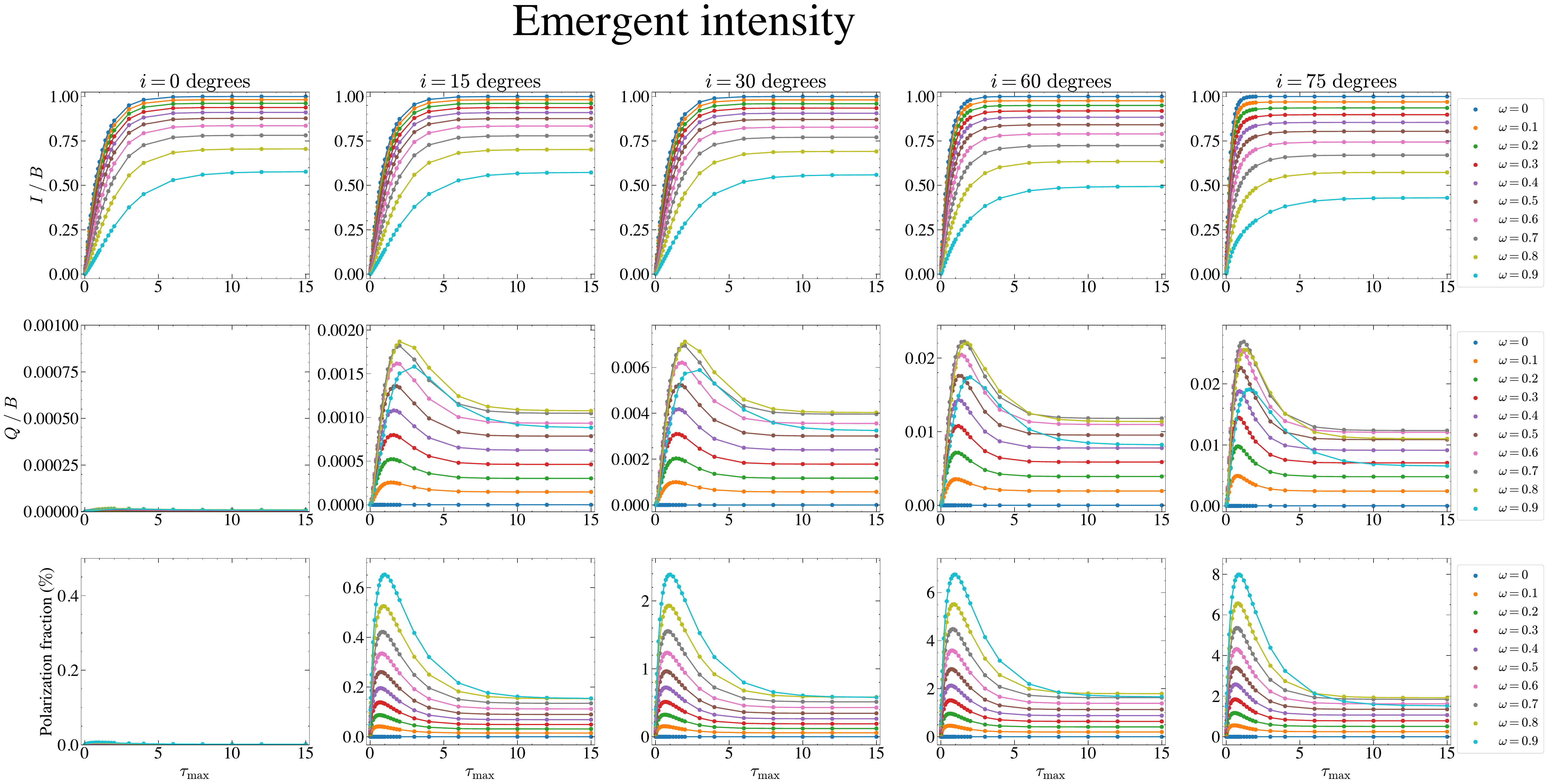}}
\caption{The $\tau_\mathrm{max}$– and $\omega$–dependence of the emergent Stokes $I$, $Q$, and polarization fraction for the Rayleigh scattering matrix at inclinations of $0\tcdegree, 15\tcdegree, 30\tcdegree, 60\tcdegree$ and $75\tcdegree$.}
\label{fig:IQPF_inc0_15_30_60_75}
\end{figure*}

Fig. \ref{fig:IQPF_inc0_15_30_60_75} presents the $\tau_\mathrm{max}$– and $\omega$–dependence of the emergent Stokes $I$, $Q$, and polarization fraction for the Rayleigh scattering matrix at $i=0\tcdegree, 15\tcdegree, 30\tcdegree, 60\tcdegree$ and $75\tcdegree$. As the inclination increases, Stokes $I$ decreases due to the surface-layer effect, whereas Stokes $Q$ and the polarization fraction increase because the radiation field becomes more anisotropic. Moreover, at an inclination of $0\tcdegree$, no scattering-induced polarization is produced. See also Section \ref{subsub:Dependence_on_inclination} and Fig. \ref{fig:Inclination_dependence} for related discussion.

\section{Comparison with analytical approximations for other inclinations} \label{Appendix:AddInc_comparison}

\begin{figure*}[h!]
\resizebox{\hsize}{!}{\includegraphics{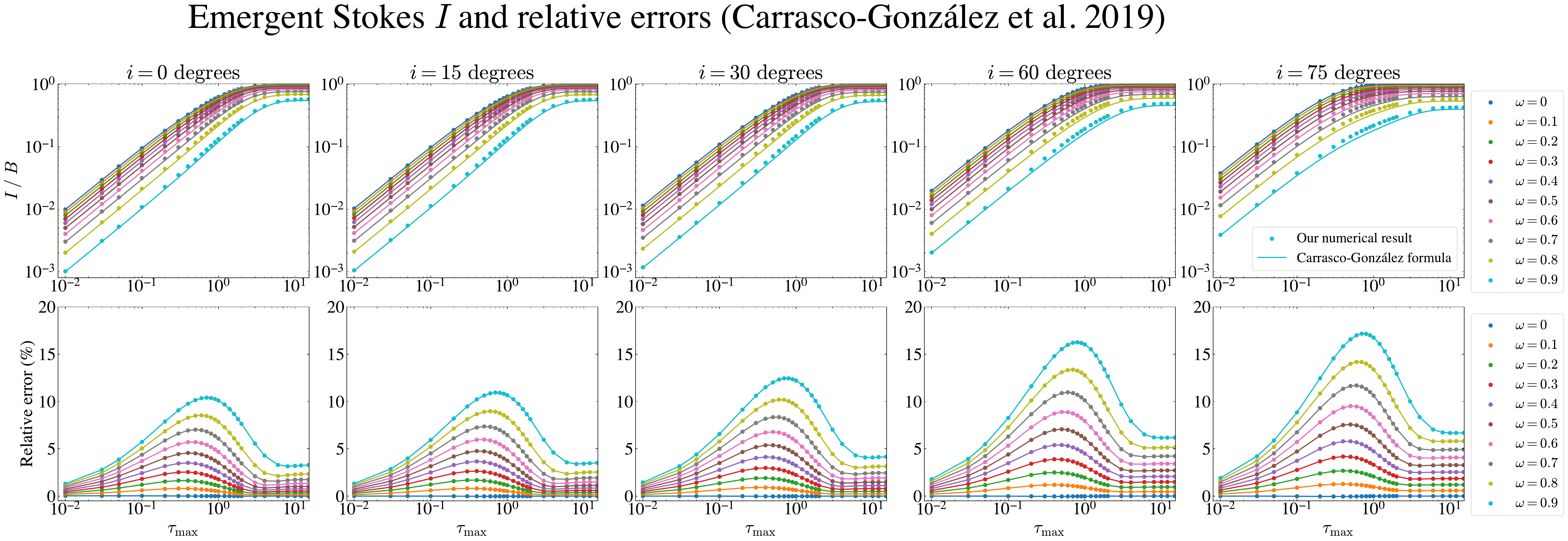}}
\caption{The upper panels compare the dependence of the emergent Stokes $I$ on $\tau_\mathrm{max}$ and $\omega$ between analytic approximations by \citet{Carrasco-Gonzalez_2019} and our numerical solutions for a slab at $i=0\tcdegree, 15\tcdegree, 30\tcdegree, 60\tcdegree$, and $75\tcdegree$, with dots indicating numerical results and solid lines showing the analytic formulae. The lower panels plot the corresponding relative error between each approximation and the numerical calculation.}
\label{fig:Relative_error_CG19}
\end{figure*}

\begin{figure*}[h!]
\resizebox{\hsize}{!}{\includegraphics{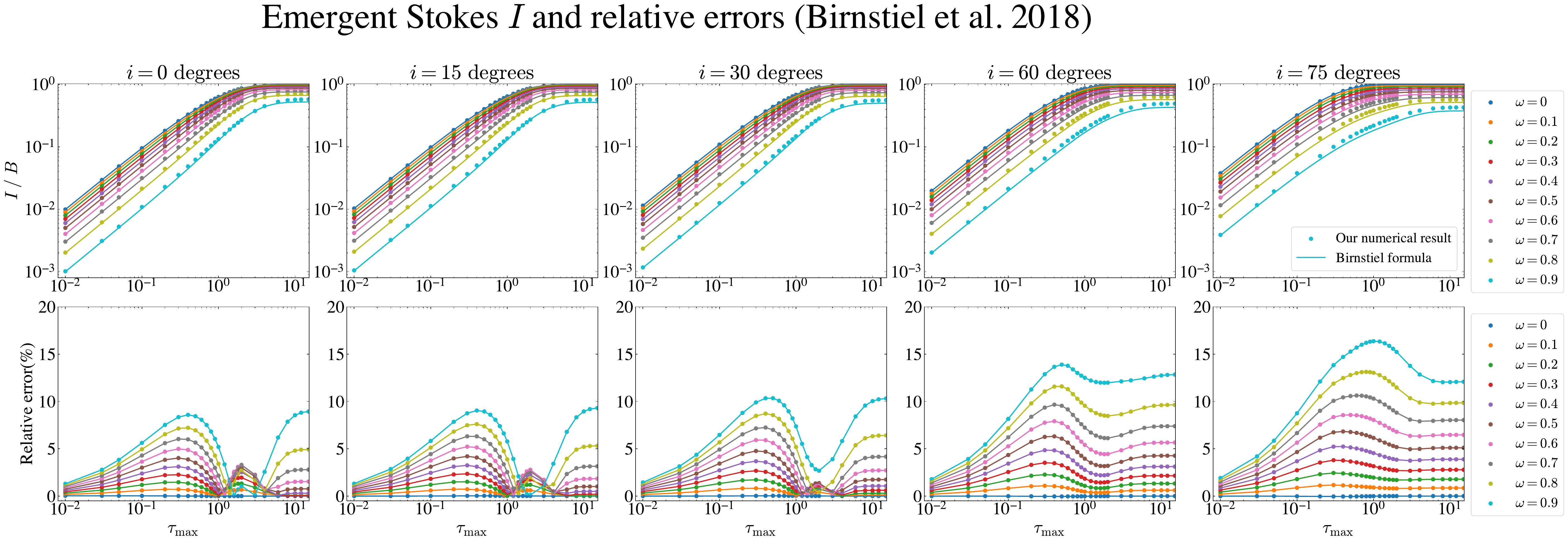}}
\caption{Same as Fig. \ref{fig:Relative_error_CG19}, but for the emergent-intensity approximation of \citet{2018ApJ...869L..45B}. The relative errors in the lower panels are shown in absolute value.}
\label{fig:Relative_error_Till18}
\end{figure*}

\begin{figure*}[h!]
\resizebox{\hsize}{!}{\includegraphics{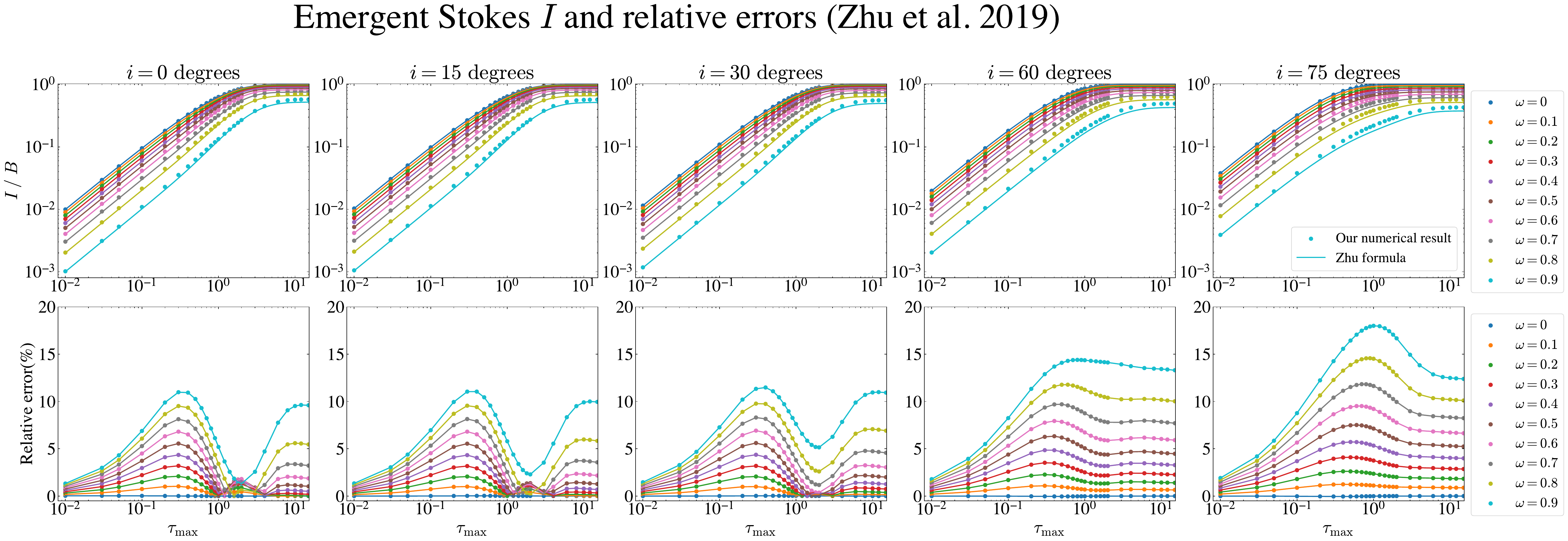}}
\caption{Same as Fig. \ref{fig:Relative_error_CG19}, but for the emergent-intensity approximation of \citet{2019ApJ...877L..18Z}. The relative errors in the lower panels are shown in absolute value.}
\label{fig:Relative_error_Zhu}
\end{figure*}

The upper panels of Fig. \ref{fig:Relative_error_CG19} compare our numerically computed emergent Stokes $I$ with the analytic approximation of \citet{Carrasco-Gonzalez_2019} at $i=0\tcdegree, 15\tcdegree, 30\tcdegree, 60\tcdegree,$ and $75\tcdegree$.
The lower panels of Fig. \ref{fig:Relative_error_CG19} show the relative error, defined as $100 \times |(I_\mathrm{num} - I_\mathrm{CG19}) / I_\mathrm{num}|$. Figs. \ref{fig:Relative_error_Till18} and \ref{fig:Relative_error_Zhu} show the corresponding comparisons with \citet{2018ApJ...869L..45B} and \citet{2019ApJ...877L..18Z}, respectively. For all three approximations, the relative error increases with inclination. Moreover, for \citet{2018ApJ...869L..45B} and \citet{2019ApJ...877L..18Z}, the sign of the discrepancy reverses around $\tau_\mathrm{max} \sim 2$, with the numerical intensity becoming smaller than the analytic prediction.

\section{Fitting parameters} \label{Appendix:AddInc_fitting}

\begin{figure*}[h!]
\resizebox{\hsize}{!}{\includegraphics{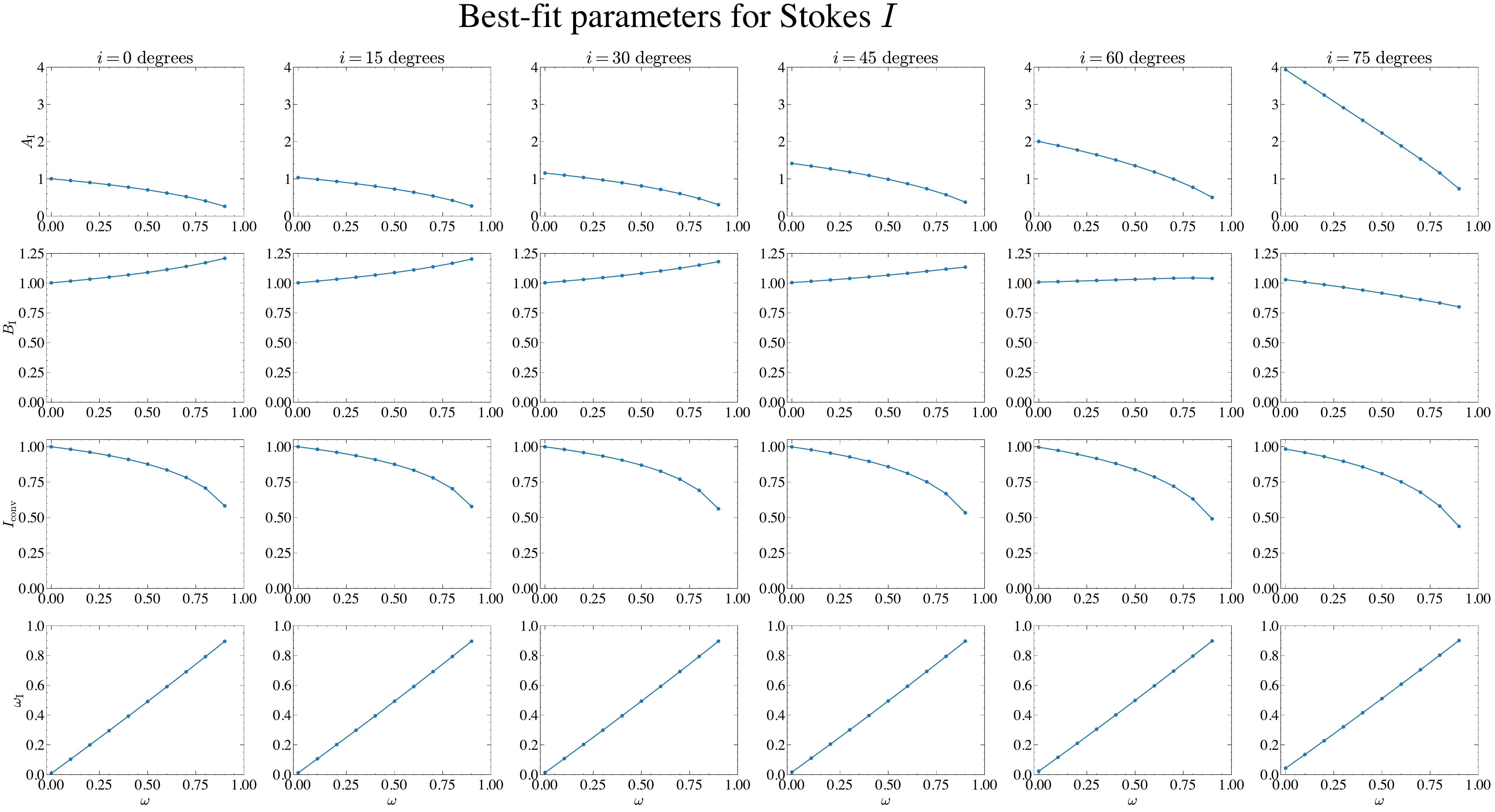}}
\caption{Dependence on albedo $\omega$ of the best-fit parameters obtained by fitting Eq. (\ref{eq:fitting_I}) to the numerically computed emergent Stokes $I$ for a plane-parallel slab inclined at $0\tcdegree, 15\tcdegree, 30\tcdegree, 45\tcdegree, 60\tcdegree$ and $75\tcdegree$.}
\label{fig:Fitting_intensity_params_addinc}
\end{figure*}

Fig. \ref{fig:Fitting_intensity_params_addinc} shows the albedo $\omega$ dependence of the best-fit parameters obtained by fitting Eq. (\ref{eq:fitting_I}) to our numerically computed emergent intensities for plane-parallel slabs at inclinations $i=0\tcdegree, 15\tcdegree, 30\tcdegree, 45\tcdegree, 60\tcdegree$ and $75\tcdegree$.
\begin{figure*}[htbp]
\resizebox{\hsize}{!}{\includegraphics{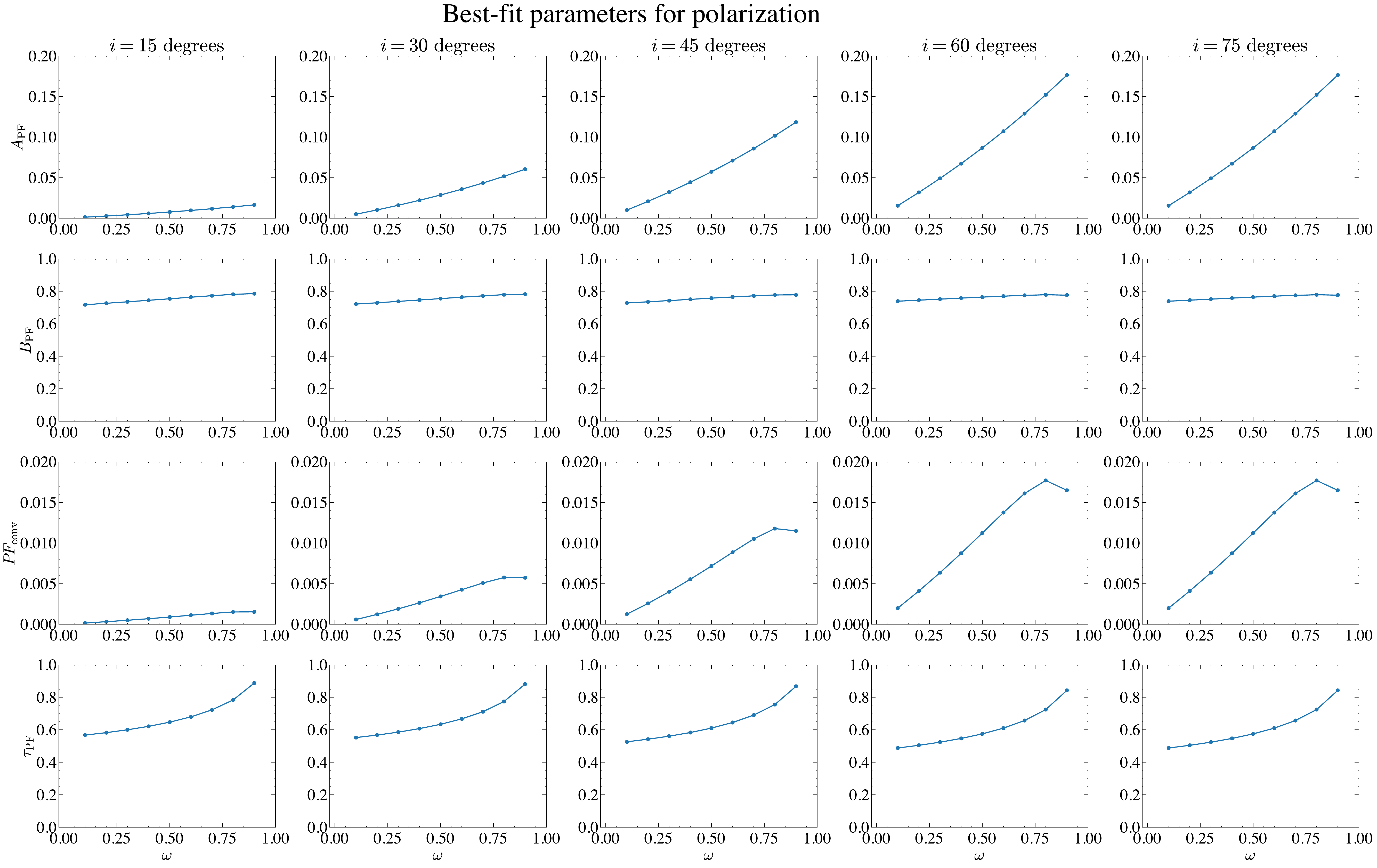}}
\caption{Dependence on albedo $\omega$ of the best-fit parameters obtained by fitting Eq. (\ref{eq:fitting_PF}) to the numerically computed emergent polarization for a plane-parallel slab inclined at $15\tcdegree, 30\tcdegree, 45\tcdegree, 60\tcdegree$ and $75\tcdegree$.}
\label{fig:Fitting_polarization_params_addinc}
\end{figure*}
Fig. \ref{fig:Fitting_polarization_params_addinc} shows, analogously to the intensity case, the $\omega$-dependence of the best-fit parameters obtained by fitting Eq. (\ref{eq:fitting_PF}) to our numerically computed emergent polarization for plane-parallel slabs at inclinations $i=15\tcdegree, 30\tcdegree, 45 \tcdegree, 60\tcdegree$, and $75\tcdegree$. Results for $i=0\tcdegree$ are omitted because no scattering-induced polarization is produced at face-on viewing.

\section{$\tau_\mathrm{peak}$ for other inclinations} \label{Appendix:AddInc_taupeak}

\begin{figure*}[htbp]
\resizebox{\hsize}{!}{\includegraphics{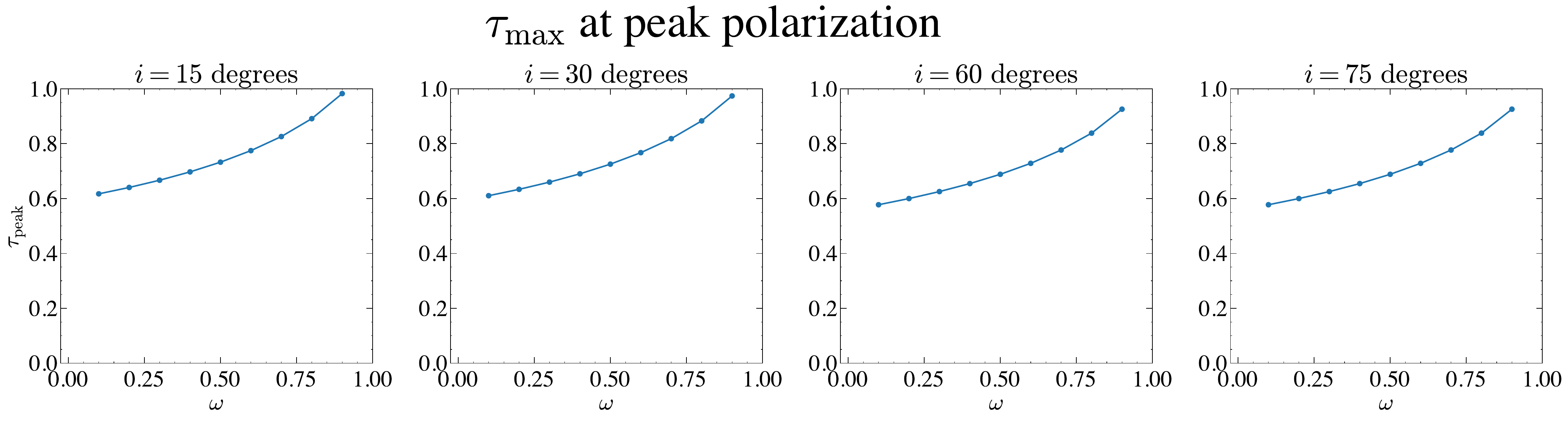}}
\caption{$\omega$-dependence of the optical depth $\tau_\mathrm{peak}$ at which the polarization fraction peaks for plane-parallel slabs at $i=15\tcdegree, 30\tcdegree, 60\tcdegree$, and $75\tcdegree$, computed using the fitting parameters from Fig. \ref{fig:Fitting_polarization_params_addinc}.}
\label{fig:Polarization_peak_addinc}
\end{figure*}

Using the fitting parameters from Fig. \ref{fig:Fitting_polarization_params_addinc}, we compute the optical depth $\tau_\mathrm{peak}$ at which the polarization fraction peaks for inclinations $i=15\tcdegree, 30\tcdegree, 60\tcdegree$, and $75\tcdegree$; the results are shown in Fig. \ref{fig:Polarization_peak_addinc}. $\tau_\mathrm{peak}$ exhibits little dependence on inclination.

\section{Estimating Mie-Scattering disk emission for other inclinations} \label{Appendix:AddInc_Mie}

\begin{figure*}[htbp]
\resizebox{\hsize}{!}{\includegraphics{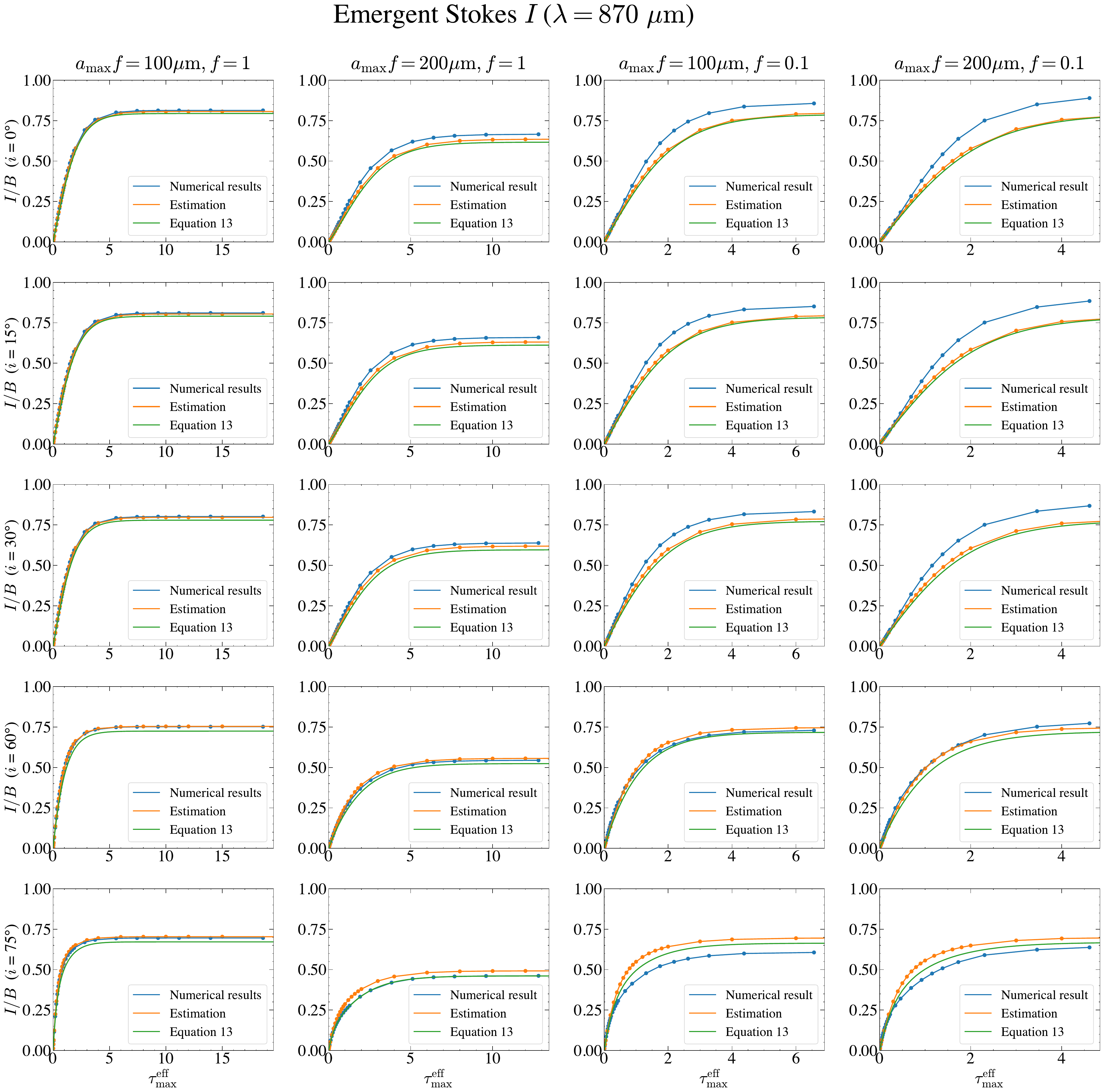}}
\caption{Comparison, for the four dust models, between the numerically computed emergent Stokes $I$ including Mie scattering and two estimates based on the effective scattering opacity, for $\lambda=870~\mathrm{\mu m}$ and $i=0\tcdegree, 15\tcdegree, 30\tcdegree, 60\tcdegree$ and $75\tcdegree$. The first estimate (orange lines) is obtained from our Rayleigh-scattering calculations using $\omega=\omega_\mathrm{eff}$ and $\tau_\mathrm{max}=\tau_\mathrm{max}^\mathrm{eff}$. The second estimate (green lines) is obtained from Eq. \ref{eq:CG19} using $\omega=\omega_\mathrm{eff}$ and $\tau_\mathrm{max}=\tau_\mathrm{max}^\mathrm{eff}$.}
\label{fig:DSHARP_vs_Rayleigh_I_addinc}
\end{figure*}

Fig. \ref{fig:DSHARP_vs_Rayleigh_I_addinc} compares, for the four dust models and for various inclinations, the numerically computed emergent Stokes $I$ including Mie scattering with two estimates based on the effective scattering opacity. The first estimate (orange lines) is obtained from our Rayleigh-scattering calculations using $\omega=\omega_\mathrm{eff}$ and $\tau_\mathrm{max}=\tau_\mathrm{max}^\mathrm{eff}$. The second estimate (green lines) is obtained from Eq. \ref{eq:CG19} using $\omega=\omega_\mathrm{eff}$ and $\tau_\mathrm{max}=\tau_\mathrm{max}^\mathrm{eff}$.
Figs. \ref{fig:DSHARP_vs_Rayleigh_I} and \ref{fig:DSHARP_vs_Rayleigh_I_addinc} show that the estimate obtained from our Rayleigh-scattering calculations using the effective scattering opacity reproduces the Mie-scattering results less accurately at low inclinations ($\sim 0\tcdegree$) and high inclinations ($\sim 75\tcdegree$) than at intermediate inclinations ($\sim 45 \tcdegree$). Therefore, when $i$ is close to $0\tcdegree$ or $75\tcdegree$, this estimate may not be accurate enough to be applied to ALMA data analysis.

In addition, Fig. \ref{fig:DSHARP_vs_Rayleigh_I_addinc} shows that, at high inclinations, the estimate based on Eq. \ref{eq:CG19} can reproduce the Mie-scattering results more accurately than the estimate based on our Rayleigh-scattering calculations. Such behavior is seen, for example, for the dust models with $a_\mathrm{max}f = 100 ~\mathrm{\mu m}$  and $200~\mathrm{\mu m}$, $f=0.1$, at $i=75\tcdegree$.

\begin{figure*}[htbp]
\resizebox{\hsize}{!}{\includegraphics{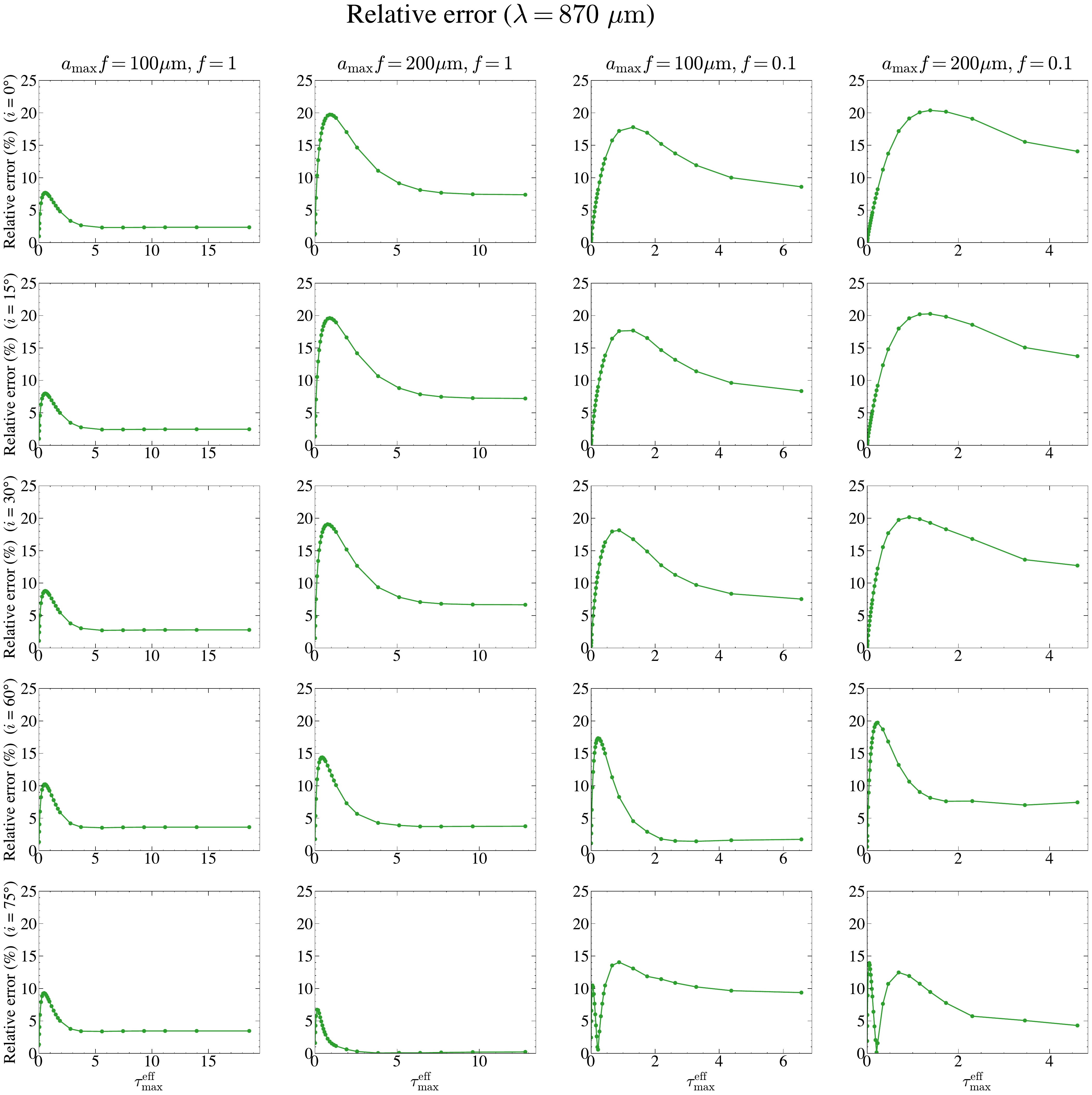}}
\caption{Relative errors, for the four dust models at $i=0\tcdegree$, $15\tcdegree$, $30\tcdegree$, $60\tcdegree$, and $75\tcdegree$, of the estimate obtained from Eq. \ref{eq:CG19} with $\omega=\omega_\mathrm{eff}$ and $\tau_\mathrm{max}=\tau_\mathrm{max}^\mathrm{eff}$ ($I_{\mathrm{Eq.} \ref{eq:CG19}}^\mathrm{eff}$) with respect to the numerically computed emergent Stokes $I$ including Mie scattering ($I_\mathrm{Mie}$), defined as $|(I_\mathrm{Mie} - I_{\mathrm{Eq.} \ref{eq:CG19}}^\mathrm{eff})/I_\mathrm{Mie}|$.}
\label{fig:DSHARP_vs_Rayleigh_I_error_addinc}
\end{figure*}

Fig. \ref{fig:DSHARP_vs_Rayleigh_I_error_addinc} shows the relative errors, for the four dust models at $i=0\tcdegree$, $15\tcdegree$, $30\tcdegree$, $60\tcdegree$, and $75\tcdegree$, of the estimate obtained from Eq. \ref{eq:CG19} with $\omega=\omega_\mathrm{eff}$ and $\tau_\mathrm{max}=\tau_\mathrm{max}^\mathrm{eff}$ with respect to the numerically computed emergent Stokes $I$ including Mie scattering. In many cases, the relative error exceeds $10\%$.



\end{appendix}

\end{document}